\documentclass{sig-alternate}

\usepackage{multirow}
\usepackage{latexsym}
\usepackage{epsfig}
\usepackage{color}
\usepackage{algorithm,algorithmic}
\usepackage{times}
\usepackage{subfigure}
\usepackage{amsmath}
\usepackage{mdwlist}

\newcommand{\rw}[2]{#2}
\newcommand{\x}[1]{}

\newcommand{\comment}[1]{}

\newsavebox{\boxone}
\newsavebox{\boxtwo}
\newsavebox{\boxthree}

\newlength{\narrow}
\newlength{\cnarrow}
\newcommand{\topline}{
  \hrule
  \vskip .5\baselineskip}
\newcommand{\bottomline}{
  \vskip 2pt
  \hrule}

\newcommand{\chbox}[2]{
  \hbox to #1{\hss\vtop{#2}\hss}}

\newcommand{\nchbox}[1]{
  \chbox{\narrow}{#1}}

\newcommand{\cnchbox}[1]{
  \chbox{\cnarrow}{#1}}

\newcommand{\fcode}[1]{
  
  \chbox{\textwidth}{\tgrind\input{#1}}}

\newcommand{\ncode}[1]{
  
  \chbox{\narrow}{\tgrind\input{#1}}}

\def\nfig#1#2#3{
  \vtop{\nchbox{#1}
  \hbox to\narrow{\parbox{\narrow}{\caption{#2}\label{#3}}}}}

\newcommand{\cncode}[1]{
  \chbox{\cnarrow}{\tgrind\input{#1}}}

\def\codefiggen[#1]#2#3#4#5#6{
  \begin{figure}[#1]
  #5
  \fcode{#2}
  \center\parbox{.9\textwidth}{\caption{#3}\label{#4}}
  #6
  \end{figure}}

\def\codefig[#1]#2#3#4{
  \codefiggen[#1]{#2}{#3}{#4}{}{}}

\def\codefigline[#1]#2#3#4{
  \codefiggen[#1]{#2}{#3}{#4}{\topline}{\bottomline}}

\def\doublefiggen[#1]#2#3#4#5#6#7#8#9{
  \begin{figure}[#1]
  #8
  \hbox to \textwidth{
  \nfig{#2}{#3}{#4}
  \hfil
  \nfig{#5}{#6}{#7}}
  #9
  \end{figure}}

\def\doublefig[#1]#2#3#4#5#6#7{
  \doublefiggen[#1]{#2}{#3}{#4}{#5}{#6}{#7}{}{}}

\def\doublefigline[#1]#2#3#4#5#6#7{
  \doublefiggen[#1]{#2}{#3}{#4}{#5}{#6}{#7}{\topline}{\bottomline}}

\def\doublecodefig[#1]#2#3#4#5#6#7{
  \doublefig[#1]{\ncode{#2}}{#3}{#4}{\ncode{#5}}{#6}{#7}}

\def\doublecodefigline[#1]#2#3#4#5#6#7{
  \doublefigline[#1]{\ncode{#2}}{#3}{#4}{\ncode{#5}}{#6}{#7}}

\newcommand{\codepair}[4]{\vbox{
  \hbox{\ncode{#1} \hfil \ncode{#3}}
  \vskip .3\baselineskip plus .3\baselineskip
  \hbox{\hbox to\narrow{#2\hfil} \hfil \hbox to\narrow{#4\hfil}}}}

\def\codepairfig[#1]#2#3#4#5#6#7{
  \begin{figure}[#1]
  \codepair{#2}{#3}{#4}{#5}
  \center\parbox{.9\textwidth}{\caption{#6}}
  \label{#7}
  \end{figure}}

\def\cncodepairfiggen[#1]#2#3#4#5#6#7{
  \begin{figure}[#1]
  #6
  \hbox{\cncode{#2}\hfil\cncode{#3}}
  \center\parbox{.9\columnwidth}{\caption{#4}\label{#5}}
  #7
  \end{figure}}

\def\cncodepairfig[#1]#2#3#4#5{
  \cncodepairfiggen[#1]{#2}{#3}{#4}{#5}{}{}}

\def\cncodepairfigline[#1]#2#3#4#5{
  \cncodepairfiggen[#1]{#2}{#3}{#4}{#5}{\topline}{\bottomline}}

\def\doublefigOnecap*[#1]#2#3#4#5{
  \begin{figure*}[#1]
  \hbox to \textwidth{
  \nchbox{#2}
  \hfil
  \nchbox{#3}}
  \caption{#4}
  \label{#5}
  \end{figure*}}

\def\doublefigOnecap[#1]#2#3#4#5{
  \begin{figure}[#1]
  \topline
  \hbox to \columnwidth{
  \cnchbox{#2}
  \hfil
  \cnchbox{#3}}
  \caption{#4}
  \label{#5}
  \bottomline
  \end{figure}}

\def\PSfig[#1]#2#3#4{
 \begin{figure}
 \centerline{\psfig{file=#2,width=\columnwidth}}
 \caption{{#3}} 
 \label{#4}
 \end{figure}}

\def\PSfiglines[#1]#2#3#4{
 \begin{figure}[#1]
 \topline
 \centerline{\psfig{file=#2,width=\columnwidth}}
 \caption{{#3}} 
 \label{#4}
 \bottomline
 \end{figure}}

\def\PSfiglinesht[#1]#2#3#4#5{
 \begin{figure}[#1]
 \topline
 \centerline{\psfig{file=#2,height=#3}}
 \caption{{#4}} 
 \label{#5}
 \bottomline
 \end{figure}}

\def\doublePSfig[#1]#2#3#4#5#6{
  \doublefigOnecap[#1]
    {\cnchbox{\psfig{file=#2,height=#4}}}
    {\cnchbox{\psfig{file=#3,height=#4}}}
    {#5}
    {#6}}

\newlength{\boxwidth}
\newcommand{\bproof}{{\bf Proof: \ }}

\newcommand{\eproof}{\mbox{$\Box$}}

\def\tabdoublecode#1#2#3#4{
 \begin{figure*}[t]
 \topline\vs{-.4}
 \hbox to \columnwidth{
 \vtop{\small
 \begin{tabbing}
 #1
 \end{tabbing}}
 \hfil
 \hfil
 \hfil
 \vtop{\small
 \begin{tabbing}
 #2
 \end{tabbing}}
 }
 \caption{#3\label{#4}}
 \bottomline
 \end{figure*}
}
\def\tabtriplecode#1#2#3#4#5{
 \begin{figure}
 \topline\vs{-.4}
 \hbox to \columnwidth{
 \vtop{\small
 \begin{tabbing}
 #1
 \end{tabbing}}
 \hfil
 \hfil
 \hfil
 \vtop{\small
 \begin{tabbing}
 #2
 \end{tabbing}}
 \hfil
 \hfil
 \hfil
 \vtop{\small
 \begin{tabbing}
 #3
 \end{tabbing}}
 }
 \caption{#4\label{#5}}
 \bottomline
 \end{figure}
}

\newtheorem{lemma}{Lemma}
\newcommand{\blemma}{\begin{lemma}}
\newcommand{\elemma}{\end{lemma}}

\newtheorem{thm}{Theorem}
\newcommand{\bthm}{\begin{thm}}
\newcommand{\ethm}{\end{thm}}

\newtheorem{defin}{Definition}
\newcommand{\bdefin}{\begin{defin}}
\newcommand{\edefin}{\end{defin}}

\newtheorem{cor}{Corollary}
\newcommand{\bcor}{\begin{cor}}
\newcommand{\ecor}{\end{cor}}

\newcommand{\vs}[1]{\vspace{#1cm}}
\newcommand{\be}{\begin{equation}}
\newcommand{\ee}{\end{equation}}

\newcommand{\bdesc}{\begin{description}}
\newcommand{\edesc}{\end{description}}
\newcommand{\benum}{\begin{enumerate}}
\newcommand{\eenum}{\end{enumerate}}
\newcommand{\bitem}{\begin{itemize}}
\newcommand{\eitem}{\end{itemize}}
\newcommand{\bcenter}{\begin{center}}
\newcommand{\ecenter}{\end{center}}
\newcommand{\btabular}{\begin{tabular}}
\newcommand{\etabular}{\end{tabular}}
\newcommand{\beqnarr}{
 \begin{eqnarray}}
\newcommand{\eeqnarr}{\end{eqnarray}}

\title {Axiomatic Ranking of Network Role Similarity\titlenote{A revised version of this paper will be published for KDD'11, August 2011.}}

\author{
\alignauthor
Ruoming Jin~~~~Victor E. Lee~~~~Hui Hong\\
       \affaddr{\mbox{ }Department of Computer Science}\\
       \affaddr{Kent State University, Kent, OH, 44242, USA}\\
       \email{\{jin,vlee,hhong\}@cs.kent.edu}
}

\begin{document}
\textwidth      6.85in
\topmargin      -0.60in
\textheight     9.35in

\hyphenation{sub-trajectory}
\hyphenpenalty=100
\tolerance=9000
\maketitle

\begin{abstract}

A key task in social network and other complex network analysis is role analysis: describing and categorizing nodes according to how they interact with other nodes.  Two nodes have the same role if they interact with {\em equivalent} sets of neighbors.
The most fundamental role equivalence is automorphic equivalence.  Unfortunately, the fastest algorithms known for graph automorphism are nonpolynomial. Moreover, since exact equivalence may be rare, a more meaningful task is to measure the role {\em similarity} between any two nodes.  This task is closely related to the structural or link-based similarity problem that SimRank attempts to solve. However, SimRank and most of its offshoots are not sufficient because they do not fully recognize automorphically or structurally equivalent nodes.
In this paper we tackle two problems.  First, what are the necessary properties for a role similarity measure or metric? Second, how can we derive a role similarity measure satisfying these properties? For the first problem, we justify several axiomatic properties necessary for a role similarity measure or metric: range, maximal similarity, automorphic equivalence, transitive similarity, and the triangle inequality.  For the second problem, we present RoleSim, a new similarity metric with a simple iterative computational method.  We rigorously prove that RoleSim satisfies all the axiomatic properties.  
We also introduce an iceberg RoleSim algorithm which can guarantee to discover all pairs with RoleSim score no less than a user-defined threshold $\theta$ without computing the RoleSim for every pair. 
We demonstrate the superior interpretative power of RoleSim on both both synthetic and real datasets.

\comment{ 
In the study of social networks, role analysis is describing and categorizing nodes according to how they interact with other nodes.  Two nodes have the same role if they interact with equivalent sets of neighbors.
The most fundamental role equivalence is automorphic equivalence.  Unfortunately, the fastest algorithms known for graph automorphism are nonpolynomial. More, since exact equivalence may be rare, a more meaningful task is to measure the role {\em similarity} between any two nodes.  This task is closely related to the structural or link-based similarity problem that SimRank attempts to solve. However, SimRank and most of its offshoots are problematic because they do not fully recognize automorphically or structurally equivalent nodes.  SimRank, then, is not a role similarity measure.
In this paper we tackle two problems.  First, what are the necessary properties for a role similarity measure or metric?  Second, how can we derive a role similarity measure satisfying these properties? For the first problem, we justify several axiomatic properties necessary for a role similarity measure or metric - range, maximal similarity, automorphic equivalence, transitive similarity, and the triangle inequality.  For the second problem, we present RoleSim, a new similarity metric with a simple iterative computational method.  We rigorously prove that RoleSim satisfies all the axiomatic properties.  We demonstrate the interpretative power of an axiomatic similarity metric by analyzing real datasets.
}

\comment{
A powerful way to classify vertices in a network is according to the similarity of their connections to others, or their role similarity.  In social network analysis, the role of an entity is its set of interactions with others.  While many role similarity measures have been offered, some of them operate counter to our intuitive idea of similarity.  The fundamental problem is twofold: there are multiple definitions of role eqivalence, and there is no universal understanding of what is means for two vertices to be more similar. To address this problem, we present a universal framework for role similarity.  First, we define a generalized equivalence, adjustable to fit common variations of meaning.  Second, we present an axiomatic definition of role similarity, which we use to evaluate some of the well-known metrics.  Third, we present a new generalized role similarity metric which satisfies the axiomatic definition.  We evaluate our new metric on several real datasets to demonstate its increased effectiveness at identifying roles in a network.
}
\end{abstract}



\section{Introduction}
\label{section:intro}

\comment{
Over the last decade, network science has developed into a rapidly growing interdisciplinary field studying complex systems comprised of multiple interacting elements.
Everything from biocellular, ecological, and neurological systems, to the World Wide Web, social networks, and economical and financial markets can be expressed and analyzed as complex networks.
Network or graph representation enables scientists to study these diverse systems in a unified framework: each system is modeled as a large set of elements (nodes) joined by non-trivial relationships (edges).
An array of tools ranging from statistical modeling, to motif discovery, to network clustering, among many others, have been developed to help scientists analyze these complex networks. 
Data mining has also contributed significantly to the network science by designing efficient and scalable frequent subgraph mining and clustering algorithms, and analyzing/modeling  large dynamic social networks, etc~\cite{}. 
}

\comment{
Network plays an increasingly important role in understanding the ubiquitous systems, ranging from biocellular, ecological, to social and economical domains. 
In such systems, the interactions (edges) between elements (nodes) enables its diverse functionality and behaviors. 
A particular important 
Understanding patterns of interaction of each individual node and their roles is of particular importance in euclidating the relationship between the network structure to system behaviors. 
Specifically, the basic question is: {\em ``How to characterize a node's pattern of interaction with its surroundings, and how similar it is the interaction pattern of other nodes?''}
In particular, {\em as each node's role in a network is defined based on who they interact with,  
how can we rank two nodes' role similarity in terms of their interaction patterns?} 
}

In social science, it is well-established that individual agents tend to play roles or assume positions within their interaction network.
For instance, in a university, each individual can be classified into the position of faculty member, administration, staff, or student. Each role may be further partitioned into sub-roles: faculty may be further classified into tenure-track or non-tenure-track positions, etc. 
Indeed, role discovering is a major research subject in classical social science~\cite{wf94}. 
Interestingly, recent studies have found not only do roles appear in other types of networks, including food webs~\cite{LBE03},
\rw{cellular networks~\cite{HolmeHuss05}, world trade~\cite{HolmeHuss05},}
{world trade~\cite{hafner2009network},}
and even software systems~\cite{DBLP:conf/icsm/DraganCM09}, but also roles can help predict node functionality within their domains. 
For instance, in a protein interaction network, proteins with similar roles tend to serve similar metabolic functions. Thus, if we know the function of one protein, we can predict that all other proteins having a similar role would also have similar function~\cite{HolmeHuss05}. 

Role is complementary to network clustering, a major tool in analyzing network structures.  
Network clustering attempts to decompose a network into densely connected components.
It produces a high level structural model consisting of a small number of ``cluster-nodes'' and the ``super-edges'' between these cluster-nodes.
Since its goal is to minimize the number of edges (interactions) between clusters, it will result in strong interactions between nodes within each cluster.
Given this, the clustering scheme inevitably overlooks and over-simplifies the interaction patterns of each node. 
For instance, each node in a cluster may take very different ``roles'': some of them may serve as the core of the clusters, some may be peripheral nodes, and some serve as the connectors to link between clusters.  
Indeed, those nodes with similar or same roles may not even directly link to each other as they may simply share similar interaction patterns. 
Furthermore, even when a network lacks modularity structure, for instance, a hierarchical structure, roles can still be applied for characterizing the interaction patterns of each node. 
To sum, ``roles'' provide an orthogonal abstraction for simplifying and highlighting the complex interactions among nodes.

A central question in studying the roles in a network system is how to define {\em role similarity}. 
In particular,
how can we rank two nodes' role similarity in terms of their interaction patterns?
Despite its vital importance for network analysis and decades of work by social scientists, joined recently by computer scientists, no satisfactory metric for role similarity has yet emerged. 
A key issue is the encapsulation of graph automorphism (and its generalization) into a role similarity metric:  {\em if two nodes are automorphically equivalent, then they should share the same role and their role similarity should be maximal}.  
From a network topology viewpoint, automorphic nodes have equivalent surroundings, so one can replace the other.
Figure~\ref{fig:equiv} illustrates a graph with nodes $S1$ and $J1$ being automorphically equivalent.
Automorphism can be further generalized in terms of {\em coloration}: assuming each node is assigned a color, then two nodes are equivalent if their neighborhoods consist of the same color spectrum~\cite{EverettBorgatti94}. 

Traditionally, the social science community has approached role analysis by defining suitable mathematical equivalence relations so that nodes can be partitioned into equivalence classes (roles).
An essential property of these equivalences is that they should positively confirm automorphic equivalence, i.e., if any two nodes are automorphic, then they are role-equivalent. (The converse is not necessarily true.)
Automorphism confirmation is an instance of verifying a solution, which is often algorithmically less complex than discovering a solution.
Therefore, even though there is no known polynomial-time algorithm for discovering graph automorphism~\footnote{The computational complexity of graph isomorphism and automorphism are still unproven to be either $P$ or $NP-Complete$.},
role equivalence algorithms~\cite{BDF92,borgatti93_alg_reg_equiv,Sparrow93_linearequiv} can still guarantee to satisfy the aforementioned automorphism confirmation property. 
These equivalence rules also directly correspond to the aforementioned coloration. 

However, by relying on strict equivalence rules, these role modeling schemes can produce only binary similarity metrics: two nodes are either equivalent (similarity $= 1$) or not (similarity $= 0$).
In real-world networks, usually only a very small portion of the node-pairs would satisfy an equivalence criteria~\cite{MacArthurSRA08} and among those, many are simply trivially equivalent (such as singletons or children of the same parent).   
In addition, strict rule-based equivalence is not robust with respect to network noise, such as false-positive or false-negative interactions. 
Thus, it is desirable in many real world applications to rank node-pairs by their degree of similarity or provide a real-valued node similarity {\em metric}.  

Several recent research works have proposed to measure real-valued structural similarity or to rank nodes' similarity based on their interaction patterns~\cite{Jeh02_simrank,Leicht06_vertexsim}.
SimRank~\cite{Jeh02_simrank} is one of the best-known such measures. 
It generates a node similarity measure based on the following principle: ``two nodes are similar if they link to similar nodes''. 
Mathematically, for any two different nodes $x$ and $y$, SimRank computes their similarity recursively according to the average similarity of all the neighbor pairs (a neighbor of $x$ paired with a neighbor of $y$). A single node has self-similarity value $1$.  This is equivalent to the probability that two simultanous random walkers, starting at $x$ and $y$, will eventually meet.  
Most of the existing node structural similarity measures~\cite{Antonellis08_simrankpp,Fogaras05_scaleSim,Li09_blocksim,Xi05_simfusion,Yin06_simtree,Zhao09_prank} are variants of SimRank. 
Though SimRank seems to capture the intuition of the above recursive structural similarity, its random walk matching does not satisfy the basic graph automorphism condition.  For example, in Figure~\ref{fig:equiv}, though $S1$ and $J1$ are automorphically equivalent, SimRank assigns them a value of 0.226.  We discuss this further in Section~\ref{lab:simrank_not}.
\comment{
~\footnote{Such discrepancy or flaw for the automorphic equivalent nodes in SimRank is known~\cite{}. 
However, the paradox that 
}.}
To our best knowledge, there is no available real-valued structural similarity measure satisfying the automorphic equivalence requirement. 
Since automorphic equivalence is a pivotal characteristic of the notion of role, its lack disqualifies these existing measures from serving as authentic role similarity measures. 
Here is a paradox: SimRank and its variants seem to implement the recursive structural similarity definition of automorphic equivalence (two nodes are similar if they link to similar nodes), yet they do not produce desired results (to assign value $1$ to those pairs).

Thus we have an open problem: {\em Can we derive a real-valued role similarity measure or ranking which complies with the automorphic equivalence requirement? }
In this paper, we develop the first real-valued similarity measure to solve this problem.
In addition, our measure is also a metric, i.e., it satisfies the triangle inequality. 
The key feature of our role similarity measure is a weighted generalization of the {\em Jaccard coefficient}
to measure the neighborhood similarity between two nodes. 
Unlike SimRank, which considers the average similarity among all possible pairings of neighbors, our measure considers only those pairs in the optimal matching of their two neighbor sets which maximizes the targeted similarity function. 
We show this approach successfully resolves the aforementioned SimRank paradox.



\comment{
The paper is organized as follows: 
Section~\ref{section:equivalence} reviews basic concepts of role equivalence. 
Section~\ref{section:similarity} presents axiomatic properties of any real-valued similarity measure, including a requirement for automorphic equivalence.
We also show that SimRank does not satisfy the automorphic equivalence requirement. 
Section~\ref{section:algorithm} describes our RoleSim measure, its computation, and its correctness with respect to the axiomatic properties. 
Section~\ref{section:experiment} provides experimental validation and evaluation of our proposed approach for ranking the the role similarity between vertex pairs. 
}

\comment{

Since exact equivalence may be rare, a more meaningful task is to rate role similarity between any two nodes.  This task is essentially the same as the structural similarity or link-based similarity problem that SimRank attempts to solve.
However, SimRank and most of its offshoots are problematic because they do not compute full similarity even when two nodes have identical sets of neighbors.  SimRank is not a role similarity measure.
In this paper we tackle two problems.  First, what are the necessary properties for a role similarity measure?  Second, how can we efficiently compute automorphic equivalence and similarity?}

\comment{
In a nut-shell, the

Discovering roles is vital for many applications.  In many business sectors, knowing how customers interact with one another is valuable marketing information.  For some communications-based businesses, the customer network {\em is} the business activity.  For example, for a telephone company, we can construct a network showing who calls whom.  Here, a role defines how a customer uses the network to call others, equivalent to a market segment. Identifying meaningful market segments and partitioning its users into these segments helps the company determine how best to price its services, to develop special offers and marketing campaigns, and to see where expansion or contraction of resources is indicated.

Role analysis can also be applied to non-social networks. In biology, different proteins can fullfil many roles within the overall functioning of an organism.  Knowing a protein's role is not only of academic interest, but it is also invaluable for understanding disorders and developing therapeutic drugs. However, it is easier to isolate and identify a protein than to discover its function.  By analyzing and comparing protein-protein interaction networks, we can gather evidence about a protein's role.

Even software engineers are seeing the value of role analysis.  In large open source development projects, the source code is written by many different authors who work independently.  Different authors may use different naming conventions for variables and objects, but some of these independently authored objects in fact perform similar functions.  Using role analysis, can we partition code into blocks of similar role?

Traditionally, the social science community has approached role analysis by first defining a suitable equivalence relation, which can be used to identify nodes that serve identical roles.  Several different equivalence relations are possible.  One of most useful is automorphic equivalence, which is related to the graph automorphism problem.  Since there is no known polynomial-time algorithm for finding a graph automorphism, solving this exactly for large networks is intractable.  It is therefore necessary to consider heuristic methods.

Since exact equivalence may be rare, a more meaningful task is to rate role similarity between any two nodes.  This task is essentially the same as the structural similarity or link-based similarity problem that SimRank attempts to solve.
However, SimRank and most of its offshoots are problematic because they do not compute full similarity even when two nodes have identical sets of neighbors.  SimRank is not a role similarity measure.
In this paper we tackle two problems.  First, what are the necessary properties for a role similarity measure?  Second, how can we efficiently compute automorphic equivalence and similarity?

Our contributions are as follows:
\begin{enumerate}
\item We propose a unified framework for describing roles and positions, using the notion of a neighbor color spectrum, which encompasses all the commonly-used forms of role equivalence.\\
\item We define a theoretically rigorous set of properties that any role similarity measure or algorithm should satisfy, based on equivalence classes and distance metrics.  We show that random walk-based similarity algorithms, including SimRank, do not satisfy these properties.\\
\item We present RoleSim, a simple algorithm for computing equivalence-compliant similarity for all node-pairs in a network.  The algorithm is adjustable to work with different ideas of dissimilarity.
\item We compare the similarity results of RoleSim with other methods for role similarity.  We also compare the results and analyze the effectiveness when different initialization schemes are employed.
\end{enumerate}

The rest of the paper is organized as follows.
Section 2 reviews the social network concepts of roles and role equivalences. In Section 3 we construct the necessary and desireable properties of a role similarity measure or algorithm.    Section 4 presents the RoleSim algorithm.  In Section 5 we describe our experimental validation methods and results.  



}

\section{Role Equivalence}
\label{section:equivalence}


\comment{ Old intro and perspective
In this section, we develop a general yet precise definition of structural role in a network based on the idea of a color spectrum.  This definition has a natural social interpretation and is easily represented in mathematical terms.  Our framework encompasses all the standard definitions of role equivalence, as well as a suggesting a new variation.

Social network analysis is a subfield of sociology where graph theory, matrix algebra, and other mathematical methods are used to study networks of social relationships.  Strictly speaking, {\em role} is the set of interactions (edges) of an actor (node or vertex), whereas a {\em position} is the set of all actors (vertices) who have the same or similar roles~\cite{wf94_role}. However, for simplicity, we will use role to refer to both a vertex's set of interactions (colored edges) and a group of similarly situated vertices, where the distinction should be clear from context.
} 

\begin{figure}[t]
\centering
\epsfig{file=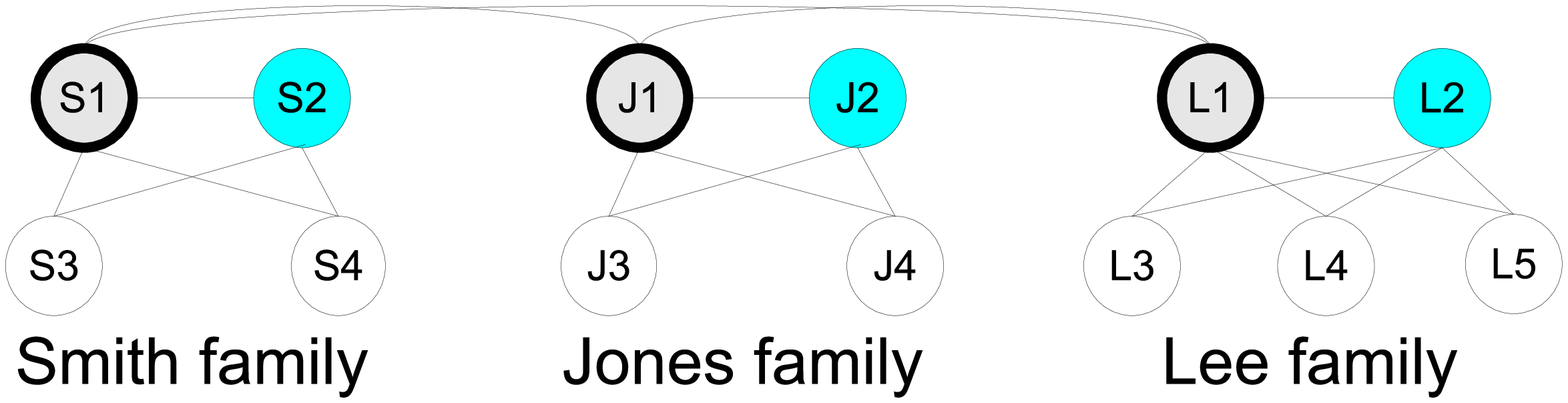,width=3.3in, height = 0.9in}
\caption{Example Graph for Equivalence Classes.} \label{fig:equiv}
\end{figure}

In social network analysis, the traditional approach for formalizing roles and role groups is to define a equivalence relation and to partition the actors into equivalence classes. Actors who fulfill the same role are equivalent. Over the years, four definitions, offering different degrees of strictness, have stood out.  These four, in decreasing strictness order, are structural equivalence, automorphic equivalence, equitable partition, and regular equivalence. Figure~\ref{fig:equiv} shows how these different definitions generate different roles from the same network. 

Let $G = (V, E)$ be a graph with vertex set $V = \{v_1,...,v_n\}$ and edge set $E$.
For any node $v \in V$, let $N(v)$ be the neighbors of $v$ and $N_v$ be the degree of $v$.\\

\begin{table}[t]
\centering
\small
\begin{tabular}{|p{.65in}|p{.74in}|p{1.48in}|} \hline
{\bf Equivalence} & {\bf Neigh. Rule} & {\bf Non-singleton Classes}  \\ \hline
Structural  & exactly same
						& \{S3,S4\},\{J3,J4\}, \{L3,L4,L5\} \\ \hline
Automorphic, Exact Color. & same number per class			
						& \{S1,J1\},\{S2,J2\}, \{S3,S4,J3,J4\}, \{L3,L4,L5\} \\ \hline
Regular     & same class
						& \{S1,J1,L1\},\{S2,J2,L2\}, \{S3,S4,J3,J4,L3,L4,L5\} \\ \hline
\end{tabular}
\caption{Equivalence Classes for Figure~\ref{fig:equiv}} \label{tab_equiv}
\end{table}

\vspace{-1.0ex}

\noindent{\bf Structural Equivalence:}
Two actors are {\em structurally equivalent} if they interact with the {\em same} set of others~\cite{Lorrain71}.  
Mathematically, $u$ and $v$ are structurally equivalent if and only if $N(u)=N(v)$.
For example, consider the extended family shown in Figure~\ref{fig:equiv}.  $S1$, $J1$, and $L1$ are siblings, $S2$, $J2$, and $L2$ are spouses, and the remaining nodes are their children. Each family's children, $\{S3,S4\}$, $\{J3,J4\}$, and $\{L3,L4,L5\}$ form a nontrivial equivalence class.  However, none of the parents can be grouped together via structural equivalence.
This equivalence model is too strict to be useful for simplifying a large network and to discover meaningful roles.  

\noindent{\bf Automorphic Equivalence:}
Two actors (nodes) $u$ and $v$ are {\em automorphically equivalent} if there is an automorphism $\sigma$ of $G$ such that $v=\sigma(u)$ ~\cite{Borgatti92_position}.
An automorphism $\sigma$ of a graph $G$ is a permutation of vertex set $V$ such that for any two nodes $u$ and $v$, $(u,v) \in E$ iff $(\sigma(u),\sigma(v)) \in E$.  In social terms, $u$ and $v$ can swap names, along with possibly some other name swaps, while preserving all the actor-actor relationships.
Let $\Gamma(G)$ be the group of all automorphisms of graph $G$. For any two nodes $u$ and $v$ in $G$, $u \equiv v$ if $u=\sigma(v)$ for some $\sigma \in \Gamma(G)$. 
Note that $\equiv$ is an equivalence relation on $V$; if $u \equiv v$ we say that $u$ is automorphically equivalent to $v$. 
The equivalence classes generated under $\Gamma(G)$ (or $\equiv$) are called orbits. The equivalence class for vertex $v \in V$ is called the orbit of $v$, and denoted as $
\Delta(v)=\{ \sigma(v) \in V, \sigma \in \Gamma(G)\}=\{u | u \equiv v\}$. 
Each orbit corresponds to a role in the automorphic equivalence. 
Understanding the  importance of automorphic equivalence and applying it to role modeling was a major breakthrough in classical social network research.
In our example Figure~\ref{fig:equiv}, from the topology alone, we cannot distinguish between the Smith family and the Jones family. The Lee family is distinct, because it has three children instead of two. Therefore, the equivalence classes are $\{S1,J1\}, \{S2,J2\}, \{S3,S4,J3,J4\}, \{L1\}, \{L2\}$, and $\{L3,L4,L5\}$.  
Interestingly, we can observe that automorphically equivalent classes must have equivalent indirect relations as well, such as equivalent in-laws and cousins. 
However, automorphic equivalence is hard to compute and still very strict. 

\noindent{\bf Exact Coloration (Equitable Partition):}
An {\em exact coloration} of graph $G$ assigns a color to each node, such that 
any two nodes share the same color iff they have the same number of neighbors of each color~\cite{everett96_exact_color}. 
Nodes of the same color form an equivalence class.
An exact coloration is also referred to as equitable partition~\cite{GR01} and graph divisor~\cite{CDSCH98} and is often applied in the vertex classification/refinement for canonical labeling of graph isomorphism test~\cite{ReadCorneil77,McKay81}.
A graph may have several exact colorations; in general we seek the fewest colors.
In our running example, the structural equivalence partitioning and the automorphic partitioning offer two different exact colorations.
Exact coloration relaxes automorphism by considering only immediate neighborhood equivalence.
Two nodes with the same color under an exact coloration may not necessarily be automorphically equivalent, but the graph automorphic equivalence does introduce an exact coloration by assigning a unique color to each orbit.
Like autmomorphic equivalence, exact coloration equivalence provides a recursive aspect to role modeling. 

\comment{
An alternate definition is that there is an isomorphism of the original graph which maps the first actor to the second one. In graph isomorphism, a function $\sigma$ maps every vertex in graph $G$ to a distinct vertex in graph $H$, such that for every edge $(u,v) \in G$, there is a distinct edge $(\sigma(u), \sigma(v)) \in H$.  There may be more than one mapping, so an equivalence class contains all vertices that are equivalent. Since we require an isomorphism for the entire graph, the equivalence holds not just for an actor's direct neighbors, but also for its indirect neighbors, e.g., friends of friends.
}

\noindent{\bf Regular Equivalence (Bisimulation):}
Two actors are {\em regularly equivalent} if they interact with the same variety of role classes, where class is recursively defined by regular equivalence~\cite{White83}.  Unlike automorphic equivalence and exact coloration, regular equivalence does not care about the cardinality of neighbor relationships, only whether they are nonzero. For example, using regular equivalence, all three families are now equivalent.  There are only three equivalence classes: $sibling-parent\{S1,J1,L1\}$, $spouse-parent\{S2,J2,L2\}$, and $child$. 
Note that under regular equivalence, any two automorphically equivalent nodes may be partitioned into the same regular equivalence class. 
In computer science, the regular equivalence is often referred to as the bisimulation, which is\x{ the} widely used in automata and modal logic~\cite{MarxM03}. 

\comment{
Most of the recent practical work has used regular equivalence, because the cardinality relaxation seems to be necessary, and algorithmic tools have been developed for researchers to use [REGE, CATREGE].  However, our contention is that cardinality does matter, but it should be a weighting factor, not an accept/reject rule.  Let us develop a clearer definition of role:
} 

\comment{ 
\subsection{Role Assignment as Coloring}
Each of these equivalences can be described in terms of what a vertex sees when it observes its neighbors. Assigning a vertex to an equivalence class is the same as generalized vertex coloring.  In the classical vertex coloring problem~\cite{West01_graphbook}, the constraint is that adjacent vertices may not be the same color. Instead, our objective is to find a coloring such that two vertices of the same color each see the same spectra (distributions of colors) when they observe their own neighborhoods.

Let $C = \{c_1,...,c_k\}$ be a set of role colors, one color for each role class.
Thus, the role modeling problem is to assign every vertex $v \in V$ a color label $c(v), c(v) \in C$.

\bdefin {\bf Neighbor spectrum. }
A neighbor spectrum of vertex $v$ is a vector describing the neighborhood of $v$, where element values are a function of vertex set $V$ and color set $C$.  The specific formulation may vary, depending on the type of analysis of interest. We use the names $\overrightarrow{N}(v)$, $\overrightarrow{A}(v)$, and $\overrightarrow{P}(v)$ for the appropriate neighbor spectra for structural equivalence, automorphic equivalence, and regular equivalence, respectively.

\noindent $\bullet$~
$\overrightarrow{N}(v)$ is the indicator vector $(n_i(v), \cdots, n_n(v))$, where $n_i(v) = 1$ if $v_i$ is a neighbor of $v$; 0 otherwise. This is the degenerate case where the spectrum describes the neighbor identities but not their colors. If we have an adjacency matrix $A_G$ representation of graph $G$, then $\overrightarrow{N}(v)$ is simply the row of $A_G$ that corresponds to node $v$.

\noindent $\bullet$~
$\overrightarrow{\alpha}(v)$ is the weighted color vector $(\alpha_1(v), \cdots, \alpha_k(v))$, where $\alpha_i(v)$ is the count of neighbors of $v$ with color $i$:
$\alpha_i(v) = \sum_{u \in N(v)} (c(u) = c_i)$.

\noindent $\bullet$~
$\overrightarrow{\rho}(v)$ is the color indicator vector $(\rho_1(v), \cdots, \rho_k(v))$, where
$\rho_1(v) = 1$ if there is some $u \in N(v), c(u) = c_i$; 0 otherwise.

\edefin

Using neighbor spectra, we can provide a general description of the Role Coloring Problem:

\bdefin {\bf Exact Role Coloring Problem. }
Given graph $G(V,E)$, the exact role coloring problem is to assign each vertex $v \in V$ a role $c(v)$ such that for any  $u,v \in V$ where $c(u) = c(v)$, their neighbor spectra are equal.
\edefin

Table \ref{spectra} summarizes the neighbor spectra used for the various role equivalence models.

\begin{table}[h]
\centering
\small
\begin{tabular}{|c|c|c|} \hline
Equivalence Type & Vector Dimensions & Element Weights  \\ \hline
Structural  & vertex set	& 1 or 0 indicator \\ \hline
Automorphic & color	set		& count \\ \hline
Regular     & color	set		& 1 or 0 indicator \\ \hline
\end{tabular}
\caption{Spectral Vectors for Equivalence Classes} \label{spectra}
\end{table}

} 


\comment{ 
\bdefin {\bf Color matrix. } Each vertex $u$ has a $n \times k$ binary-valued {\em color matrix} $C(v)$, where $c(v)_{ij} = 1$ if vertex $v_i$ is a neighbor of $u$ and has color $r_j$. Otherwise $c(v)_{ij} = 0$.  If the graph is directed, then we instead form a $2n \times k$ matrix, where the first $n$ are for in-neighbors and the second $n$ are for out-neighbors.  
\edefin

There is an interesting graphic interpretation for the color matrix.  Assume we form a bipartite graph with vertices $(V \cup R)$.  For each neighbor $v_i$ of $u$, include an edge $(v_i, r(v_i))$.  Then matrix $C(v)$ is the adjacency matrix for this bipartite graph.  Figure \ref{fig_spectra}(a) shows a sample color matrix and (b) its corresponding bipartite graph.

\begin{figure}[bt]
\centering
\epsfig{file=Figures/spectra.eps,width=6cm,height=4cm}
\caption{Neighbor Color Matrix and Projected Spectra.} \label{fig_spectra}
\end{figure}

\bdefin {\bf Neighbor spectrum. }
A neighbor spectrum $P(u)$ is a projection of the color matrix onto either the vertex space or the color space.  The vertex space projection forms an $n$-length ($2n$ for directed graphs) vector, where $p(u)_i = 1$ iff $v_i$ is a neighbor of $u$.  The color space projection forms a $k$-length vector, where $p(u)_j = \sum_{i=1}^n c(u)_{ij}$.  That is, it is the number of neighbors with color $r_j$
\edefin

Figure \ref{fig_spectra}(c) shows the vertex projection spectrum and (d) the color projection spectrum that correspond to the color matrix in part (a). The weights in the color spectrum indicate the degree of intensity of each color, so it is analogous to the light spectrum of a physical object. As we will see soon, it is in our interest to generalize our definition of spectrum slightly:

\bdefin {\bf Generalized Neighbor spectrum. }
A generalized neighbor spectrum $\overrightarrow{N}(u)$ is a non-decreasing function $f$ on a projection of the color matrix.  Given an direct projection $P(u)$, then the generalized spectrum is $\overrightarrow{N}(u)$, where $\overrightarrow{N}(u)[i] = f(p(u)_i)$.
\edefin

This general role equivalence model encompasses structural equivalence, automorphic equivalence, and regular equivalence, depending on how the spectra are defined.  For structural equivalence, the individual identity of each neighbor is important, so we use the vertex space projection. For automorphic equivalence, we use the color space projection.  

} 

\comment{
\subsection{Spectral Equivalence}
Our model suggests a new type of equivalence, which we call spectral equivalence. Two actors are {\bf spectrally equivalent} if they interact with the same proportions of classes.  In this case, we specify $f$ to normalize the spectral vector: $N = f(P) = P/\|P\|$.}

\comment{
{\bf Optimal Role Coloring Problem}\\
Let $G = (V,E)$ be a directed graph.  Assign each vertex $v$ a role $c(v)$ such that for any two vertices $u$ and $v$ where $c(u) = c(v)$,\\
(1) their I-spectra $C(I_u)$ and $C(I_v)$ are similar, and \\
(2) their O-spectra $C(O_u)$ and $C(O_v)$ are similar,\\
(3) the number of roles is minimized,\\
(4) the mutual similarity with each color set $\kappa(c_i) = \{v ] in V, c(v) = c_i\}$ is minimized.

Referring to this definition, it is easy to see how role discovery differs from other kinds of network pattern analysis.  In component discovery[], one seeks subgraphs whose vertices are highly interconnected; however, density of connections is not a consideration for roles.  In role analysis, we want to understand all the different groups that a role tends to connect to; internal connections are no more or less important than other group-group connections. In frequent subgraph pattern mining[], we look for isomorphically equivalent subgraphs that occur frequently.  For this problem, any subgraph may be considered, but for role analysis, we must look at a vertex's entire neighborhood.  We cannot focus on only a subset of neighbors, as pattern mining does. More importantly, in frequent pattern mining, vertices can be unlabeled, and a pattern can be fully described independent of its position in the graph.  On the contrary, a role is a label, whose value depends on a vertex's position within the content of the entire graph.
} 

\comment{
Probably cut this:\\
We point out that there is no unique regular equivalence role assignment for a graph, because regular equivalence does not demand parsimony.  Consider the clique of a complete graph. Due to symmetry, we could say that every vertex is equivalent, so there is only one role.  Conversely, we could color every vertex differently and say that there are $n$ different roles with one member each. Consequently, the idea of an optimal assignment hinges on finding the smallest number of roles such that each actor's actual role is sufficient similar to others in its cohort.

Several algorithms have been produced to measure similarity of vertices.  However, none of these follow the exact needs of role similarly.  For example, several authors have noted the similarity between the SimRank metric [] and the definition of regular equivalence and declared them to be the same or sufficiently so.  In SimRank, the similarity between two vertices $u$ and $v$ is defined as
The similarity between vertices $u$ and $v$ is
\[ S(u,v) = \frac{1}{|N(u)||N(v)|} \sum_{x \in N(u)} \sum_{y \in N(v)} S(x,y) \]
with the overriding rule that $(v,v) = 1$.  However, SimRank diverges from true role similarity in several respects: weighting, path length, and initialization.  We provide further detail and improvements in the next section.

[Leicht] uses the relation $S_{ij} = \phi \sum_v A_{iv}S_{vj} + \psi \delta_{ij}$.  That is, they say $i$ is similar to $j$ if $i$'s neighbors are similar to $j$.  This definition inserts a one-sided direct adjacency into the computation, whereas there should be none according our understanding of role.

Rather than offering yet another similarity metric, we offer an ensemble of clustering methods to produce a consensus role grouping.  
}

\section{Axiomatic Role Similarity}
\label{section:similarity}

An equivalence relation, however, tells us nothing about non-equivalent items.  Using our example, the intuitive and real-world need is for a measure that not only recognizes automorphic equivalence, such as Smith child/spouse/parent to Jones child/spouse/parent, but also tell us that a Lee child/spouse/parent has strong similarity to either a Lee or Smith child/spouse/parent.
Over the years, several methods have been developed for addressing various link-based similarity problems (co-citation~\cite{Small73_cocitation}, coupling~\cite{Kessler63_coupling}, SimRank ~\cite{Jeh02_simrank}). 
Recently, several researchers have tried to apply these measurements to role modeling~\cite{Leicht06_vertexsim,Zhao09_prank}. 
However, none of these encompass the aforementioned automorphic equivalence property and thus are inadequate for measuring role similarity. 
To deal with this shortcoming and to clarify the problem, we first identify a list of axiomatic properties that all role similarity measures should obey.

\comment{ 
  However, role similarity is distinct from these problems because it originates from and must preserve a defining equivalence relation, which none of these other methods encompass. Recently, some researchers have mistakenly tried to apply SimRank-type algorithms to the role similarity problem. Our aim is clear up some of the confusion by identifying some axiomatic properties that all role similarity measures should obey.

The purpose of a similarity metric is to rate how much two items diverge from perfect equivalence.  Clearly, we first need to define what sort of equivalence is being using (structural, automorphic, spectral, regular), and then we would like some idea about distance or divergence. We begin by reviewing the definition of an equivalence relation:

\bdefin {\bf Equivalence Relation. }
A relation $R \subseteq A\times A$ on set $A$ is an equivalence relation if for all $a,b,c \in A$:\\
1. Reflexivity: $(a,a) \in R$\\
2. Symmetry: $(a,b) \in R$ iff $(a,b) \in R$\\
3. Transitivity: if $(a,b)$ and $(b,c) \in R$, then $(a,c) \in R$
\edefin
Notation: we use $a\equiv b$ to signify that $(a,b) \in R$.


Based on the properties of equivalence relations and distance metrics, we deduce the following properties for role similarity.
}  

\bdefin ({\bf Axiomatic Role Similarity Properties})
Given a graph $G=(V,E)$, any $sim(a,b)$ that measures the neighbor-based role similarity between vertices $a$ and $b$ in $V$ should satisfy properties P1 to P5:
\begin{itemize*}
\item P1) Range: $0 \leq sim(a,b) \leq 1$, for all $a$ and $b$.
\item P2) Symmetry: $sim(a,b) = sim(b,a)$.
\item P3) Automorphism confirmation: If $a\equiv b$, $sim(a,b) = 1$.
\item P4) Transitive similarity: If $a \equiv b$, $c \equiv d$, then $sim(a,c)=sim(a,d)=sim(b,c)=sim(b,d)$.
\item P5) Triangle inequality: $d(a,c) \leq d(a,b) + d(b,c)$, where
distance $d(a,c)$ is defined as $1 - sim(a,c)$.
\end{itemize*}
Any node similarity measure satisfying the first four conditions (without triangle inequality) is called an {\bf admissible role similarity measure}.
Any node similarity measure satisfying all five conditions is an {\bf admissible role similarity {\em metric}}. 
If the converse of the automorphic confirmation property is also true (if $sim(a,b)=1$, then $a \equiv b$), then the node similarity measure(metric) is an {\bf ideal role similarity measure(metric)}.
 
\label{def:role_sim_metric}
\edefin

Property 1 describes the standard normalization where 1 means fully similar and 0 means completely dissimilar (i.e., the two neighborhoods have nothing in common). Property 2 indicates that similarity, like distance, must be symmetric.
Property 3 expresses our idea that fully similar means automorphically equivalent.
Property 4 claims that the similarity between two nodes is equal to the similarity between equivalent members of the first two node's respective equivalence classes.
In other words, we can simply define the similarity for the orbits, i.e., $sim(\Delta(u),\Delta(v))=sim(u,v)$.
This guarantees consistency of values at an orbit-level.
Property 5 assumes the measure is metric-like, i.e., satisfying the triangle inequality.
This is much stronger than transitivity, enforcing an {\em ordering} of values.
Indeed, the only condition which excludes $d(a,b)=1-sim(a,b)$ from being a strict distance metric is the automorphic equivalence (it allows the distance between two different nodes to be $0$). 
In addition, note that Property 5 implies Property 4. 

\blemma ({\bf Transitive Similarity})
\label{transitivity}
For any $a,b \in V$ and $c,d \in V$, if $a \equiv b$ and $c \equiv d$, then $sim(a,c)=sim(a,d)=sim(b,c)=sim(b,d)$.
\elemma
\bproof 
From triangle inequality, we have $d(a,c) \leq d(a,b)+d(b,c) \leq d(b,c)$ and $d(b,c) \leq d(b,a)+d(a,c) \leq d(a,c)$ ($d(a,b)=0$). 
Thus, $d(a,c)=d(b,c)$. 
Similarly, $d(a,d)=b(b,d)$, $d(c,a)=d(d,a)$, and $d(d,a)=d(d,b)$. 
Put together, we have $sim(a,c)=sim(a,d)=sim(b,c)=sim(b,d)$. 
\eproof

However, since most similarity measures do not necessarily satisfy the triangle inequality, we explicitly include Property 4 as one of the axiomatic properties.  Further, Property 3 is an essential criterion which distinguishes the role similarity measure from other existing measures.  As we discussed earlier, the automorphic equivalence can be relaxed to exact coloration or regular equivalence.  In this case, we may replace Property 3 accordingly.  Our work will focus on the automorphic equivalence though it can handle its generalization as well.  

\bthm ({\bf Generalized Transitive Similarity}) 
\label{GTS}
For any two pairs of nodes $a,b \in V$, $c,d \in V$, if $sim(a,b)=1$ and $sim(c,d)=1$, then, 
their cross similarities are all equal, i.e., $sim(a,c)=sim(a,d)=sim(b,c)=sim(b,d)$.
\ethm
\bproof 
From the triangle inequality, we have $d(a,c) \leq d(a,b)+d(b,c) \leq d(b,c)$ and $d(b,c) \leq d(b,a)+d(a,c) \leq d(a,c)$ ($d(a,b)=0$). 
Thus, $d(a,c)=d(b,c)$. 
Similarly, $d(a,d)=b(b,d)$, $d(c,a)=d(d,a)$, and $d(d,a)=d(d,b)$. 
Put together, we have $sim(a,c)=sim(a,d)=sim(b,c)=sim(b,d)$. 
\eproof

Thus, if we partition the nodes into equivalence classes where similarity equals $1$, we can simply record the similarity values between equivalent classes.  
Let $\Delta(x)$ and $\Delta(y)$ be the equivalence classes for node $x$ and $y$, respectively. 
Then, we can define $sim(\Delta(x),\Delta(y))=sim(x,y)$.

\subsection {\bf Binary-Valued Role Similarity Measures}
\bthm({\bf Binary Admissibility})
\label{binaryadmissible}
Given any equivalence relation that also satisfies automorphism confirmation (P3), its binary indicator function is an admissible similarity {\em metric}.
\ethm
\bproof
Binary values satisfy the Range(P1). Any equivalence relation satisfies symmetry (P2) and transitivity (P4), by definition. For triangle inequality( P5), consider all possible cases:\\
Binary values satisfy the Range(P1). Any equivalence relation satisfies symmetry (P2) and transitivity (P4), by definition. For triangle inequality( P5), consider all possible cases:\\
\noindent Case 1: All in the same class: $0 \leq 0 + 0$\\
\noindent Case 2: All in different classes: $1 \leq 1 + 1$\\
\noindent Case 3: $a$ and $c$ in the same class: $0 \leq 1 + 1$\\
\noindent Case 4: $b$ and one other in the same class: $1 \leq 0 + 1$\\
\eproof

Note that automorphic equivalence, regular equivalence, and exact coloration all satisfy P3, so they are admissible metrics. In addition, the binary similarity measure introduced by {\em automorphic equivalence} is an {\em ideal} role similarity metric.
\comment{ 
The binary similarity measures introduced by {\em automorphic equivalence}, {\em exact coloration}, and {\em regular equivalence}, i.e., if two nodes are equivalent, then, their similarity is $1$, otherwise $0$, are admissible role similarity measures. 
[
\textcolor{red}{\bf (JOURNAL)}: In addition, the binary similarity measure introduce by {\em automorphic equivalence} is also an ideal role similarity measure.
]
\ethm
The proof is straightforward and thus omitted.
} 
Though these binary similarity measures are admissible, they provide no meaningful information about cross-class similarities, because they set $sim(\Delta(x),\Delta(y))=0$ if $\Delta(x)\neq \Delta(y)$.  We would like a real-valued measure that ranks the degree of role similarity.

Before presenting our proposed real-valued role similarity metric for network roles, we first examine some similarity measures proposed in earlier works.  We will see that these do not satisfy our required properties.

\comment{ 
\bproof
By maximal similarity, $a \equiv b$ and $c \equiv d$. Then,
\begin{align*}
(a,c) &= (b,c) \mbox{~(transitivity)} \\
      &= (c,b) \mbox{~(symmetry)} \\
	  &= (d,b) \mbox{~(transitivity)} \\
	  &= (b,d) \mbox{~(symmetry)}
\end{align*} \eproof
} 

\comment{ 
Property 4 (transitivity) extends symmetry from individuals to classes, insuring some degree of consistency among all the measurements.  Imagine if $a \equiv b$, but we do not have this property, so that $sim(a,c) \neq sim(b,c)$.  The inequality means that we can somehow distinguish between $a$ and $b$.  But that would contradict our knowledge that $a \equiv b$.

The properties above imply that we can measure similarity at the role level. That is, for any representative $x$ of role $X$ and any representative $y$ of role $Y$, we should assign the same value for $sim(x,y)$, meaning we could simply speak of $sim(X,Y)$.

\bthm {\bf Class-level similarity.}
For any $a,b \in$ role $X$ and $c,d \in$ role $Y$, $sim(a,c) = sim(b,d)$.
\label{thm:class_sim}
\ethm
\bproof
By maximal similarity, $a \equiv b$ and $c \equiv d$. Then,
\begin{align*}
(a,c) &= (b,c) \mbox{~(transitivity)} \\
      &= (c,b) \mbox{~(symmetry)} \\
	  &= (d,b) \mbox{~(transitivity)} \\
	  &= (b,d) \mbox{~(symmetry)}
\end{align*} \eproof
} 

\comment{ 
}

\subsection {\bf SimRank is NOT Admissible}
\label{lab:simrank_not}

\begin{figure}[t]
\centering
	\subfigure[Structural equivalence]{\label{fig:simrank_subequiv}\includegraphics[width=1.8in,height=0.7in]{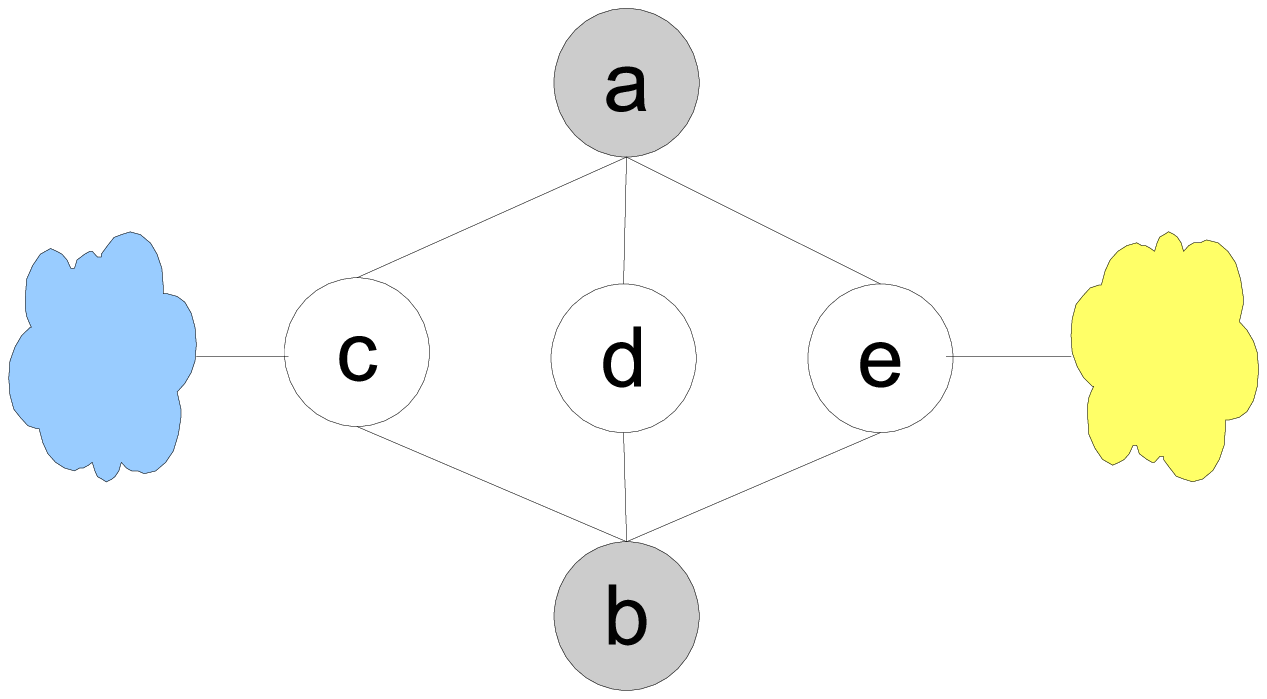}}
    \subfigure[Odd Distance]{\label{fig:simrank_odd}\includegraphics[width=1.4in,height=0.7in]{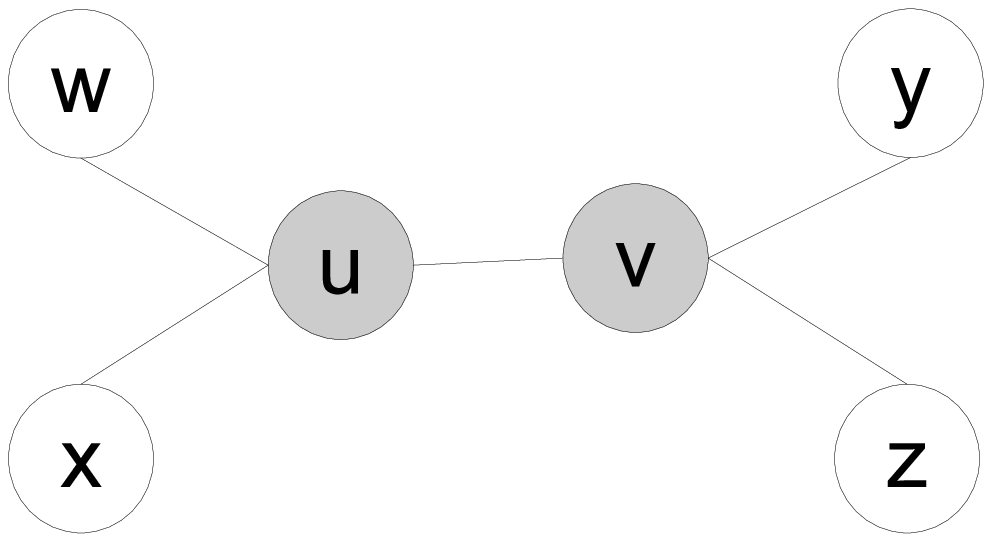}}
	\caption{Problematic configurations for SimRank}
	\label{fig:simrank}
\end{figure}

The SimRank~\cite{Jeh02_simrank} similarity between nodes $u$ and $v$ is the average similarity between $u$'s neighbors and $v$'s neighbors:
{\small
\begin{align*}
SR(u,v) &=	\frac{(1 - \beta)}{|N(u)||N(v)|} \sum_{x \in N(u)} \sum_{y \in N(v)} SR(x,y), \mbox{for } u \neq v, \\
SR(v,v) &=	 1,
\end{align*}
}
where $\beta$ is a decay factor, $0 < \beta < 1$, so that the influence of neighbors decreases with distance.
The original SimRank measure is for directed graphs. 
Here, we focus on its undirected version, though our comments also hold for the directed version.
SimRank values can be computed iteratively, with successively iterations approaching a unique solution, much as PageRank~\cite{Page99_pagerank} does.
\bthm
SimRank is not an admissible role similarity measure.  
\ethm
\bproof
We give examples where property 3 (automorphic equivalence) does not hold.
In Figure~\ref{fig:simrank_subequiv}, $a$ and $b$ have the same neighbors.  By even the strictest definition (structural equivalence), $a$ and $b$ have the same role.  However, since SimRank's {\em initial} assumption is that there is no similarity among $c$, $d$, and $e$, when it computes the average similarity of $a$ and $b$'s neighbors, it will never discover their equivalence.  Assuming the best case where $c$,$d$, and $e$ are structurally equivalent and using the recommended $\beta=0.15$, $SR(a,b)$ converges to only 0.667. If the neighbors are not equivalent, $a$ to $b$ should still be equivalent, but SimRank gives an even lower value.
SimRank has an another problem (Figure~\ref{fig:simrank_odd}) when there is an odd distance between two nodes.  Nodes $u$ and $v$ are automorphically equivalent, but because there are no nodes that are an equal distance from both $u$ and $v$, $SimRank(u,v) = 0$!

We note that other variants of SimRank~\cite{Antonellis08_simrankpp,Fogaras05_scaleSim,Li09_blocksim,Xi05_simfusion,Yin06_simtree,Zhao09_prank} also do not meet the automorphic equivalence property for to similar reasons. 
More discussion of these variants can be found in the Appendix.

\comment{
However, this does not mean that the neighbors $c$, $d$, and $e$ are equivalent to one another.  Since SimRank averages the neighbor similarity, in general it will NOT detect structural equivalence.  If we assume $SR(c,d) = SR(d,e) = SR(c,e) = 0.5$, then $SR(a,b) = \frac{6(0.5)+3(1)}{9} = 2/3$.  Strangely, more neighbors in common leads to a lower SimRank value.
} 
\comment{ 
} 

\comment{ 
In this case, $SR^{(t)}(u,v)$ is the value computed after $t$ steps.  
$SR^{(0)}$ is initialized to the $N\times N$ identity matrix I.  Moreover, there is simple random walk interpretation:  $SR(u,v)$ is the probability that two independent random walkers will meet, if they start at vertices $u$ and $v$ and walk along the graph's edges.
} 

\comment{ 
SimRank in its original form is not a general purpose role similarity metric because it considers only in-neighbors, whereas social roles depend on all neighboring interactions.  It is trivial to add a term for out-neighbors as well~\cite{Holme05_roleYeast,Zhao09_prank}.
[] and [] have given definitions in terms of expected $f$-meeting distance, where $f$ indicates that actual distances are remapped according to the function $f(d) = \beta^d$.\
}

\comment{ 
We provide a more straightforward interpretation, using the actual graph and actual distances:
\bthm
Assume there are two random walkers, who take independent but synchronous steps {\em backwards} on the graph's edges.  During each step, each walker will decide to leave the graph with probability $1 - \sqrt{\beta}$. If the two walkers start at $u$ and $v$, then the $t$-step SimRank value $s^t(u,v)$ is the probability that the random walkers will meet in no more than $t$ steps.  The stationary value $s(u,v)$ is the probability that they will ever meet.
\ethm
The proof is given in Appendix 1.
} 

\comment{ 3 problematic features of Sim-Rank
\noindent{\bf Structural Equivalence:} 
For any two nodes, which are structurally equivalent, i.e., sharing the same neighbors, and their degree is larger than one, their $SR \neq 1$. 
Indeed, the more neighbors they have, the {\em lower} the SimRank value.  
Consider Figure~\ref{fig:simrank_subequiv}.  Vertices $a$ and $b$ have the same two neighbors: $c$ and $d$ (ignoring $f$ for now). $SR(c,d)$ must be less than $1$, because $c$ and $d$ have different neighbors.

\begin{align*}
SR(a,b) &=  \frac{\beta}{4}[SR(c,c) +  SR(c,d) + SR(d,c) + SR(d,d)] \\
		&=  \frac{\beta(1 + SR(c,d))}{2} < 1
\end{align*}
because $\beta < 1$ and $SR(c,d) < 1$.

If $a$ and $b$ have an additional neighbor $f$, then the SimRank value will drop further, which we show in Section~\ref{section:experiment}.
Indeed, for structurally equivalent nodes, $ SR < 1/|N(v)|$, which is the probability the two independent walkers can meet.

\noindent{\bf Self-Similarity:}
SimRank scores emanate from self-equivalence: for any node $v$, $SR(v,v) = 1$.
However, its formula based on average neighbor-pair similarity cannot naturally maintain self-equivalence. 
Instead, the algorithm must manually reset $SR(v,v) = 1$ at each iteration.  Without this reset, the iterative computation would degrade the $SR(v,v)$ values. 

This rejects the notion of role equivalence, where any two vertices with equivalent surroundings should have full similarity, regardless of their proximity to one another.   Moreover, SimRank cannot even naturally maintain self-similarity. Since the random walk and decay factor both tend to degrade scores, the  algorithm is forced to manually reset $SR(v,v) = 1$ at each iteration.

\noindent{\bf Odd-Length Paths:}
Due to the combination of self-equivalence-only and paired random walk, SimRank requires that 
the two walkers travel the same distance to reach a common meeting point.  
It ignores paths where one walker takes an extra step. In Figure~\ref{fig:simrank_odd}, since walkers starting at $3$ and $4$ will never meet if they must take the same number of steps (all paths from $3$ to $4$ are odd-lengthed), $SR(3,4) = 0$!
} 



\comment{
\subsection{Admissible Similarity Measures}
As shown above, SimRank does not satisfy the properties for a similarity metric, failing the maximal similarity rule.  Here we evaluate some admissible similarity metrics and consider which would be best for role similarity.  For our application, we need a measure that compares two neighbor spectra, where a spectrum is either an 1-0 vector of nodes (structural equivalence) or a weighted vector of colors (other equivalences).

Previous authors~\cite{Fogaras05_scalesim} have attempted to modify the SimRank formula with the Jaccard coefficient.  The Jaccard coefficient compares two (unweighted) sets, measuring their set intersection as a fraction of their set union:
\vspace{-1pt}
\[ \mbox{Jaccard}(U,V) = |U \cap V| / |U \cup V| \]
\vspace{-1pt}
It is easy to see that the Jaccard Coefficient satisfies the range, symmetry and maximal similarity properties.
However, since the Jaccard coefficient operates only on sets or unweighted vectors, its usefulness is limited to either structural or regular equivalence.

Another metric, widely used for similarity search among text documents, is the cosine similarity metric.  Given two nodes (documents) described as weighted vectors in a $k$-dimensional feature space, $u = (u_1,\cdots,u_k)$ and $v = (v_1,\cdots,v_k)$, then their cosine similarity is:
\vspace{-1pt}
\[ \mbox{Cosine}(u,v) = u \cdot v / ||u|| ||v|| \]
\vspace{-1pt}
The feature space dimensions used in document similarity correspond nicely to our role equivalence classes in role analysis.  Cosine similarity also satisfies our range, symmetry, and maximal equivalence properties. There is one small behavior, however, that many not be what we are looking for.  As the name suggests, cosine similarity is related to the angle between the two object vectors.  If one object changes in magnitude only which the other remains unchanged, the angle does not change:
\vspace{-1pt}
\[
\mbox{Cosine}(\alpha u,v) = \frac{\alpha u \cdot v}{||\alpha u|| ||v||}
								= \frac{\alpha (u \cdot v)}{\alpha ||u||||v||}
								=  \mbox{Cosine}(u,v)
\]
\vspace{-1pt}
This is perfect for our new equivalence relation, spectral equivalence.  However, it is not appropriate for automorphic equivalence, because it no longer enforces a 1-to-1 mapping of $u$ to $v$'s neighbors.  How can we preserve the benefits of cosine similarity, yet also make it sensitive to the ratio between $u$ and $v$?

A related metric, the Tanimoto coefficient~\cite{Tanimoto}, holds the answer.  Used primarily in structural analysis of chemical compounds, this coefficient has been called a generalization of Jaccard and of Cosine:
\vspace{-1pt}
\[ \mbox{Tanimoto}(u,v) = \frac{u \cdot v}{||u||^2 + ||v||^2 - u \cdot v} \]
\vspace{-1pt}
Like cosine similarity, it makes use of the dot product. Also, if the vectors are unweighted (i.e. indicators), then the formula is equal to the Jaccard coefficient. Let us formally verify that it satisfies our properties:\\

\bthm
The Tanimoto coefficient applied to the neighbor spectra is an admissible role similarity measure.
\ethm
\bproof
~We show that it satisfies the four properties:\\
a. Symmetry: Since $u \cdot v = v \cdot u$, $\frac{u \cdot v}{||u||^2 + ||v||^2 - u \cdot v}$ is symmetric.\\
b. Maximal Similarity: If $u = v$, then $Tanimoto(u,v) = \frac{v\cdot v}{v\cdot v + v\cdot v - v\cdot v} = 1$.\\
c. Range: Each term in a spectrum vector is nonnegative, because it is a count of edges. Thus, $u \cdot v$ is always nonnegative.  Then, $min(Tanimoto) = 0$.  We can show by taking partial derivatives that the maximum value occurs when $u=v$.  We just showed that the value here is 1; therefore, $0 \leq Tanimoto(u,v) \leq 1$.\\
d. Transitive Similarity: If $N^a = N^b$, then
\begin{align*}
sim(a,c) &= \frac{N^a \cdot N^c}{||N^a||^2 + ||N^c||^2 - N^a \cdot N^c} \\
		&=  \frac{N^b \cdot N^c}{||N^b||^2 + ||N^c||^2 - N^b \cdot N^c} = sim(b,c)
\end{align*}
\eproof

\subsection{Choosing an Equivalence Relation}
Earlier, we presented four different equivalence relations, any of which could be used for a role similarity and discovery task.  However, we argue that for general applications, automorphic equivalence is the most useful.  Recall our earlier example with the families in Figure~\ref{fig:equiv}. On one hand, structural equivalence is too strict. For example, it does not support the idea of a "parent" class, because different parents have different children.  On the other hand, regular equivalence is too loose if our intent is to measure degrees of similarity.  Regular equivalence was devised so that we could have a "parent" equivalence class, even if families have different numbers of children.  However, if we go beyond strict equivalence and want to measure similarity, then we should retain the standard that two families are fully similar only if they have exactly the same configuration.  This leaves us with either automorphic or spectral equivalence.  Spectral equivalence may be useful in some applications where proportions are more important than absolute quantities.  For example, the ratio of in-neighbors to out-neighbors may be the focus of study. However, a node that has 2:1 in- vs. out- neighbors may be very different from one that has 200:100.  Therefore, in general, we recommend automorphic equivalence as the most informative equivalence for role similarity analysis.

  

To summarize, the Tanimoto coefficient is an admissible metric for role similarity based on automorphic equivalence because it satisfies our required axiomatic properties, achieving full similarity only when neighbor spectrum vectors are equal.

\comment{ 
(2) SimRank looks at specific neighbors while role depends only on class of neighbors.  A simple example illustrates the problem.  Suppose $u$ has three in-neighbors $a$, $b$, and $c$.  Vertex $v$ has the same in-neighbors.  Our intuitive understanding of structural similarity says that $u$ and $v$ are perfectly similar.  However, SimRank will only give them a value $\beta /3$.  In fact, the SimRank value decreases as the in-degree increases.  Some authors [] have addressed this problem by replacing $1/|I(u)||I(v)|$ with the Jaccard coefficient $|I(u) \cap I(v)|/|I(u) \cup I(v)|$.
\\
(3) SimRank assumes the root source of similarity is self-similarity, as encoded in $s(v,v) = 1$.  Social role, however, assumes no connection at all, direct or indirect, between two actors with the same roll.  We could have two separate networks representing two companies.  Each network has an actor who plays to role of president, but there is no path between from one to the other.

} %

}

\section{RoleSim: A Real-Valued\\ Admissible Role Similarity}
\label{section:algorithm}

To produce an admissible real-valued role similarity measure, we face two key challenges: 
First, it is computationally difficult to satisfy the automorphic equivalence property. Though not proven to be NP-complete, the graph automorphism problem has no known polynomial algorithm~\cite{Fortin96_isomorph}. 
Second, all the existing real-valued role similarity measures have problems dealing with even simple conditions such as structural equivalence (Subsection~\ref{lab:simrank_not}).
To meet these challenges, we take the following approach:
Given an initial simplistic but admissible role similarity measurement for any pair of nodes in a graph, refine the measurement by expressing the similarity in terms of neighboring values, while maintaining the automorphic and structural equivalence properties.
In the following, we formally introduce RoleSim, the first admissible real-valued role similarity measure (metric) and its associated properties.

	\comment{
	Finally, more recent applications on node similarity analysis deal with
	{\em weighted graph}, where each edge is associated with a weight~\cite{}. 
	Can we generalize the role similarity measure to the weighted graph?
	} 
\comment {
To meet these challenges, we seek inspiration from the following two basic questions:
1) Since the regular equivalence binary indicator function is an admissible similarity measure, can we use it as a basis for a real-valued role similarity measure? 
Note that regular equivalence tends to be much easier to compute~\cite{borgatti93_alg_reg_equiv}.
2) Assuming we already have an admissible role similarity measure for any pair of nodes in the graph, can we express this as a function of their neighbor's similarities?
Specifically, such measure should correctly assign value to automorphically equivalent and structurally equivalent nodes.
} 

\subsection{RoleSim Definition}

\comment{ 
The second phase is an iterative computation that uses the (Tanimoto) similarity metric to improve upon the similarity score computed previously.  A feature of this iterative process is that it will preserve the full similarity score of any automorphic nodes.  Moreover, it retains the desireable properties of SimRank: it follows a recursive definition and it lends itself to straightforward computation.

As we noted in our SimRank case study, a random walk expectation value (summing all possible pairings of neighbors) does not properly describe role equivalence.  Equivalence arises from aligned role-role matching, not random meetings, between the neighbors.  This matching is exemplified by the dot product $N^u \cdot N^v$ in our similarity metric, where the $i^{th}$ term in vector $N^v$ is the weight of role $i$ among node $v$'s neighbors.  However, there are two problems: (1) in a complex network, there may be very few strictly equivalent nodes; in other words, the number of roles is very large. (2) The typical recursive challenge: we are defining roles in terms of as-yet-unknown roles, so we cannot know the correct matching. For the first problem, since the number of neighbors is relatively small, we chose to express the neighbor spectrum $N^v$ in terms of individual nodes instead of roles.  Then, the weights in $N^v$ refer not to the cardinality of role membership but to a possible edge weight $(v\rightarrow i)$ in the original network.

This leads to a solution to the second problem: if $Sim(u,v)$ is relatively high, then $u$ and $v$ may be in the same eqivalence class.  If $Sim(u,v)$ is relatively low, then they probably are not.  Therefore, a reasonable estimate for $u \cdot v$ is $MaxMatch(u,v)$, where

\beqnarr
MaxMatch(u,v) =  max_\sigma(\sum_{x \in N^u} RoleSim(x, \sigma(x)))
\eeqnarr

where  $\sigma$ is a bijective mapping between $N^u$ and $N^v$.  This is equivalent to the maximal weighted bipartite matching problem.  We can define a {\em Matching} $M(u,v) = \{(x,y)|x \in N^u, y = \sigma(x))\}$, the set of all pairings.
} 

Given a graph $G=(V,E)$, the RoleSim measure realizes the recursive node structural similarity principle ``two nodes are similar if they relate to similar objects'' as follows. 
\bdefin({\bf RoleSim metric}) 
Given two vertices $u$ and $v$, where $N(u)$ and $N(v)$ denote their respective neighbohoods and $N_u$ and $N_v$ denote their respective degrees, then $RoleSim(u,v)=$
{\small
\beqnarr
\label{eqn:rolesim}
(1-\beta) \max_{M(u,v)} \frac {\sum_{(x,y) \in M(u,v)} RoleSim(x,y)}{N_u+N_v-|M(u,v)|}+\beta 
\eeqnarr
}
where $x \in\! N(u)$, $y \in\! N(v)$, and $M(u,v)$ is a {\bf matching} between $N(u)$ and $N(v)$, i.e., 
$M(u,v)=\{(x,y)| x \in N(u), y \in N(v), \mbox{ and no other} (x^\prime, y^\prime) \in M(u,v), \mbox{ s.t. }, x=x^\prime \mbox{ or } y=y^\prime\}$. 
The parameter $\beta$ is a decay factor, $0 < \beta <1$.  
\edefin

\begin{figure}[hbt]
\centering
\epsfig{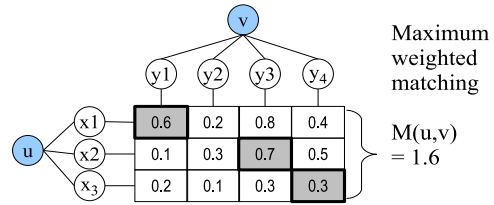}
\caption{RoleSim(a,b) based on similarity of their neighbors}
\label{fig:neigh_match}
\end{figure}

The decay factor, similar to the one used in PageRank~\cite{Page99_pagerank}, both dampens the recursive effect and guarantees a minimal RoleSim score of $\beta$.
We will sometimes abbreviate $RoleSim(u,v)$ as $R(u,v)$.  $\mathbf{R}$ refers to the entire matrix of values.
Figure~\ref{fig:neigh_match} illustrates the matching process. The $(x,y)$ grid is the subset of the RoleSim matrix of values corresponding to the pairings of neighbors of these two vertices. 
A matching selects one cell per row and column.  If the number of rows differs from the number of columns, then the matching size is limited to $|M(u,v)| = min(N_u, N_v)$.    A maximal matching is a matching where the total value of selected cells is maximum.
In contrast, SimRank computes the average of every cell in the neighbor grid.

\subsubsection{Relation to Jaccard Coefficient}
RoleSim is built on top of a natural generalization of the Jaccard coefficient, which measures the similarity between two sets $A$ and $B$ as $J(A,B)=\frac{|A \cap B|}{|A \cup B|}$.
The Jaccard coefficient has been used previously to measure node-node similarity based on their neighborhood commonality~\cite{Fogaras05_scaleSim}. 
In our generalization, however, sets $A$ and $B$ do not necessarily share any common element; instead, there is a matching $M$ between {\em similar} elements in $A$ and $B$, i.e., $(a,b) \in M, a \in A, b \in B$. 
Let $r(a,b) \in [0,1]$ record the similarity between $a$ and $b$.

\bdefin({\bf Generalized Jaccard Coefficient})
The generalized Jaccard coefficient measures the similarity between two sets $A$ and $B$ under matching $M$, defined as  
\small{
\beqnarr
\label{eqn:gjc}
J(A,B|M)=\frac{\sum_{(a,b) \in M} r(a,b)}{|A|+|B|-|M|} 
\eeqnarr
}
\edefin 

The original Jaccard coefficient is a special case which uses the following matching $M$: Let $r(x,y) = 1$ if $x=y$; otherwise $0$.  Then define $M=\{r(x,x) | x \in A, x\in B\}$.
Thus, the generalized Jaccard coefficient $J(A,B|M)$ reduces to $J(A,B)$.
Comparing Eq. \eqref{eqn:rolesim} and \eqref{eqn:gjc}, we see that the heart of $RoleSim(u,v)$ is equivalent to the maximum of the generalized Jaccard coefficient between $N(u)$ and $N(v)$, among all matchings $M(u,v)$. Then, $RoleSim(u,v)=$
\beqnarr
(1-\beta) \max_{M(u,v)} J(N(u),N(v)|M(u,v)) + \beta
\eeqnarr 

\subsubsection{Relation to Weighted Matching}

The definition and significance of the RoleSim for any node pair $(u,v)$ is closely related to {\em maximal weighted matching}. 
For any nodes $u$ and $v$ in graph $G$, define a weighted bipartite graph $(N(u) \cup N(v), N(u) \times N(v))$ , with each edge $(x,y) \in N(u) \times N(v)$ having weight $RoleSim(x,y)$. 
Let the total weight of neighbor matching $M(u,v)$ between $u$ and $v$ be 
$w(M(u,v))=\sum_{(x,y)\in M(u,v)} RoleSim(x,y)$.  
Let $\mathcal{M}$ be the maximal weighted matching for $(N(u) \cup N(v), N(u) \times N(v))$.  It is clear that
\beqnarr
w(\mathcal{M})=\max_{M(u,v)} w(M(u,v)).
\label{eqn:max_weight_match}
\eeqnarr
Using this, we can represent $RoleSim(u,v)$ in terms of maximal weighted matching $\mathcal{M}$.
In Figure~\ref{fig:neigh_match}, the shaded cells represent the maximal matching: $0.7+0.6+0.3=1.6$.  


\bthm({\bf Maximal Weighted Matching})
\label{MWM}
The RoleSim between nodes $u$ and $v$ corresponds linearly to the maximal weighted matching $\mathcal{M}$ for the bipartite graph $(N(u) \cup N(v), N(u) \times N(v))$, with each edge $(x,y) \in N(u)\times N(v)$ having the weight $RoleSim(x,y)$: 
{\small
\beqnarr
RoleSim(u,v)=(1-\beta)\frac{w(\mathcal{M})}{\max{ (N_u, N_v)}}+\beta 
\label{eqn:rs_is_mwm}
\eeqnarr
}
\ethm
\bproof
We need to show that Equations~\eqref{eqn:rolesim} and~\eqref{eqn:rs_is_mwm} are equivalent.
Without loss of generality, let $N_u\geq N_v$. 
First, we show that {\em the cardinality of the maximal weighted matching $|\mathcal{M}| = \min{(N_u,N_v)}=N_v$ }. It cannot be greater, because there are insufficient elements in $N_v$.
It cannot be smaller, because if it were, there must exist an available edge between an uncovered node in $N_u$ with one in $N_v$.  Adding this edge would increase the matching (every edge has weight $\geq \beta$). If $|\mathcal{M}| = \min{(N_u,N_v)}$, it follows that
${N_u+N_v-|M|} = \max{ (N_u, N_v)}$.  Thus, the denominators in Equations~\eqref{eqn:rolesim} and~\eqref{eqn:rs_is_mwm} are constant and identical.  It is then a trivial observation that the numerators are in fact the same.  Therefore, the maximal value for the entire Equation~\eqref{eqn:rolesim} is the same as the value in ~\eqref{eqn:rs_is_mwm}.
\eproof

	\comment{
	Second, let use assume that there is another matching $M$,
	such that $w(M) < w (\mathcal{M})$ but 
	$ \frac{w(M)}{N_u+N_v-|M|}=\max_{M(u,v)} \frac{w(M)}{N_u+N_v-|M|}
	> \frac{w(\mathcal{M})}{N_u}$. 
	For matching $M$, $|M| \leq N_v$. 
	Now consider two cases: 
	1) if $|M|=N_v$, then, $\frac{w(M)}{N_u}<\frac{w(\mathcal{M})}{N_u}$; 
	2) if $|M|<N_v$, then, $\frac{w(M)}{N_u+N_v-|M|}<\frac{w(M)+\beta}{N_u+N_v-(|M|+1)}\leq \frac{w(\mathcal{M})}{N_u}$. 
	In both cases, we found conflicting. 
	Therefore, the maximal weighted matching $\mathcal{M}$ would produce the RoleSim score. 
	\eproof
} 

	\comment{ 
	In certain degree, this measure can also be looked as a weighted Jaccard coefficient for matching scenario: 
	assuming that any pair $(x,y) \in N(u)\times N(v)$ in a matching can be treated as being {\em equivalent}, and also the edge weights are all $1$, then, \\
		{\small
		\beqnarr
		\max_{M(u,v)} \frac {\sum_{(x,y) \in M(u,v)} RoleSim(x,y)}{\max{(N_u,N_v)}}
		= \frac{\min{(N_u,N_v)}}{\max{(N_u,N_v)}} =\frac{|N(u) \cap N(v)|}{|N(u) \cup N(v)|}  \nonumber 
		\eeqnarr
		}
	}
	
Theorem ~\ref{MWM} not only shows the key equilibrium for the role similarities RoleSim between pairs of nodes in a graph $G$, but shows that each iteration can be computed using existing maximal matching algorithms.


\subsection{RoleSim Computation}
\label{computation}

RoleSim values can be computed iteratively and are guaranteed to converge, just as in PageRank and SimRank.  First we outline the procedure. In the next section, we prove that the calculated values comprise an admissible role similarity metric.

\noindent{\bf Step 1:} Let the initial matrix of RoleSim scores $\mathbf{R^0}$ be any set of admissible scores between any pair of nodes in $G$.  

\noindent{\bf Step 2:} Compute the $k^{th}$ iteration $\mathbf{R^k}$ scores from the $(k-1)^{th}$ iteration's values, $\mathbf{R^{k-1}}$. Specifically, for any nodes $u$ and $v$, 
{\small
\beqnarr
\label{iterative}
\mathbf{R^k}(u,v)=(1-\beta) \max_{M(u,v)} \frac {\sum_{(x,y) \in M(u,v)} \mathbf{R^{k-1}}(x,y)}{N_u+N_v-|M(u,v)|}+\beta 
\eeqnarr 
}
Based on Theorem~\ref{MWM}, we compute Equation~\eqref{iterative} by finding the maximal weighted matching in the weighted bipartite graph
$( N(u) \cup N(v), N(u) \times N(v))$ with each edge $(x,y) \in N(u) \times N(v)$ having weight $\mathbf{R^{k-1}}(x,y) )$. 

\noindent{\bf Step 3:} Repeat Step $2$ until $\mathbf{R}$ values converge for each pair of nodes in $G$. 

\comment{ 
There is an extremely simply initialization scheme.  Partition nodes according to node degree.
For all nodes $u$,$v$ in the same partition (same degree), $R^0(u,v) = 1$.  For all pairs $u$,$w$ in different partitions, $R^0(u,w) = 0$.  The reader can verify that this satisfies all the required role similarity properties.
}

\bthm({\bf Convergence}) 
For any admissible set of RoleSim scores $RoleSim^0$, 
the iterative computational procedure for RoleSim converges, i.e., for any $(u,v)$ pair,
\beqnarr
\lim_{k \rightarrow \infty} RoleSim^k(u,v) = RoleSim(u,v)
\eeqnarr
\label{thm:convergence} 
\ethm
This can be proven by showing that the maximum absolute difference between any $\mathbf{R^k}(u,v)$ and $\mathbf{R^{k+1}}(u,v)$ is monotonically decreasing.  The proof is in the Appendix.
  
Unlike PageRank and SimRank which converge to values independent of the initialization, the convergent RoleSim score is sensitive to the initialization.  That is, different initial values may generate different final RoleSim values. 
Rather than being a disadvantage, this is actually the key to coping with the graph automorphism complexity,
by allowing the ranking to utilize prior knowledge (the equivalence relationship) 
of the network topological structure.

\comment{
efficient algorithms [?],
[?] are available to discover the exact coloration and regular
equivalence in a graph, which guarantees to group automorphic
equivalent vertices together. Interestingly, these
algorithms typically rely on an iterative refinement process
by starting with certain initial coloration (partition) and then
apply the equivalence rule to further split the partition. In
our problem, such iterative refinement is not necessarily
required as the RoleSim computation actually involves a
similar procedure (Step 2&3). Given this, we can simply
employ a degree-based initialization, which also provides
initial partition for the equivalence approach: partition the
nodes based on their degrees, i.e., for any node v, we define
the equivalent classes as [v] = {u|Nu = Nv}. Note that
such a simple partition actually induces an admissible role
similarity. This is because if two nodes are automorphically
equivalent, then they must have the same degree.
}

\subsection{Admissibility of RoleSim}
\label{admissibility}

Here, we present one of the key contributions of this paper: the axiomatic admissibility of RoleSim.  If the initial computation is admissible, and \rw{if the interative}{because the iterative} computation of Equation~\eqref{eqn:rs_is_mwm} maintains admissibility (i.e., is an invariant transform of the axiomatic properties), then the final measure is admissible.

\bthm({\bf Invariant Transformation})  
\label{invariant}
If the $k^{th}$ iteration $RoleSim^k$ is an admissible role similarity metric, then so is $RoleSim^{k+1}$.
\ethm 

Properties $1$ (Range) and $2$ (Symmetry) are trivially invariant,
so we will focus on Properties $3$ (Automorphic Equivalence), $4$ (Transitive Similarity), and $5$ (Triangle Inequality). 

\blemma ({\bf Automorphism Confirmation Invariance})
\label{lem:invariant3}
If the $k^{th}$ iteration $RoleSim^k$ satisfies Axiom $3$ (Automorphism Confirmation), then so does $RoleSim^{k+1}$.
\elemma
\bproof
For nodes $u \equiv v$, there is a permutation $\sigma$ of vertex set $V$, such that $\sigma(u)=v$, and any edge $(u,x) \in E$ iff $(v, \sigma(x)) \in E$.
This indicates that $\sigma$ provides a one-to-one equivalence between nodes in $N(u)$ and $N(v)$.
Also, $u$ and $v$ have the same number of neighbors, i.e., $N_u=N_v$. 
So, it is clear that the maximal weighted matching  $\mathcal{M}$ in the bipartite graph $(N(u) \cup N(v), N(u) \times N(v))$ selects $N_u = N_v$ pairs of weight 1 each.
Thus, $RoleSim^{k+1}(u,v)=(1-\beta)\frac{w(\mathcal{M})}{\max{(N_u,N_v)}}+\beta=1$.
\eproof

\blemma ({\bf Transitive Similarity Invariance})
\label{lem:invariant4} 
If the $k^{th}$ iteration $RoleSim^k$ satisfies Axiom $4$ (Transitive Similarity), then so does $RoleSim^{k+1}$.
\elemma
\bproof
We know for any $a \equiv b$, $c \equiv d$, $RoleSim^{k}(a,c) = RoleSim^{k}(b,d)$.
Denote the maximal weighted matching between $N(a)$ and $N(c)$ as $\mathcal{M}$.
Since there is a one-to-one equivalence correspondence $\sigma$ between $N(a)$ and $N(b)$
and a one-to-one equivalence correspondence $\sigma^\prime$ between $N(c)$ and $N(d)$, 
we can construct a matching $\mathcal{M}^\prime$ between $N(b)$ and $N(d)$ as follows: $\mathcal{M}^\prime=\{(\sigma(x),\sigma^\prime(y))| (x,y) \in \mathcal{M}\}$. 
Since the transitive similarity property holds for $RoleSim^{k}$, we have 
$RoleSim^{k}(x,y)=RoleSim^{k}(\sigma(x),\sigma^\prime(y))$. 
Thus, $w(\mathcal{M}^\prime)=w(\mathcal{M})$, and

{\small
\begin{align*}
(1-\beta)\frac{w(\mathcal{M})}{\max{(N_a,N_c)}}+\beta
							&= (1-\beta)\frac{w(\mathcal{M}^\prime)}{\max{(N_b,N_d)}}+\beta  \\
 RoleSim^{k+1}(a,c) 	&= RoleSim^{k+1}(b,d).
\end{align*}
} \eproof

\blemma ({\bf Triangle Inequality Invariance}) 
\label{lem:invariant5}
If the $k^{th}$ iteration $RoleSim^k$ satisfies Axiom $5$ (Triangle Inequality), then so does $RoleSim^{k+1}$.
\elemma
\bproof
For iteration $k$, for any nodes $a$, $b$, and $c$, $d^k(a,c) \leq d^k(a,b)+d^k(b,c)$, where $d^k(a,b)=1 - RoleSim^k(a,b)$. We must prove that this inequality still holds for the next iteration: 
$d^{k+1}(a,c) \leq d^{k+1}(a,b)+d^{k+1}(b,c)$. 

Observation: 
{\em if there is a matching $M$ between $N(a)$ and $N(c)$ which satisfies
$1-((1-\beta)\frac{w(M)}{N_c}+\beta) \leq d^{k+1}(a,b)+d^{k+1}(b,c)$, then $d^{k+1}(a,c) \leq d^{k+1}(a,b)+d^{k+1}(b,c)$.}
This is because $\frac{w(M)}{N_c}  \leq \frac{w(\mathcal{M})}{N_c}$, where $\mathcal{M}$ is the maximal weighted matching between $N(a)$ and $N(c)$, and thus, 
$1-((1-\beta)\frac{w(M)}{N_c}+\beta) \geq 1-((1-\beta)\frac{w(\mathcal{M})}{N_c}+\beta)=d^{k+1}(a,c)$. 

We break down the proof into three cases:\\
\noindent Case 1. ($N_b \leq N_a \leq N_c$),\mbox{  } Case 2. ($N_a \leq N_b \leq N_c$), and\\
\noindent Case 3. ($N_a \leq N_c \leq N_b$).\\

\noindent{\bf Case 1 ($N_b \leq N_a \leq N_c$):}
Since $N_b$ is smallest, $|\mathcal{M}(a,b)|=|\mathcal{M}(b,c)|=N_b$. 
Define matching $M$ between $N(a)$ and $N(c)$ as
$M=\{(x,z) | (x,y) \in \mathcal{M}(a,b) \wedge (y,z) \in \mathcal{M}(b,c)\}.$
Then using our observation above: 
{\small
\begin{align*}
d&^{k+1}(a,b)+d^{k+1}(b,c)- (1-(1-\beta)\frac{w(M)}{N_c}-\beta)  \\
 = & \,(1-\beta)[-\frac{w(\mathcal{M}(a,b))}{N_a} - \frac{w(\mathcal{M}(b,c))}{N_c} + \frac{w(M)}{N_c}] + 1 - \beta \\
 = & \,(1-\beta)[\frac{N_b-w(\mathcal{M}(a,b))}{N_a} - \frac{N_b}{N_a} + \frac{N_b-w(\mathcal{M}(b,c))}{N_c} \\
   &- \,\frac{N_b}{N_c} - \frac{N_b-w(M)}{N_c} + \frac{N_b}{N_c}] + 1 - \beta  \\
\geq& \,(1-\beta) [1-\frac{N_b}{N_a} + \frac{\sum_{(x,y)\in \mathcal{M}(a,b)} (1-R^k(x,y))}{N_c} \\
   & + \,\frac{\sum_{(y,z) \in \mathcal{M}(b,c)}(1-R^k(y,z))}{N_c} - \frac{\sum_{(x,z) \in M} (1-R^k(x,z))} {N_c}] \\
\geq& \,(1-\beta)[\frac{\sum_{(x,y,z)} (d^k(x,y)+d^k(y,z)-d^k(x,z))}{N_c}] \geq 0 \\
   & \mbox{where }(x,y) \in \mathcal{M}(a,b), (y,z) \in \mathcal{M}(b,c), (x,z) \in M
\end{align*}

} 

\noindent{\bf Cases 2 and 3} can be proven by a similar technique; the details are in the Appendix.

\comment{
Due to its complexity, this proof is in the Appendix.  To get a general idea of our approach, we prove 3 separate cases.
1. ($N_b \leq N_a \leq N_c$),
2. ($N_a \leq N_b \leq N_c$), and
3. ($N_a \leq N_c \leq N_b$).
In each case, we note the required size of each pair-wise matching, which forces every neighbor of the smallest neighborhood to participate in the matchings with both other neighborhoods.  Figure~\ref{fig:triangle} illustrates this idea.
\begin{figure}[hbt]
\centering
\epsfig{file=Figures/triangle.eps,width=3.3in, height = 1.6in}
\caption{Triangle Inequality.}
\label{fig:triangle}
\end{figure}
}

By combining the admissible initial configurations given in Sec~\ref{sec:init} with Theorem~\ref{invariant} on invariance, we have shown that the iterative RoleSim computation generates a real-valued, admissible role similarity
measure.

\bthm({\bf Admissibility})
\label{admissible} 
If the initial $RoleSim^0$ is an admissible role similarity measure, 
then at each $k$-th iteration, $RoleSim^k$ is also admissible. When RoleSim computation converges, the final measure $\lim_{k \rightarrow \infty} RoleSim^k$ is admissible.
\ethm



\subsection{Initialization}
\label{sec:init}
According to Theorem~\ref{admissible}, an initial admissible RoleSim measurement $\mathbf{R^0} = I(\cdot)$ is needed to generate the desired real-valued role similarity ranking. 
What initial admissible measures or prior knowledge should we use?
We consider three schemes:
\benum
\setlength{\itemsep}{0pt}
\setlength{\itemindent}{0ex}
\item {\bf ALL-1 }: $I(u,v) = 1$ for all $u,v$.
\item {\bf Degree-Binary (DB)}: If two nodes have the same degree ($N_u = N_v$), then $I(u,v) = 1$; otherwise, $0$.
\item {\bf Degree-Ratio (DR)}: $I(u,v) = (1-\beta)\frac{min(N_u,N_v)}{max(N_u,N_v)}+\beta$.
\eenum

These schemes come from the following observation: {\em nodes that are automorphically equivalent have the same degree}. Basically, equal degree is a necessary but not sufficient condition for automorphism.
This observation is key to RoleSim: degree affects both the size of a maximal matching set and the denominator of the Jaccard Coefficient. 

\bthm ({\bf Admissible Initialization})
\label{Admissible Initialization}
ALL-1, Degree-Binary, and Degree-Ratio are all admissible role similarity measures.  Moreover, Degree-Binary and ALL-1 are admissible  role similarity {\em metrics}.
\ethm
\bproof
It is easy to see that ALL-1 degenerately satisfies all the axioms of a role similarity metric.
We focus on the two degree-based schemes.
Clearly, they satisfy Range(P1) and Symmetry(P2).
If $N_u = N_v$, then $I(u,v) = 1$, so they both satisfy Automorphism Confirmation (P3). 
For transitive similarity (P4), we only need to show that $I(u,v)$ depends only on class membership (Theorem \ref{GTS}).
For these schemes, class is defined by degree, and the measurement clearly depends only on degree.
Finally, because Degree-Binary and ALL-1 are binary indicators of equivalence, Theorem \ref{binaryadmissible} states that they are metrics.
\eproof

Note that SimRank's initialization ($SimRank^0(u,v) = 1$ iff $u = v$) is NOT admissible, because it does exactly the wrong thing: setting the initial value of any potentially equivalent nodes to 0.
SimRank iterations try to build up from zero.  However, due to its problems with structural equivalence and odd-length paths that we noted, SimRank will never increase the value enough to discover equivalent pairs that were neglected at the start.

In addition, we make the following interesting observations on the different initialization schemes. 

\blemma
\label{all1DR}
Let $\mathbf{R^1}(ALL-1)$ be the matrix of  RoleSim values at the first iteration after $\mathbf{R^0}=\mathbf{1}$ (All-1 initialization). 
Let $\mathbf{R^0}(DR)$ be the matrix  of RoleSim initialized by the Degree-Ratio (DR) scheme.
Then, $\mathbf{R^1}(ALL-1)=\mathbf{R^0}(DR)$. 
\elemma

This lemma can be easily derived by following the definition of RoleSim formula. 
Basically, the Degree-Ratio (DR) is exactly equal to the RoleSim state one iteration after ALL-1 initialization. 
Thus, ALL-1 and DR generate the same final results.  
The simple formula for DR is much faster than neighbor matching, so DR is essentially one iteration faster.
On the other hand, we may consider the simple ALL-1 scheme to be sufficient, since it works as well as the more sophisticated DR. 
Especially, after the simple initialization, RoleSim's maximal matching process automatically discriminates between nodes of different degree and continues to learn differences among neighbors as it iterates.
Also, both ALL-1 and DR initialization have the following convergence property:

\bthm({\bf Monotone Convergence}) If ALL-1 initialization is used, each RoleSim value is monotonically decreasing (or non-increasing): $\mathbf{R^{k+1}}(u,v) \leq \mathbf{R^k}(u,v)$ for all $k$.
\ethm
\bproof
At any iteration, the RoleSim value for any $(u,v)$ is the maximal matching of its neighbors. The value can increase only if some neighbor matchings increase.  If no value increased in the previous iteration, then no value can increase in the current iteration.  In the first iteration after ALL-1, clearly no value increases.  Therefore, no value ever increases.
\eproof

Indeed, this monotone convergence property can be generalized into the following format: 
{\em if $\mathbf{R^1} \leq \mathbf{R^0}$ (for any ($u,v$) pair,  $\mathbf{R^1}(u,v) \leq \mathbf{R^0}(u,v)$), then we have $\mathbf{R^{k+1}} \leq \mathbf{R^k}$}.
Note that the Degree-Binary (DB) initialization scheme does not have this property. 
In our experiments, we will further empirically study these initialization schemes.

\comment{
After initialization, RoleSim's maximal matching process automatically discriminates between nodes of different degree, and continuing to learn differences among neighbors as it iterates.  So, even the ALL-1 scheme is sufficient.  However, note that Degree-Ratio (DR) is exactly equal to the RoleSim state one iteration after ALL-1 initialization, so ALL-1 and DR generate the same final results.  The simple formula for DR is much faster than neighbor matching, so DR is essentially one iteration faster.
We have both practical and theoretical advantages for ALL-1 and DR.  In our experiments, we will look for possible advantages to Degree-Binary.}


\subsection{Computational Complexity}
\label{complexity}


Given $n$ nodes, we have $O(n^2)$ node-pair similarity values to update for each iteration.  For each node-pair, we must perform a maximal weighted matching. 
For weighted bipartite graph $(N(u) \cup N(v), N(u) \times N(v))$, the fastest algorithm based on augmenting paths (Hungarian method) can compute the maximal weighted matching in $O( x (x \log x + y))$, where $x=|N(u) \cup N(v)|$ and 
$y=|N(u)| \times |N(v)|$.

A fast greedy algorithm offers a $\frac{1}{2}$-approximation of the globally optimal matching in $O(y \log y)$ time~\cite{Avis83}.  If an equivalence matching exists
(i.e., $w(\mathcal{M})=\max{ (N_u, N_v)}$), the greedy method will find it.  This is important, because it means that a greedy RoleSim computation still generates an admissible measure.
Using greedy neighbor matching, the overall time complexity of RoleSim is $O(kn^2 d^\prime)$, where \rw{}{$k$ is the number of iterations and } $d^\prime$ is the average of $y \log y$ over all vertex-pair bipartite graphs in $G$. 
The space complexity is $O(n^2)$.

\comment{ 
Theorem~\ref{binaryadmissible} suggests any binary similarity measure based on strict automorphic equivalence, exact coloration, or regular equivalence can serve as $RoleSim^0$.
However, we need not be so precise.
In fact, the simplest admissible scheme is $RoleSim^0(u,v) = 1$ for all $u$ and $v$!  
The all-1 scheme satisfies the axioms of an admissible role similarity measure and require no initial computation. 
Surprisingly, even though it provides no obvious  knowledge for the node similarity ranking (treating any pair equally), 
the RoleSim iterative computation can automatically learn the graph topology and discriminate the non-equivalent pairs.
}
\comment{
To see why this works, the following observation can provide some insights.  First: nodes that are automorphically equivalent have the same degree. 
Second: the maximal matching in RoleSim behaves so that the greater the difference in degree, the lower the RoleSim value.  Only pairs with the same degree can potentially have a RoleSim value of 1.
Another simple admissible initialization is based on the node degree: $RoleSim^0(u,v) = 1$ iff $|N_u| = |N_v|$; that is, to partition nodes based on degree.

It is not hard to see that RoleSim's maximal matching process can naturally learn such degree difference rather easily. 
In addition, our experiments show the ranking difference with these different initialization methods are actually quite small.  
Therefore, we employ the all-1 scheme for our RoleSim initialization.
}

\comment{
 the will provide all the topological knowledge that we need, reducing the value of the non-equivalent pairs appropriately.
}

\comment{
\noindent{\bf Initialization:}
In RoleSim computation, we need provide an initial RoleSim score which is admissible. 
So what initial admissible measures or prior knowledge should we use? 
According  to Theorem~\ref{binaryadmissible}, theoretically, any of the binary similarity measure based on the automorphic equivalence, exact coloration, and even regular equivalence can serve as the initial RoleSim score, $R^0$. 
Especially, efficient algorithms~\cite{borgatti93_alg_reg_equiv,ReadCorneil77} are available to discover the exact coloration and regular equivalence in a graph, which guarantees to group automorphic equivalent vertices together. 
Interestingly, these algorithms typically rely on an iterative refinement process by starting with certain initial coloration (partition) and then apply the equivalence rule to further split the partition.
In our problem, such iterative refinement is not necessarily required as the RoleSim computation actually  involves a similar procedure (Step $2 \& 3$). 
Given this, we can simply employ a degree-based initialization, which also provides initial partition for the equivalence approach:  
{\em partition the nodes based on their degrees, i.e., 
for any node $v$, we define the equivalent classes as $[v]=\{u| N_u=N_v\}$. }
Note that such a simple partition actually induces an admissible role similarity. 
This is because {\em if two nodes are automorphically equivalent, then they must have the same degree.}
} 

\comment{ 
Given $n$ nodes, we have $O(n^2)$ node-pair similarity values to update for each iteration.  For each node-pair, we must perform a maximal weighted matching. 
For weighted bipartite graph $(N(u) \cup N(v), N(u) \times N(v))$, the fastest algorithm based on augmenting paths (Hungarian method) can compute the maximal weighted matching in $O( x (x \log x + y))$, where $x=|N(u) \cup N(v)|$ and 
$y=|N(u)| \times |N(v)|$.

A fast greedy algorithm offers a $\frac{1}{2}$-approximation of the globally optimal matching in $O(y \log y)$ time~\cite{Avis83}.  If an equivalence matching exists
(i.e., $w(\mathcal{M})=\max{ (N_u, N_v)}$), the greedy method will find it.  This is important, because it means that a greedy RoleSim computation still generates an admissible measure.
Using greedy neighbor matching, the overall time complexity of RoleSim is $O(kn^2 d^\prime)$, where $d^\prime$ is the average of $y \log y$ over all vertex-pair bipartite graphs in $G$. 
The space complexity is $O(n^2)$.
} 

\comment{
When $n$ increases, the space complexity is becoming a bottleneck for the RoleSim to process the large graphs. 
We introduce two key techniques to deal with the memory bottleneck: 
1) According to Theorem~\ref{GTS}, if we group those nodes with their pair-wise similarity being equal to one, then, we all nodes from one group $c(i)$ have the same similarity with respect to all nodes in another group $c(j)$, i.e., 
$sim(x,y) = sim(c(i),c(j)), \forall x \in c(i), y \in c(j)$.  
Generalizing this, we can see that if $sim(x,y) \approx 1$, then 
Following this property together with the triangle inequality, 
Utilizing this property, 

Therefore, we do not need to store a $n\times n$ similarity matrix, only $|K|\times |K|$, $K$ is the set of all equivalence classes.  We can choose to limit the number of classes to $k$, merging similar classes, to get an approximate result. 

}

\comment{
Recall that The~\ref{binaryadmissible} confirms that the binary indicator functions for automorphic equivalence and regular equivalence both are admissible similarity measures. 
Other simple and efficient algorithms is also available. 
According
Later in Subsection~\ref{complexity}, we will discuss an even simpler initialization scheme.  

However, how can we find the desired initial admissible measures?
Recall that Lemma~\ref{binaryadmissible} confirms that the binary indicator functions for automorphic equivalence and regular equivalence both are admissible similarity measures. 
Efficient algorithms~\cite{borgatti93_alg_reg_equiv} are available to discover the regular equivalence in a graph. 
Other simple and efficient algorithms is also available. 
For instance, consider a degree-based approach mentioned in Section~\ref{computation}: 
{\em partition the nodes based on their degrees, i.e., 
for any node $v$, we define the equivalent classes as $[v]=\{u| N_u=N_v\}$. }
This is because {\em if two nodes are automorphically equivalent, then they must have the same degree.}
Further, if two vertices are automorphically equivalent, then there is a correspondence of their neighbors which will match degrees also.  So, we can subdivide each partition to enforce a stricter equivalence.  For each node, generate a neighbor degree vector $D^1(N^v) = (d(N^v_1), d(N^v_2),\cdots,d(N^v_{N_v}))$, where the elements are sorted by decreasing magnitude.  For example, in Figure~\ref{fig:equiv}, $d(N^{S1}) = (d(L1),d(J1),d(S2),d(S3),d(S4)) = (6,5,3,2,2)$.  If two nodes do not have equal $D_1$ vectors, they cannot be automorphically equivalent. 
In general, we can perform $D^k$ initialization to depth $k$.  Instead of computing a sorted vector,~\cite{Sparrow93_linearequiv} computes a real value in $|N_v|$ time.  
Using hashing, we can determine which nodes have the same degree spectra in near-linear time.
}

\comment{ 
\subsubsection{$k$-Block Computation}
A more aggressive scalability approach is to consider each initialization partition as a solid block and then to perform only block-block similarity iterative computations.

{\bf Block-block similarity}: To perform RoleSim at the block-level, we need a description of the "mean" block-level neighborhood: Compute the neighbor color spectrum of each node, that is, a weighted vector where the $i^{th}$ element is the number of neighbors which belong to role $i$.  For each block $B_i$: compute the mean neighbor color spectrum, akin to a cluster's centroid.  For each pair of blocks, perform a RoleSim iterative computation, but using fractional greedy matching instead of discrete greedy matching.  This is similar to the fractional backpack problem. To compute $RoleSim^{i+1}(B1,B2)$:\\
\noindent(1) Initialize WeightAvailable $w = min(deg(B1),deg(B2))$ and Matching $m = 0$.\\
\noindent(2) Select the unused neighbor-pair with the highest $RoleSim^i$ value.  The first choice will be one of the diagonals, with value 1.\\
\noindent(3) Now, allocate as much fractional weight from the neighborhoods that we can.  If step (2) selected neighbor-pair $(B_x,B_y)$, then subtract from element $x$ of $B1$ and from element $y$ of $B2$ an amount equal to $min(min(B1_x, B2_y),w)$.  If there was a tie in step 2, we want to break the tie by selecting the highest result for step (3).\\
\noindent(4) $m \leftarrow m + min(B1_x, B2_y,w) \times RoleSim^i(B_x,B_y)$;\\
$w \leftarrow w - min(B1_x, B2_y,w)$\;\
\noindent(5) Eliminate $(B_x,B_y)$ from consideration.  Repeat steps (2) -(4) until all $w = 0$.\\

Example: we are matching two blocks which have neighbor spectra $N_{B1} = (0.9, 1.5, 1.2, 0.4)$ and $N_{B2} = (1.2, 0.6, 1.2)$.  Note that the total of each block's neighbor weights equals the characteristic node degree of its members.\\
\noindent(Step 1) The degrees of $B1$ and $B2$ are 4 and 3, respectively, so $w = 3$.\\
\noindent(Step 2) For the 1st iteration, we could select any of the diagonal neighbor-pairs (1,1), (2,2), or (3,3).\\
\noindent(Step 3) The three choices from step 2 would allow us to allocate weights of $min(0.9,1.2), min(1.5,0.5), min(1.2,1.2)$, respectively.  The best choice is (3,3) with weight 1.2.\\
\noindent(Step 4) Updated neighbor spectra $N_{B1} = (0.9, 1.5, 0.0, 0.4)$,
$N_{B2} = (1.2, 0.6, 0.0)$,\\
$m \leftarrow m + min(B1_x, B2_y,w) \times RoleSim^i(B_x,B_y) = 0 + 1.2 \times 1 = 1.2$;\\
$w \leftarrow w - min(B1_x, B2_y,w) = 3 - 1.2 = 1.8$;\\

{\bf Within-block similarity}: For each block $B_i$, compute a neighbor spectrum variance:
$\frac{\sum_{u \in B_i} (Spec(u) - Spec(\bar{B})^2}{|B_i|}$.  If the variance is sufficiently high(?), split the block into two, using K-means or a simple spectral method.
} 

\comment{
, then iteratively compute for both within-block and between-block similarity.  A similar idea was proposed for SimRank improvement~\cite{Li09_blocksim}.  If $B(u)$ is the block to which node $u$ belongs, then the estimated node-node similarity $R(u,v) \approx R(u,B(u))R(B(u),B(v))R(v,B(v))$.  If the initial partition and between-block similarity is a admissible similarity measure, the subsequent iterations of between-block similarity will also be admissible (Proof omitted due to space limitation).
}

\comment{
If $m$ is the number of blocks with roughly equal-sized blocks, then the time drops to $O(k\bar{d}^2(m^2 + n^2/m))$.  We can use the $D0$ or $D1$ degree-based initialization to partition the network.  \cite{Li09_blocksim} shows that if optimal $m^\ast = (2n)^{3/2}/2$ is used, then the time drops to $O(k\bar{d}^2 n^{4/3})$.  The total space requirements drop similarly.  However, since each block can be processed independently, the main memory requirement drops even farther to $O(n^2/m^2) = O(n^{2/3})$.
}

\comment { 



\noindent 3. Set $R^0(x,y) = \frac{min(N_x,N_x)}{max(N_x,N_y)}$  So, if $x$ and $y$ are in the same initialization class, similarity = 1.  Otherwise, it is based on the ratio of their sizes.\\
\noindent 4. \\

\noindent 5. Intra-block RoleSim iteration: $R^{i+1}(x,y) = \frac{\mathcal{M}(B(x),B(y))}{max(B(x),B(y))}$\\

} 

\comment{
\subsection{Computational Complexity}
The computational cost of RoleSim is quite close to the original SimRank algorithm~\cite{}. 

The size of our matching problem is $d$, the average node degree. The original Kuhn-Munkers algorithm can solve this in $O(d^4)$ time~\cite{???}.  In our case, we match neighbor sets, which are relatively small, but the matching matrix is dense, not sparse. Then we must perform this matching for each pair of nodes.  However, a simple greedy matching algorithm achieves a $\frac{1}{2}$-approximation of the optimal matching in linear time, if the weights are already sorted~\cite{???}. Thus, the greedy approach operates in $O(n^2d\log d)$ time.
} 

\comment{
\bthm({\bf Ideal Role Similarity Metric}) 
\label{ideal}

\ethm

\comment{
where the iterative process converges to stationary similarity values.  The complete algorithm is shown in Algorithm~\ref{alg:RoleSim}.

\input{Figures/alg_RoleSimilarity.tex}

\bthm{\bf Admissibility of RoleSim.} At the conclusion of any number of iterations of the repeat loop in the GreedyRoleSim algorithm, the computed similarity values satisfy the required properties for a Role Similarity Metric (Definition~\ref{def:role_sim_metric}).
\label{thm:rolesim}
\ethm
\bproof~First we show that the initialization phase satisfies the properties.  Then we show that if the $k^{th}$ iteration $RoleSim_k$ is an admissible similarity metric, then so is $RoleSim_{k+1}$.\\
\blemma
$RoleSim_0$ is an admissible role similiarity metric.
\label{lem:rolesim0}
\elemma
(a) Domain: All $RoleSim_k$ values are either 0 or 1.\\
(b) Symmetry: If nodes $a$ and $b$ belong to the same superclass, then both $RoleSim_0(a,b)$ and $RoleSim_0(b,a)$ equal 1.  Likewise, if they are not in the same set, $RoleSim_0(a,b)$ and $RoleSim_0(b,a)$ are set to 0.\\
(c) Maximal similarity: If $a \equiv b$, then they are in the same initialization superclass, e.g. they have the same degree.  As just stated above, they are given value 1, which is maximal.\\
(d) Transitive similarity: If $a \equiv b$, then either a third node $c$ is in the same superclass or it is not.  If it is, then $RoleSim_0(a,c) = Role_Sim_0(b,c) = 1$.  If it is not, then $RoleSim_0(a,c) = Role_Sim_0(b,c) = 0$.\\
\eproof

\blemma
\label{lem:rolesimk}
\elemma
(a) Domain: We assume all edge weights are nonnegative and all initial RoleSim values are nonnegative, so according to Eqn.~(\ref{eqn:rolesim}), all RoleSim values are nonnegative.  Note that if all the maximal $RoleSim_k$ values equal 1, then the equation degenerates to the cosine similarity index, which have values in [0,1].  Therefore, $RoleSim_{k+1}$ values are bounded between 0 and 1.\\
(b) Symmetry: Eqn.~(\ref{eqn:rolesim}) is clearly symmetric.\\
(c) Maximal similarity: 
(d) Transitive Similarity: 
\eproof

Since Lemmas~\ref{lem:rolesim0} and \ref{lem:rolesimk} are true, Theorem~\ref{thm:rolesim} is true.
\eproof
}


}

\comment{
We differ from SimRank on two key points.  First, $SimRank(a,b)$ sums the similarities of all possible pairings of a neighbor of $a$ with a neighbor of $b$.  This is the paired random-walk model.  It has been shown~\cite{Fogaras05_scalesim,Lin09_matchsim} that even if they have identical sets of neighbors, independent random walkers do not know how to align their neighbors so as to match.  Similar to MatchSim~\cite{Lin09_matchsim}, we employ a maximal matching.  Second, SimRank initially sets $SimRank^{(0)}(a,b) = 0$ if $a \neq b$, and $SimRank^{(0)} = 1$.  We call this the most-pessimistic start.  Instead, we start from the opposite perspective, an opmitistic start.  We suggest several quick heuristics for partitioning the network such that every true automorphic class is contained entirely within one of the initial partitions.  Then, $RoleSim^{(0)} =1$ if $a$ and $b$ are in the same partition.  This guarantees that every pair that is automorphically equivalent (plus some false positive equivalences) are discovered at the beginning.  As our algorithm progresses, the farther a pair is from being equivalent, the more their score will tend to degrade.  However, if in fact they are equivalent, their score will not degrade at all.
}



\comment{

We have described a theoretically sound and desireable similarity metric, but we are left with a significant practical problem. Solving graph automorphism has NP time complexity (though it is not proven to be NP-complete)~\cite{Fortin96_isomorph}; therefore, for large networks, there is no guaranteed computation of equivalence classes this is both efficient and correct. To address this challenge, we present a two-phase Estimation-Improvement heuristic procedure for efficiently computing pairwise role similarities.  In the first phase, we partition the nodes into equivalence superclasses.  Each of the actual but unknown automorphic equivalence classes is guaranteed to be a subset of one of the partitions.  In other words, each partition is the union of one or more equivalence classes.  We use these partitions to assign initial similarity scores.  In the second phase, we compute updated similarity scores by applying our similarity measure to the neighborhoods of each pair of nodes.  Since we do not know the actual equivalence classes, we cannot know with certainty the "colors" of the neighbor spectra.  Therefore, to compute the similarity, we apply a "best-match" technique.

\subsection{Equivalence Initial Estimation}
The goal of this phase is to efficiently partition the nodes into supersets of the unknown equivalence classes.  By doing this, we can guarantee that every pair of vertices that is automorphically equivalent will be initialized with similarity 1.  Inevitably we might also assign full similarity to some pairs that are not equivalent (false positives) but we will have no false negatives.  We can easily create equivalence superclasses by employing the following property:

\blemma
Any two vertices that do not share a necessary attribute of equivalence cannot be in the same equivalence class.
\elemma

We propose three initialization schemes:
\begin{enumerate}
\vspace{-3pt}
\item Degree-based
\vspace{-6pt}
\item K-Core based
\vspace{-6pt}
\item Direction-based
\end{enumerate}

\subsubsection{Degree-Based Initialization}

If two nodes are automorphically equivalent, then they must have the same degree.
Thus, a trivial scheme, which we call $D^0$ initialization, simply partitions vertices according to degree.  Node-pairs in the same partition are assigned  similarity of 1. However, if two vertices are automorphically equivalent, then there is a correspondence of their neighbors which will match degrees also.  So, we can subdivide each partition to enforce a stricter equivalence.  For each node, generate a neighbor degree vector $D^1(N^v) = (d(N^v_1), d(N^v_2),\cdots,d(N^v_d))$, where the elements are sorted by decreasing magnitude.  For example, in Figure~\ref{fig:equiv}, $d(N^{S1}) = (d(L1),d(J1),d(S2),d(S3),d(S4)) = (6,5,3,2,2)$.  If two nodes do not have equal $D_1$ vectors, they cannot be automorphically equivalent. In general, we can perform $D^k$ initialization to depth $k$.  Instead of computing a sorted vector,~\cite{Sparrow93_linearequiv} computes a real value in $|d|$ time.  Using hashing, we can determine which nodes have the same degree spectra in near-linear time.

\begin{algorithm}
{\small
\caption{$DegreeInit(G(V,E), maxDepth)$}
\label{alg:degree_init}
\begin{algorithmic}[1]

\STATE Initialize: for all nodes $u,v: S(u,v) \leftarrow 0; d \leftarrow maxDepth$;

\IF{$maxDepth > 0$}
	\STATE $CompareNeighborDegrees(V,maxDepth)$;
\ELSE
	\STATE Partition nodes according to degree;
	\STATE for all $u,v$ in the same partition: $S(u,v) \leftarrow 1$;
\ENDIF
\RETURN $S$;

\STATE $CompareNeighborDegrees(V,depth) \{$
	\FORALL{$u \in V$}
		\STATE $D(u) \leftarrow multiset\{d(x)|x \in N(u)\}.sorted$;
	\ENDFOR
	\STATE Partition nodes according to equivalent $D$ values;
	\IF{$depth = 1$}
		\STATE for all $(u,v)$ in the same partition, including $(v,v)$: $S(u,v) \leftarrow 1$;
	\ELSE
		\FORALL{node partition $V_i$ of $V$}
			\STATE $CompareNeighborDegrees(V_i, depth-1)$;
		\ENDFOR
	\ENDIF
\STATE \};

\end{algorithmic}
}
\end{algorithm}


\subsubsection{K-Core-based Initialization}
A $k$-core is a maximal induced subgraph such that every vertex in the $k$-core has edges to at least $k$ other members of the $k$-core~\cite{Seidman83_kcore}. This is a much stricter initialization scheme because it enforces a connectivity requirement which is not required for equivalence.  Our early studies show that this is not an effective initialization scheme for role similarity.  For example, for Figure~\ref{fig:equiv}, since every node has least degree 2, the 2-core is the entire network.  To construct the 3-core, we first filter out all the children. When only the parents are left, their degree drops below 3 also, so the 3-core is empty.  There is no meaningful partitioning of the network.





\subsubsection{Direction-based Initialization}
If a graph has directed edges, then two nodes can be equivalent only if the connections to their neighbors have the same directionality.  Thus, edge directionality can be used to partition the network, much as degree is used above.
For each node, form a 2-tuple which counts the number of in-neighbors and out-neighbors: $S^0 = (|I(u)|,|O(u)|)$.  Nodes which have identical tuples are equivalent. Since degree $d(u) = |I(u)| + |O(u)|$, this is a straightforward extension of degree-based initialization.



}

\section{Iceberg RoleSim Computation}
\label{iceberg}

Node similarity ranking in general is computationally expensive because we need to compute the similarity for ${n \choose 2} = O(n^2)$ node-pairs. 
A graph with $100,000$ nodes needs about $40$GB memory to simply maintain the similarity values, assuming $8$ bytes per value. 
Indeed, this is a major problem for almost all node similarity ranking algorithms. 
However, in most applications, we are interested only in the {\em highest} similarity pairs, which typically compose only a very small fraction of all pairs. 
Thus, in order to improve the scalability of RoleSim, we ask the following question: 
{\em Can we identify the high-similarity pairs without computing all pair similarities?}
Formally, we consider the following question: 

\bdefin{({\bf Iceberg RoleSim})} 
Given a threshold $\theta$, the Iceberg RoleSim problem is to discover all $(u,v)$ pairs for which 
$RoleSim(u,v) \geq \theta$ and then approximate their RoleSim scores. 
\edefin 

The goal is to identify and compute those high-similarity pairs without materializing the majority of the low similarity pairs. 
To solve {\em Iceberg RoleSim}, we consider a two-step approach: 
1) use pruning rules to rule out pairs whose similarity score must be less than $\theta$; 
and 2) apply RoleSim iterative computation to the remaining candidate pairs.  
Since RoleSim computation must match all neighbor-pairs ($N(u) \times N(v)$) of a candidate pair ($u,v$), we have to handle neighbor-pairs (such as $x,y$) which are not themselves candidate pairs. 
Here, we employ upper and lower bounds for estimating RoleSim values for the non-candidate pairs.


\noindent{\bf Upper and Lower Bound for RoleSim:} 
\blemma
\label{lem:theta_similarity}
Given nodes $u$, $v$ and without loss of generality, $N_u \geq N_v$, 
if $N_v \leq \theta N_u$,
then similarity $R(u,v) \leq (1-\beta)\theta+ \beta$.   
\elemma
\bproof
$R(u,v)	=	(1-\beta)\frac{w(\mathcal{M})}{N_u} + \beta	\leq	(1-\beta)\frac{N_v}{N_u} + \beta$ \eproof


\comment{ 

} 

\comment{
\blemma
\label{lem:similarity_range}
For $N_u \geq N_v$, since matching $0 \leq w(\mathcal{M}) \leq N_v$, then $R(u,v)$ is in the range $[\beta, (1-\beta)\frac{N_v}{N_u} + \beta]$.
\elemma

\bdefin{\bf $\theta$-similarity:}
Without loss of generality, let $N_u \geq N_v$.
Define $\theta^\prime = (\theta - \beta)/(1 - \beta)$.
If $N_v \geq \theta^\prime N_u$, then their $\theta$-similarity $S_\theta(u,v)=1$; otherwise, $S_\theta(u,v)=0$.
\edefin

We use these lemmata to generate bounded estimates of the RoleSim iterative values, and aggressively prune away those pairs whose RoleSim cannot be greater than or equal to threshold $\theta$, without knowing their exact $RoleSim$ scores.
}

Given this, assuming $N_u \geq N_v$, since matching $0 \leq w(\mathcal{M}) \leq N_v$, then $R(u,v)$ is in the range $[\beta, (1-\beta)\frac{N_v}{N_u} + \beta]$.
Furthermore, to facilitate our discussion, we further define $\theta^\prime = (\theta - \beta)/(1 - \beta)$.
Now, we introduce the following {\em pruning rules} to filter out those pairs whose RoleSim cannot be greater than or equal to threshold $\theta$, without knowing their exact $RoleSim$ scores (Without loss of generality, let $N_u \geq N_v$): 

\begin{enumerate}
\setlength{\itemsep}{2pt}
\setlength{\parskip}{0pt}
\setlength{\parsep}{0pt}
\item If $N_v < \theta^\prime N_u$, then $R(u,v) < \theta$
\item If maximal matching weight $w(\mathcal{M}) < \theta^\prime N_u$,  then $R(u,v) < \theta$
\item Assume neighbor lists $N(u)$ and $N(v)$ are sorted by degree, with $d^u_1$ and $d^v_1$ being the first items.
The maximum possible similarity of this pair is $m_{11} = (1-\beta)\frac{min(d^u_1,d^v_1)}{max(d^u_1,d^v_1)} + \beta$.
If the shorter list has the smaller degree ($d^v_1 \leq d^u_1$), and if $m_{11} + N_v - 1 < \theta^\prime N_u$, then $R(u,v) < \theta$.
\end{enumerate}

Rule 1 is just a restatement of Lemma \ref{lem:theta_similarity}.  Rule 2 is based on the upper bound of RoleSim value. 
Rule 3 requires more explanation: continuing from Rule 2, we begin to consider all the pairings of neighbors.  Because $N_v$ is the shorter list, every member must contribute to the final matching.  Either $m_{11}$ will be in the matching or not. If it is, then an upper bound for $\mathcal{M}$ is if every remaining pair has weight 1, yielding $m_{11} + (N_v - 1)$.  Additionally, because the lists are sorted, $d^v_1/d^u_1 \geq d^v_1/d^u_x$, for $x > 1$. So, if $m_{11}$ is too small to satisfy Rule 2, then all pairings using $d^v_1$ are too small.  This rule allows us to shortcircuit the full neighbor matching 

We now outline our approach, which is formalized in Algorithm~\ref{alg:IcebergRoleSim}. To generate the initial iceberg hash map, we sort nodes by degree (line 3) and sort each node's list of neighbors, by degree (lines 4 to 6).  The first sort allows us to consider only those node-pairs that are sufficiently similar in degree (line 8, pruning rule 1). We compute the estimated similarity for the first pair of neighbors. Note that this estimatation formula is the same as Degree-Ratio initialization. If this weight is below the limit defined in Rule 3, we terminate this pair's candidacy and move on (lines 9 to 12). Otherwise, compute the remainder of neighbor-pair initial similarities, and perform a maximal matching.  If the matching weight exceeds the $\theta^\prime$ minimum bound (Rule 2), then this node-pair and its similarity are inserted into the hash table (lines 13 to 16).  After \rw{interating}{iterating} though all qualified node-pairs, we have our full hash table.  We now perform RoleSim iterations, but only on members of the table, which is orders of magnitude smaller than a complete similarity matrix.
\comment{ 
}
When a non-candidate pair's value is needed (as a neighbor-pair of a candidate pair), we apply the following estimate based on its lower and upper bound (assuming $N_u \geq N_v$): 
\[\tilde{R}(u,v) = \alpha(1-\beta)\frac{N_v}{N_u} + \beta,  \mbox{ where } 0 \leq \alpha \leq 1.\]
In the experimental evaluation, we will empirical study the effect of $\alpha$ on the estimation accuracy. 




\begin{algorithm}
\caption{IcebergRoleSim($G(V,E)$, $\theta$, $\beta$, $\alpha$)}
\label{alg:IcebergRoleSim}
\begin{algorithmic}[1]
\STATE $H \leftarrow$ empty hash table indexed by node-pair ID $(u,v)$;
\STATE $d(v) \leftarrow$ degree of $v$;
\STATE Sort vertices $V$ by degree;
\FORALL {$v \in V$}
	\STATE $D^v = \{d^v_1, d^v_2, \cdots, d^v_{d(v)}\} \leftarrow$ degrees of neighbors of $v$, sorted by increasing order;
\ENDFOR
\FORALL { $u \in V$}
	\FORALL { $v \in V, \theta^\prime d(u) \leq d(v) \leq d(u)$ (Rule 1)}
		\STATE $m_{11} \leftarrow (1-\beta)\frac{min(d^u_1,d^v_1)}{max(d^u_1,d^v_1)} + \beta$;
		\IF {$d^v_1 \leq d^u_1$ and $N_v - 1 + M_{11} < \theta^\prime N_u$}
			\STATE Skip to the next $v$; (Rule 3)
		\ENDIF
		\STATE Compute maximal matching weight $w(\mathcal{M})$;
		\IF {$w(\mathcal{M}) \geq \theta^\prime d(u)$ (Rule 2)}
			\STATE Insert $H(u,v) \leftarrow (1-\beta)w(\mathcal{M})/d(u) + \beta$;
		\ENDIF
	\ENDFOR
\ENDFOR
\STATE Perform iterative RoleSim on $H$. For neighbor pairs $\notin H$, use $\tilde{R}(x,y) = \alpha(1-\beta) N_x/N_y+\beta$
\end{algorithmic}
\end{algorithm}

\comment{
For IcebergRoleSim, the effectiveness of the pruning is graph-dependent. The overall complexity depends on the aggressiveness of the pruning.  Aggressive pruning will take more time in initialization, resulting in faster time and less memory for Hash-RoleSim.

Let $h_0 << n^2 $ be the number of node pairs that are $\theta$-similar. Let $|H| < h_0 $ be the fraction of node-pairs that are candidates after degree-matching pruning.  $\hat{d}$ = average of $max(N_u,N_v)$ over all candidate pairs:\\
\noindent Sort degree vectors (Steps 4 - 6): $O(n \cdot \hat{d}log\hat{d})$		\\
\noindent Inner loop (Steps 11-18): $O(\hat{d})$		\\
\noindent Outer loops (Steps 8,9) times inner loop: $O(h \cdot \hat{d})$	\\
\noindent Hash-RoleSim (Step 24): $(k \cdot |H| \hat{d})$		\\
}

\comment{ 

Since the space complexity is O($n^2$), it needs at least 40GB memory to work on a graph of size 100,000, which is unacceptable for normal computer configuration.
 
A more aggressive scalability approach is to consider each initialization partition as a solid block and then to perform only block-block similarity iterative computations.

{\bf Why Block-block similarity}: The scalability issue is that naive Role-Sim algorithm suffers a big memory problem. Since the space complexity is O($n^2$), it needs at least 40GB memory to work on a graph of size 100,000, which is unacceptable for normal computer configuration. While we can not compute the similarity for any pair in a large graph, it's only possible to compute the similarity on a high level.

In theorem 1, if the similarities for any two pairs is almost the same, we can claim the cross similarities are all equal. With this insight, we can group the nodes into blocks and compute the similarity on block level.

{\bf How to do Block-block similarity}: To perform RoleSim at the block-level, we need a description of the "mean" block-level neighborhood: Compute the neighbor color spectrum of each node, that is, a weighted vector where the $i^{th}$ element is the number of neighbors which belong to role $i$.  For each block $B_i$: compute the mean neighbor color spectrum, akin to a cluster's centroid.  For each pair of blocks, perform a RoleSim iterative computation, but using fractional greedy matching instead of discrete greedy matching.  This is similar to the fractional backpack problem. To compute $RoleSim^{i+1}(B1,B2)$:\\
\noindent(1) Initialize WeightAvailable $w = min(deg(B1),deg(B2))$ and Matching $m = 0$.\\
\noindent(2) Select the unused neighbor-pair with the highest $RoleSim^i$ value.  The first choice will be one of the diagonals, with value 1.\\
\noindent(3) Now, allocate as much fractional weight from the neighborhoods that we can.  If step (2) selected neighbor-pair $(B_x,B_y)$, then subtract from element $x$ of $B1$ and from element $y$ of $B2$ an amount equal to $min(min(B1_x, B2_y),w)$.  If there was a tie in step 2, we want to break the tie by selecting the highest result for step (3).\\
\noindent(4) $m \leftarrow m + min(B1_x, B2_y,w) \times RoleSim^i(B_x,B_y)$;\\
$w \leftarrow w - min(B1_x, B2_y,w)$\;\
\noindent(5) Eliminate $(B_x,B_y)$ from consideration.  Repeat steps (2) -(4) until all $w = 0$.\\

\comment{
Example: we are matching two blocks which have neighbor spectra $N_{B1} = (0.9, 1.5, 1.2, 0.4)$ and $N_{B2} = (1.2, 0.6, 1.2)$.  Note that the total of each block's neighbor weights equals the characteristic node degree of its members.\\
\noindent(Step 1) The degrees of $B1$ and $B2$ are 4 and 3, respectively, so $w = 3$.\\
\noindent(Step 2) For the 1st iteration, we could select any of the diagonal neighbor-pairs (1,1), (2,2), or (3,3).\\
\noindent(Step 3) The three choices from step 2 would allow us to allocate weights of $min(0.9,1.2), min(1.5,0.5), min(1.2,1.2)$, respectively.  The best choice is (3,3) with weight 1.2.\\
\noindent(Step 4) Updated neighbor spectra $N_{B1} = (0.9, 1.5, 0.0, 0.4)$,
$N_{B2} = (1.2, 0.6, 0.0)$,\\
$m \leftarrow m + min(B1_x, B2_y,w) \times RoleSim^i(B_x,B_y) = 0 + 1.2 \times 1 = 1.2$;\\
$w \leftarrow w - min(B1_x, B2_y,w) = 3 - 1.2 = 1.8$;\\
} 

{\bf Within-block similarity}: For each block $B_i$, compute a neighbor spectrum variance:
$\frac{\sum_{u \in B_i} (Spec(u) - Spec(\bar{B})^2}{|B_i|}$.  If the variance is sufficiently high(?), split the block into two, using K-means or a simple spectral method.

\comment{
, then iteratively compute for both within-block and between-block similarity.  A similar idea was proposed for SimRank improvement~\cite{Li09_blocksim}.  If $B(u)$ is the block to which node $u$ belongs, then the estimated node-node similarity $R(u,v) \approx R(u,B(u))R(B(u),B(v))R(v,B(v))$.  If the initial partition and between-block similarity is a admissible similarity measure, the subsequent iterations of between-block similarity will also be admissible (Proof omitted due to space limitation).
}

\comment{
If $m$ is the number of blocks with roughly equal-sized blocks, then the time drops to $O(k\bar{d}^2(m^2 + n^2/m))$.  We can use the $D0$ or $D1$ degree-based initialization to partition the network.  \cite{Li09_blocksim} shows that if optimal $m^\ast = (2n)^{3/2}/2$ is used, then the time drops to $O(k\bar{d}^2 n^{4/3})$.  The total space requirements drop similarly.  However, since each block can be processed independently, the main memory requirement drops even farther to $O(n^2/m^2) = O(n^{2/3})$.
}

\noindent 3. Set $R^0(x,y) = \frac{min(N_x,N_x)}{max(N_x,N_y)}$  So, if $x$ and $y$ are in the same initialization class, similarity = 1.  Otherwise, it is based on the ratio of their sizes.\\
\noindent 4. \\

\noindent 5. Intra-block RoleSim iteration: $R^{i+1}(x,y) = \frac{\mathcal{M}(B(x),B(y))}{max(B(x),B(y))}$\\

{\bf Cache for accuracy}: The block-block similarity might suffer the inherited accuracy problem, since it works on a high level other than the single vertex level. How can we reduce the error rate or improve the accuracy for block-block similarity when compared to naive Role-Sim algorithm? One observation is that we care those pairs with high similarities instead of those with low similarities. 

With this observation, we can try to record only high similarities in a cache and use them for the next iteration, without computing the average value with them. In the next iteration, we should firstly check the cache for high similarities and then go to block-block matrix for low similarities.

{\bf Speedup for scalability}: Another problem for scalability involves time. It might take a long time to get results for large graphs.

\input{Figures/alg_Block.tex}

\input{Figures/alg_Shell.tex}

} 

\section{Experimental Evaluation}
\label{section:experiment}

In this section we experimentally investigate the ranking ability and performance of the RoleSim algorithm for computing role similarity metric values.  We compare RoleSim to several state-of-the-art node similarity algorithms, analyze the effect of different initialization schemes, and measure the scalability of Iceberg RoleSim.
Specifically, we focus on the following questions:
\benum
\setlength{\itemsep}{1pt}
\item How do different initialization schemes perform in terms of their final RoleSim score and computational efficiency? 
\item Do node-pairs with high RoleSim scores actually have similar network roles? For any two nodes known to have similar network roles, do they receive high role similarity scores? 
\item How much less memory and time does Iceberg RoleSim use, and how closely does its rankings match standard RoleSim's?
\eenum
Clearly, the ideal validation study requires an explicit role model and role similarity measure, which often do not exist.
In the following study, we utilize a well-known role-related random graph model and external measures of real datasets which provide strong role indication for these evaluations. 

We set $\beta = 0.1$ for both RoleSim and SimRank, defining convergence to be when values change by less than 1\% of their previous values.   We ran several RoleSim tests with both exact matching and greedy matching.  The results were nearly identical ($> 90\%$ of cells have no difference; maximum difference was small), so we focus on greedy matching from here on. We implemented the algorithms in C++ and ran all large tests on a 2.0GHz Linux machine with dual-core Opteron CPU and 4.0GB RAM.

For our tests, we use three types of graphs:\\
\noindent$\bullet$ {\bf BL}: the probabilistic block-model~\cite{wang87_stochastic}, where each block is generally considered to be corresponding to a role~\cite{White76_blockmodel}. Here, nodes are partitioned into blocks. Each node in block $i$ has probability $p_{ij}$ of linking to each node in block $j$. Thus, the underlying block-model may serve as the ground-truth for testing role similarity. \\
\noindent$\bullet$ {\bf SF}: Large Scale-Free random graphs \footnote{http://pywebgraph.sourceforge.net/} are used for testing scalability of the Iceberg RoleSim computation. \\
\noindent$\bullet$ Real-world networks, with a measureable feature similar to social role, are used for validating RoleSim performance. 

\comment{ 
\subsection{Case Study with Simple Synthetic Network}

We start with a small network (Figure~\ref{fig:synth_example3}) which embodies an explicit role model.  It features two similar but not identical modules, each with three roles, indicated by the color shading.  We test this using four similarity measures: RoleSim, SimRank, SimRank++~\cite{Antonellis08_simrankpp}, and PSimRank~\cite{Fogaras05_scaleSim}.

\begin{figure}[bt]
\centering
\epsfig{file=Figures/synth_example3.eps,scale=0.9,width=2.8in, height = 1.1in}
\vspace*{-2.0ex}
\caption{Demonstration Network}
\label{fig:synth_example3}
\end{figure}

Figure~\ref{fig:rs_vs_sr} summarizes how RoleSim and SimRank generate different similarity results.  White equals 1.0; black equals the minimum similarity value for that test. In \ref{fig:synth_rs_class} and \ref{fig:synth_sr_class}, the rows and columns group automorphically equivalent nodes together, evidenced by RoleSim's white blocks (\ref{fig:synth_rs_class}). The uniform block structure also demonstrates transitive similarity. For example, each of $\{2,3,4,5\}$ has the same similarity to each of $\{11,12,13\}$.  RoleSim scores equivalent nodes correctly.

For SimRank, however, the automorphic blocks (e.g., $\{2,3,4,5\}$) are a mix of white and various shades of gray.  In fact, the SimRank value within a role block (say, $SR(2,5) = .42$), may be no better than the similarity between roles ($SR(2,9) = .46$). Therefore, SimRank fails to meaningfully distinguish automorphically equivalent node-pairs.  Since it fails automorphism, it also fails transitivity (the interior of blocks are not all one color).

If we rearrange the rows and columns to separate the left-right modules (\ref{fig:synth_rs_clust} and \ref{fig:synth_sr_clust}), we see that in this example, SimRank is focusing more on clustering than role similarity.

\comment{
}

\begin{figure}[t]
\centering
	\subfigure[RoleSim, ordered by class]{\label{fig:synth_rs_class}\includegraphics[width=1.6in,height=1.45in]{Figures/synth_rs_class.eps}}
	\subfigure[SimRank, ordered by class]{\label{fig:synth_sr_class}\includegraphics[width=1.6in,height=1.45in]{Figures/synth_sr_class.eps}}
	\subfigure[RoleSim, ordered by cluster]{\label{fig:synth_rs_clust}\includegraphics[width=1.6in,height=1.45in]{Figures/synth_rs_clust.eps}}
	\subfigure[SimRank, ordered by cluster]{\label{fig:synth_sr_clust}\includegraphics[width=1.6in,height=1.45in]{Figures/synth_sr_clust.eps}}
	\vspace*{-1.0ex}
	\caption{Comparing RoleSim and SimRank Results}
	\label{fig:rs_vs_sr}
\end{figure}

\comment{
The SimRank values are not as well-behaved. First, SimRank does not detect any automorphism.  Without automorphic equivalence, we cannot extend this to full transitive similarity: it does not form a pure block model.  The highest ranked similarity is among $\{11,12,13\}$.  This is reasonable, since they either share the same neighbors or the same neighbors of neighbors. However, one would expect the pair $(2,3)$ to be similar as least as strongly.  Of their four neighors, two are identical $\{0,9\}$, one is each other, and the other is an automorphically equilvalent pair $\{4,5\}$ which in fact are a mirror image of themselves.  Howevre, their SimRank value of 0.45 is tied for only the $7^{th}$ highest ranked pairing.  Strangely, core node 1's similarity to outer group $\{11,12,13\}$ is higher, at 0.48.  RoleSim gave this pairing its lowest ranking.  Finally, the similarity between 0 and 1 is near the bottom of the ranking scale.  This is because 0 and 1 are direct neighbors, but SimRank can only compare elements that are an even-numbered distance apart.
}

} 

\comment{
\begin{table}[t]
	\centering
	\tiny
	\begin{tabular*}{0.47\textwidth}{c|p{4pt} p{4pt} p{4pt} p{4pt} p{4pt} p{4pt} p{4pt} p{4pt} p{4pt} p{4pt} p{4pt} p{4pt} p{4pt} p{4pt}} \hline
	id	&	0	&	1	&	2	&	3	&	4	&	5	&	6	&	7	&	8	&	9	&	10	&	11	&	12	&	13	\\ 	\hline
	\hline																														
	0	&	1	&	.53	&	.64	&	.64	&	.64	&	.64	&	.63	&	.63	&	.63	&	.46	&	.46	&	.33	&	.33	&	.33	\\ 	\hline
	1	&	.53	&	1	&	.68	&	.68	&	.68	&	.68	&	.56	&	.56	&	.56	&	.39	&	.39	&	.55	&	.55	&	.55	\\ 	\hline
	2	&	.64	&	.68	&	1	&	1	&	1	&	1	&	.37	&	.37	&	.37	&	.55	&	.55	&	.39	&	.39	&	.39	\\ 	\hline
	3	&	.64	&	.68	&	1	&	1	&	1	&	1	&	.37	&	.37	&	.37	&	.55	&	.55	&	.39	&	.39	&	.39	\\ 	\hline
	4	&	.64	&	.68	&	1	&	1	&	1	&	1	&	.37	&	.37	&	.37	&	.55	&	.55	&	.39	&	.39	&	.39	\\ 	\hline
	5	&	.64	&	.68	&	1	&	1	&	1	&	1	&	.37	&	.37	&	.37	&	.55	&	.55	&	.39	&	.39	&	.39	\\ 	\hline
	6	&	.63	&	.56	&	.37	&	.37	&	.37	&	.37	&	1	&	1	&	1	&	.33	&	.33	&	.46	&	.46	&	.46	\\ 	\hline
	7	&	.63	&	.56	&	.37	&	.37	&	.37	&	.37	&	1	&	1	&	1	&	.33	&	.33	&	.46	&	.46	&	.46	\\ 	\hline
	8	&	.63	&	.56	&	.37	&	.37	&	.37	&	.37	&	1	&	1	&	1	&	.33	&	.33	&	.46	&	.46	&	.46	\\ 	\hline
	9	&	.46	&	.39	&	.55	&	.55	&	.55	&	.55	&	.33	&	.33	&	.33	&	1	&	1	&	.68	&	.68	&	.68	\\ 	\hline
	10	&	.46	&	.39	&	.55	&	.55	&	.55	&	.55	&	.33	&	.33	&	.33	&	1	&	1	&	.68	&	.68	&	.68	\\ 	\hline
	11	&	.33	&	.55	&	.39	&	.39	&	.39	&	.39	&	.46	&	.46	&	.46	&	.68	&	.68	&	1	&	1	&	1	\\ 	\hline
	12	&	.33	&	.55	&	.39	&	.39	&	.39	&	.39	&	.46	&	.46	&	.46	&	.68	&	.68	&	1	&	1	&	1	\\ 	\hline
	13	&	.33	&	.55	&	.39	&	.39	&	.39	&	.39	&	.46	&	.46	&	.46	&	.68	&	.68	&	1	&	1	&	1	\\ 	\hline
	\end{tabular*}
	\caption{RoleSim test (\small D0 init., exact matching, $\beta=0.1$)} \label{tab:test_synth_rolesim}
\end{table}
}
\comment{
\begin{table}[t]
	\centering
	\tiny
	\begin{tabular*}{0.47\textwidth}{c|p{4pt} p{4pt} p{4pt} p{4pt} p{4pt} p{4pt} p{4pt} p{4pt} p{4pt} p{4pt} p{4pt} p{4pt} p{4pt} p{4pt}} \hline	
	id	&	0	&	1	&	2	&	3	&	4	&	5	&	6	&	7	&	8	&	9	&	10	&	11	&	12	&	13	\\ 	\hline
	\hline																														
	0	&	1	&	.20	&	.40	&	.40	&	.40	&	.40	&	.20	&	.20	&	.20	&	.46	&	.46	&	.16	&	.16	&	.16	\\ 	\hline
	1	&	.20	&	1	&	.22	&	.22	&	.22	&	.22	&	.40	&	.40	&	.40	&	.18	&	.18	&	.48	&	.48	&	.48	\\ 	\hline
	2	&	.40	&	.22	&	1	&	.45	&	.49	&	.42	&	.14	&	.14	&	.14	&	.45	&	.45	&	.13	&	.13	&	.13	\\ 	\hline
	3	&	.40	&	.22	&	.45	&	1	&	.42	&	.49	&	.14	&	.14	&	.14	&	.45	&	.45	&	.13	&	.13	&	.13	\\ 	\hline
	4	&	.40	&	.22	&	.49	&	.42	&	1	&	.45	&	.14	&	.14	&	.14	&	.45	&	.45	&	.13	&	.13	&	.13	\\ 	\hline
	5	&	.40	&	.22	&	.42	&	.49	&	.45	&	1	&	.14	&	.14	&	.14	&	.45	&	.45	&	.13	&	.13	&	.13	\\ 	\hline
	6	&	.20	&	.40	&	.14	&	.14	&	.14	&	.14	&	1	&	.48	&	.48	&	.13	&	.13	&	.46	&	.51	&	.46	\\ 	\hline
	7	&	.20	&	.40	&	.14	&	.14	&	.14	&	.14	&	.48	&	1	&	.48	&	.13	&	.13	&	.46	&	.46	&	.51	\\ 	\hline
	8	&	.20	&	.40	&	.14	&	.14	&	.14	&	.14	&	.48	&	.48	&	1	&	.13	&	.13	&	.51	&	.46	&	.46	\\ 	\hline
	9	&	.46	&	.18	&	.45	&	.45	&	.45	&	.45	&	.13	&	.13	&	.13	&	1	&	.40	&	.12	&	.12	&	.12	\\ 	\hline
	10	&	.46	&	.18	&	.45	&	.45	&	.45	&	.45	&	.13	&	.13	&	.13	&	.40	&	1	&	.12	&	.12	&	.12	\\ 	\hline
	11	&	.16	&	.48	&	.13	&	.13	&	.13	&	.13	&	.46	&	.46	&	.51	&	.12	&	.12	&	1	&	.54	&	.54	\\ 	\hline
	12	&	.16	&	.48	&	.13	&	.13	&	.13	&	.13	&	.51	&	.46	&	.46	&	.12	&	.12	&	.54	&	1	&	.54	\\ 	\hline
	13	&	.16	&	.48	&	.13	&	.13	&	.13	&	.13	&	.46	&	.51	&	.46	&	.12	&	.12	&	.54	&	.54	&	1	\\ 	\hline
	\end{tabular*}
	\caption{SimRank test (\small D0 init., exact matching, $\beta=0.1$)} \label{tab:test_synth_simrank}
\end{table}
} 
\subsection{Comparing Initialization}

In Section~\ref{sec:init} we discussed that Degree-Ratio initialization generates the same results as ALL-1 by shortcutting the first iteration.  This reduces the computation time by roughly $10\%$. Now we ask: Does Degree-Binary initialization (DB, binary indicator \rw{equaling}{which equals} 1 when degrees $N_u = N_v$) give \rw{good}{similar} results, quickly?
\x{We define a "good" result as having the same similarity {\em rankings} as ALL-1.}

We ran RoleSim using both ALL-1 and DB on 12 graphs, some scale-free and some block-model, having 500 to 10,000 nodes, and edge densities from 1 to 10. We then converted values to percentile ranking, where $100\%$ means the highest value and $50\%$ is the median value.  Test results are summarized in Table \ref{tab:init}.  The high correlation coefficient means the rankings are virtually identical\rw{}{, so the rankings are not very sensitive to the initialization method}. Moreover, DB took $20\%$ from $68\%$ less time to converge. Overall, $DB$ seems to be the preferred initialization scheme in terms of computational efficiency. 
Thus, we adopt it for the rest of the experiments. 

\begin{table}[tb]
	\centering
	\small
	\begin{tabular}{|c|c|c|c|c|} \hline
	Relative to	All-1	&	\multicolumn{3}{|c|}{Degree-Binary} 	& Degree-	\\
	Initialization		&		Min			&	Avg.	&	Max		& Ratio		\\ \hline
	Diff. in percentile rank 	&	0.14\%		&	0.38\%	&	11.17\%	&	none	\\
	Pearson correl. coeff.		&	0.9994	&	0.9998	&	0.9999	&	1		\\
	Relative execution time		&	0.32	&	0.52	&	0.80	&	$\approx 0.9$	\\
	Relative \# iterations		&	0.38	&	0.58	&	0.88	&	1 fewer	\\ \hline
	\end{tabular}
	\caption{Comparison of Initialization Methods}
	\label{tab:init}
\end{table}

\subsection{General Role Detection}
How well does RoleSim discover roles in complex graphs? 
Specifically, given a ground truth knowledge of roles, do nodes having similar roles have high scores? 
To answer this question, we generated probabilistic block-model graphs, where blocks behave like "noisy" roles, due to sampling variance.
We generated graphs with $N = 1000$ nodes and either 3 or 5 blocks. We varied the edge density $\frac{|E|}{|V|}$, with higher densities for graphs with more blocks. The size of each block and the $p_{ij}$ values were randomized; we generated 3 random instances for each graph class.  We compared RoleSim to the state-of-the-art SimRank, SimRank++~\cite{Antonellis08_simrankpp}, and P-SimRank~\cite{Fogaras05_scaleSim}.

For each measure and trial, we ranked the similarity scores. This serves to normalize the scoring among the four measures.  Then, for each graph, we computed the average ranking of all pairs of nodes within the same block.  We then averaged the three trials for each graph class.

\begin{figure}[tb]
\centering
	\includegraphics[width=1.1\columnwidth,height=1.4in]{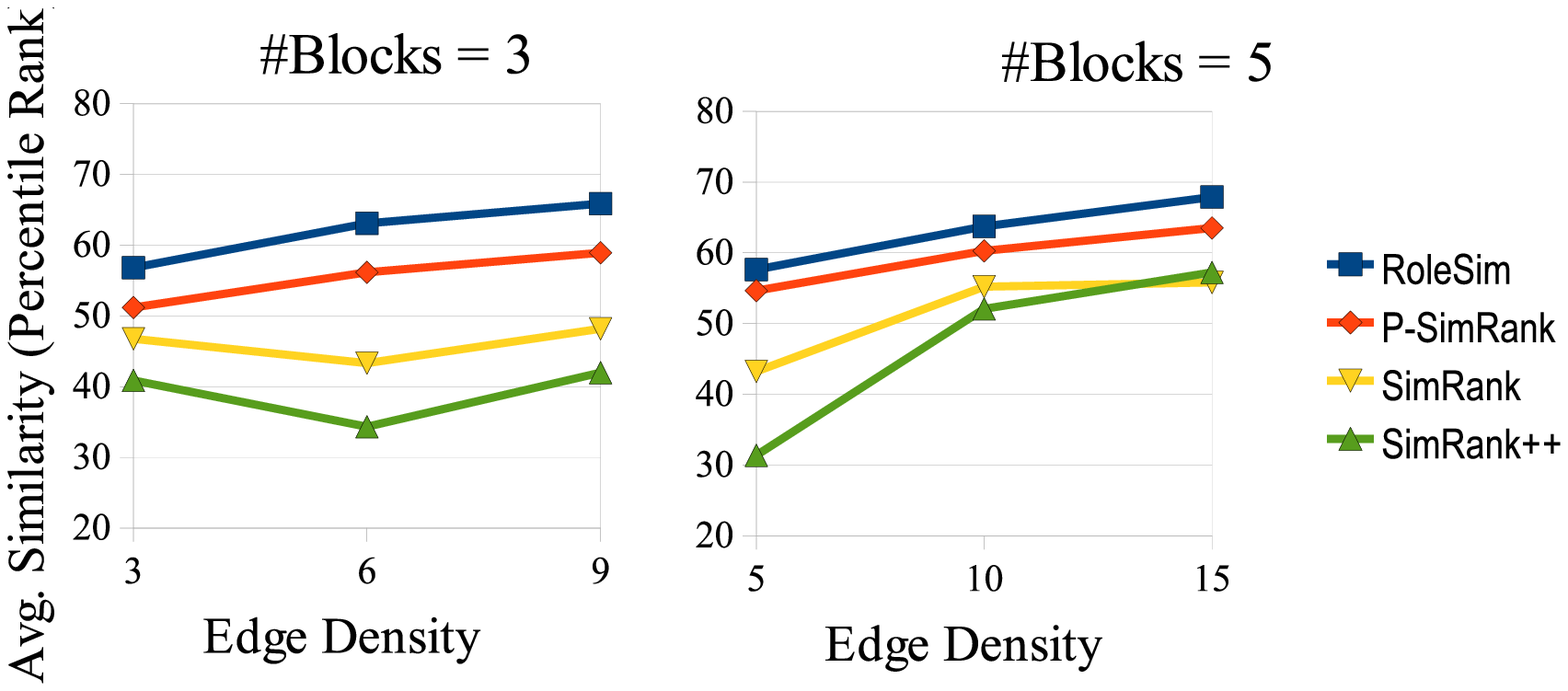}
	\caption{Avg. similarity ranking for nodes in the same block}
	\label{fig:block-within}
\end{figure}

Our results (Figure \ref{fig:block-within}) show that RoleSim outperforms all other algorithms across all the tested conditions.  None of the algorithms score perfectly, due to the inherent edge distribution variance of the probabilistic model.  P-SimRank is better than SimRank, perhaps because it uses Jaccard Coefficient weighting, a step towards our RoleSim approach.
Accuracy takes time. SimRank and SimRank++ run at the same speed.  P-SimRank is about 1.5 to 2 times slower, and standard RoleSim is about twice as slow as SimRank.

\subsection{Real Dataset: Co-author Network}
We applied RoleSim and the best alternative measure, P-SimRank, to a real-world network having an external role measure.
Our first dataset~\cite{Tang08_socnet} is a co-author network of 2000 database researchers.
Two authors are linked if they co-authored a paper from 2003 to 2008.  We pruned the network to the largest connected component (1543 nodes, 15483 edges).  An author's role depends recursively on the number of connections to other authors, and the roles of those others.  Hence, it measures collaboration.  We use the G-index as a proxy measure for co-author role (H-index provides similar results and thus is omitted here).
The G-index measures the influence of a scientific author's publications, its value being the largest integer $G$ such that the $G$ most cited publications have at least $G^2$ citations.  While G-index and co-author role are not precisely the same, G-index score is influenced strongly by the underlying role.  High impact authors tend to be highly connected, especially with other high impact authors.  If a paper is highly cited, this boosts the score of every co-author.  Thus, we expect that if two authors have similar G-index scores, their node-pair is likely to have a high role simlarity value.
To normalize RoleSim, P-SimRank, and G-index values, we converted each raw value to a percentile rank.

\begin{figure}[t]
\centering
	\subfigure[Top Coauthors]{\label{fig:coauthor-topK}\includegraphics[width=1.6in,height=1.8in]{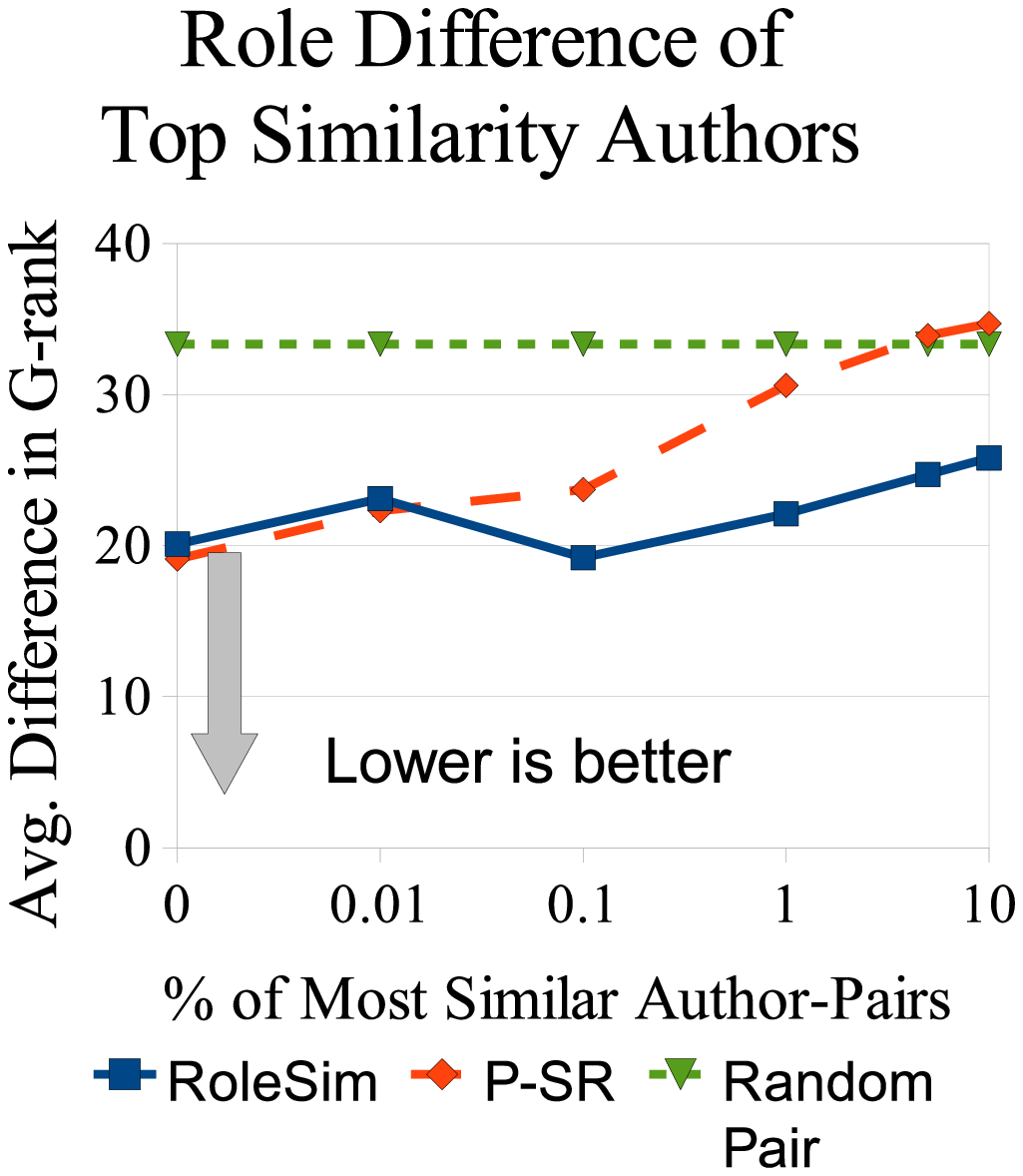}}
	\subfigure[Top Internet nodes]{\label{fig:internet-topK}\includegraphics[width=1.6in,height=1.8in]{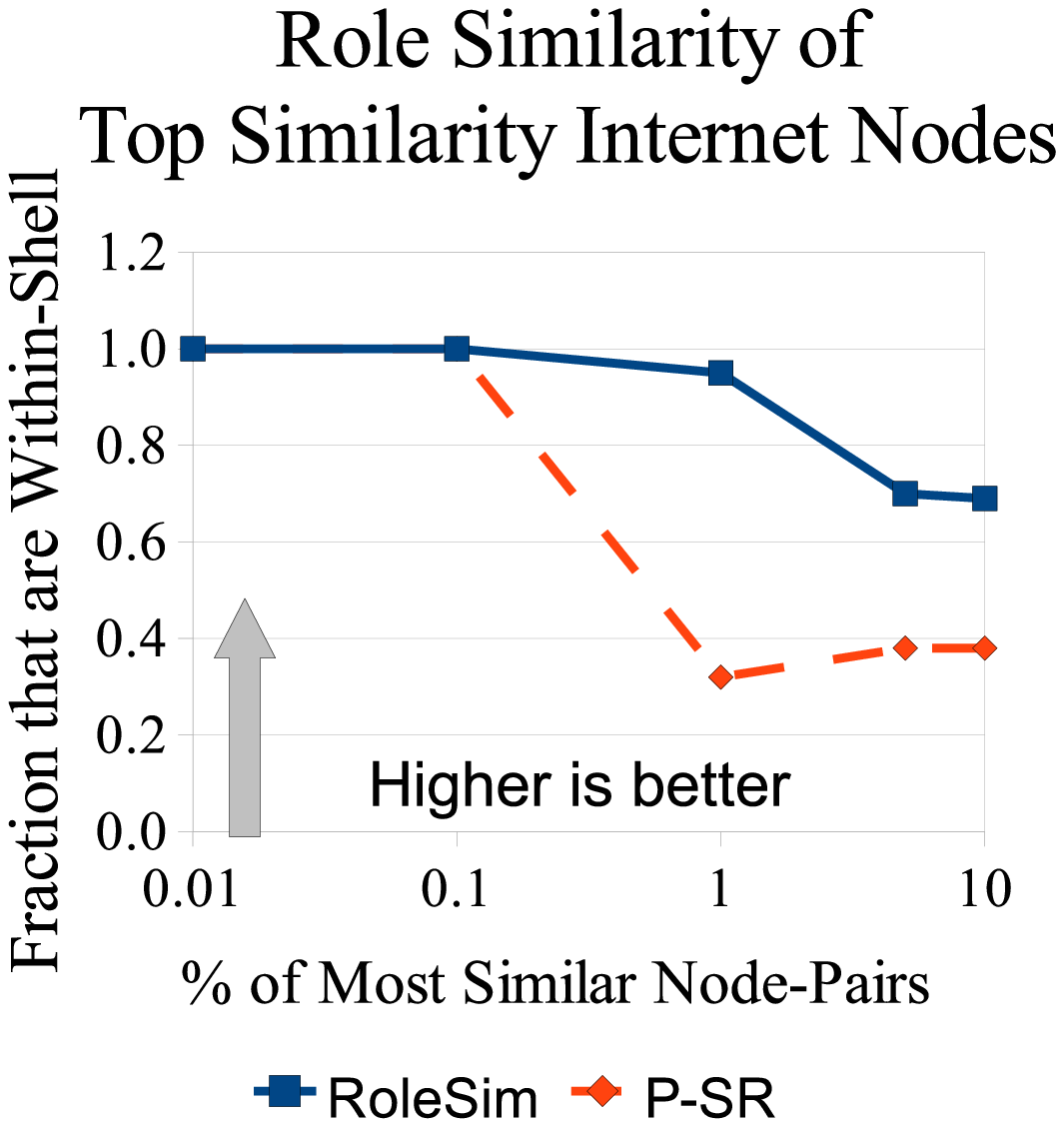}}
	\vspace*{-1.0ex}
	\caption{Similarity of Nodes for Top Ranked Node-Pairs}
	\label{fig:topK}
\end{figure}

Figure~\ref{fig:coauthor-topK} addresses our second validation question (high rank$\rightarrow$ similar roles?). For the top ranked 0.01\% of author-pairs, their difference in G-index ranking is about 20 points, for both RoleSim and P-SimRank, well below the random-pair value of 33.  A below-average difference confirms that the authors are relatively similar.  However, as we expand the search towards 10\%, RoleSim continues to detect authors with similar authorship performance, while P-SimRank converges to random scoring. 


\begin{figure}[bt]
\centering
\epsfig{file=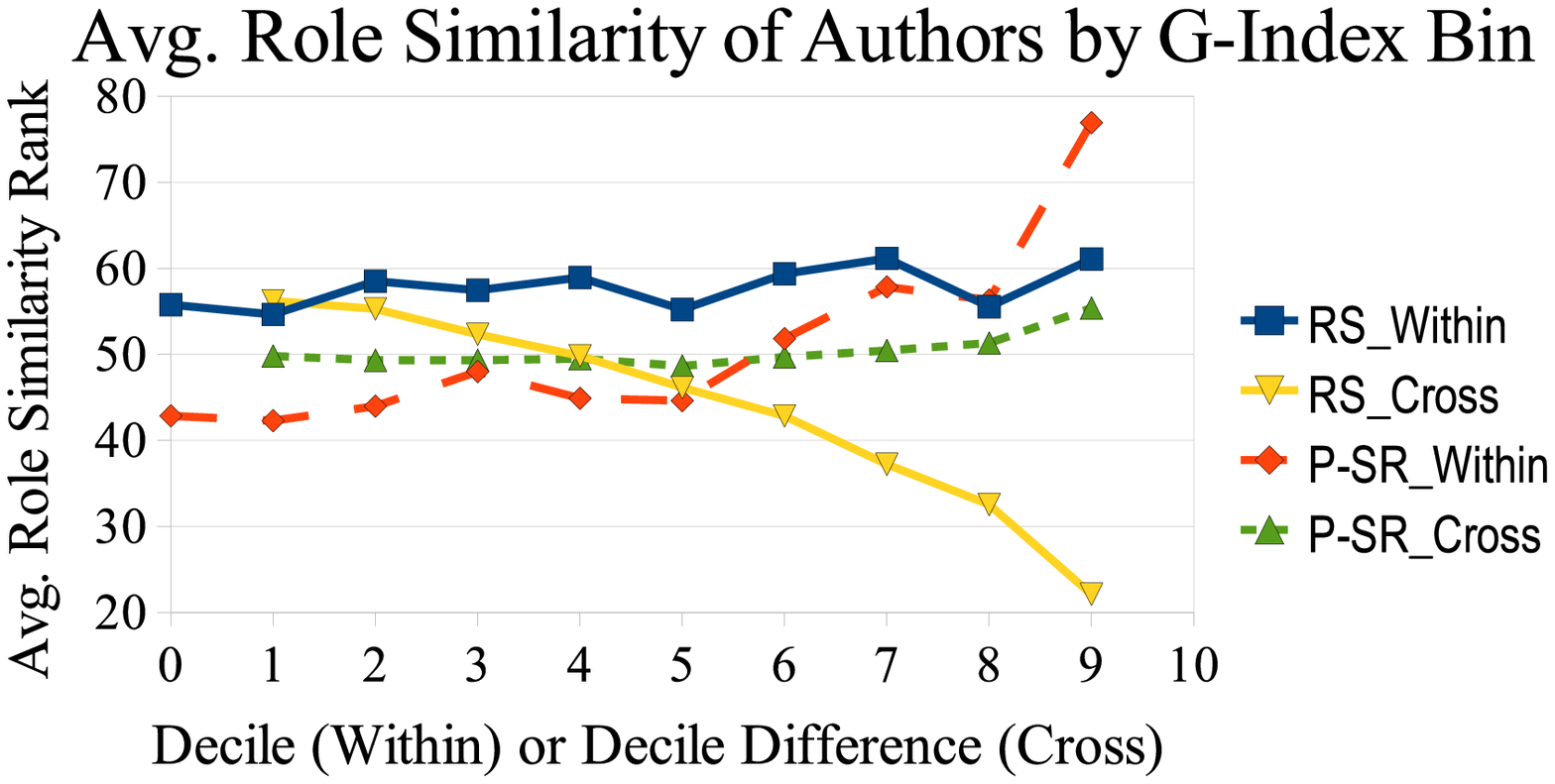,width=3.3in, height = 1.5in}
\caption{Similarity of Authors Binned by K-index}
\vspace*{-3.0ex}
\label{fig:coauthor_bins}
\end{figure}

To validate $role\rightarrow rank$ performance, we binned the authors into 10 roles based on G-index value (bottom 10\%, next 10\%, etc.).  For every pair of authors within the same role decile, we looked up role similarity percentile rank and computed an average per bin.  We also computed averages for pairs of authors not in the same bin (dissimilar roles).  Figure~\ref{fig:coauthor_bins} shows our results.  The average within-bin RoleSim value is consistently between 55\% and 60\%, better than the random-pair score of 50, and independent of whether the G-index is high or low.  It performs equally well for all roles.  P-SimRank within-bin scores (dashed line), however, are inconsistent.  Performance of P-SimRank is worse than random for low G-scores, perhaps due to low density of links in the network.  For the cross-bin data, the X-axis is the difference in decile bins for the two authors in a pair. The falling line of RoleSim indicates that role similarity correctly decreases as G-index scores become less similar.  For P-SimRank, however, the cross-bin scores (dashed line) hover around 50, equivalent to random scoring.

\subsection{Real Dataset: Internet Network}
Our second dataset is a snapshot of the Internet at the level of autonomous systems (22963 nodes and 48436 edges), as generated by~\cite{newman_internet}. Several studies have confirmed that the Internet is hierarchically organized, with a densely connected core and stubs (singly-connected nodes) at the periphery~\cite{Tauro01_jellyfish,carmi07_kshelldecomp}. A node's position within the network (proximity to the core) and its relation to others (such as density of connections) affects its efficiency for routing and its robustness. Inspired by ~\cite{carmi07_kshelldecomp}, we use $K$-shells to delineate roles.

The $K$-core of a graph is the induced subgraph where every node connects to at least $K$ other nodes in the subgraph.  If $K^\prime > K$, then the $K^\prime$-core must be an induced subgraph of the $K$-core.  The $K$-shell is defined as the 'ring' of nodes that are included in a graph's $(K-1)$-core but not its $K$-core.  In other words, we can decompose a graph into a set of nested rings, becoming denser as we move inward.

Using K-shells as our roles, we perform tests and analyses similar to those of the coauthor network.
In Figure~\ref{fig:internet-topK} we see that both measures do well for the top 0.1\%, but P-SimRank's falters significantly when the range is expanded to the top 1\%.

\begin{figure}[bt]
\centering
\epsfig{file=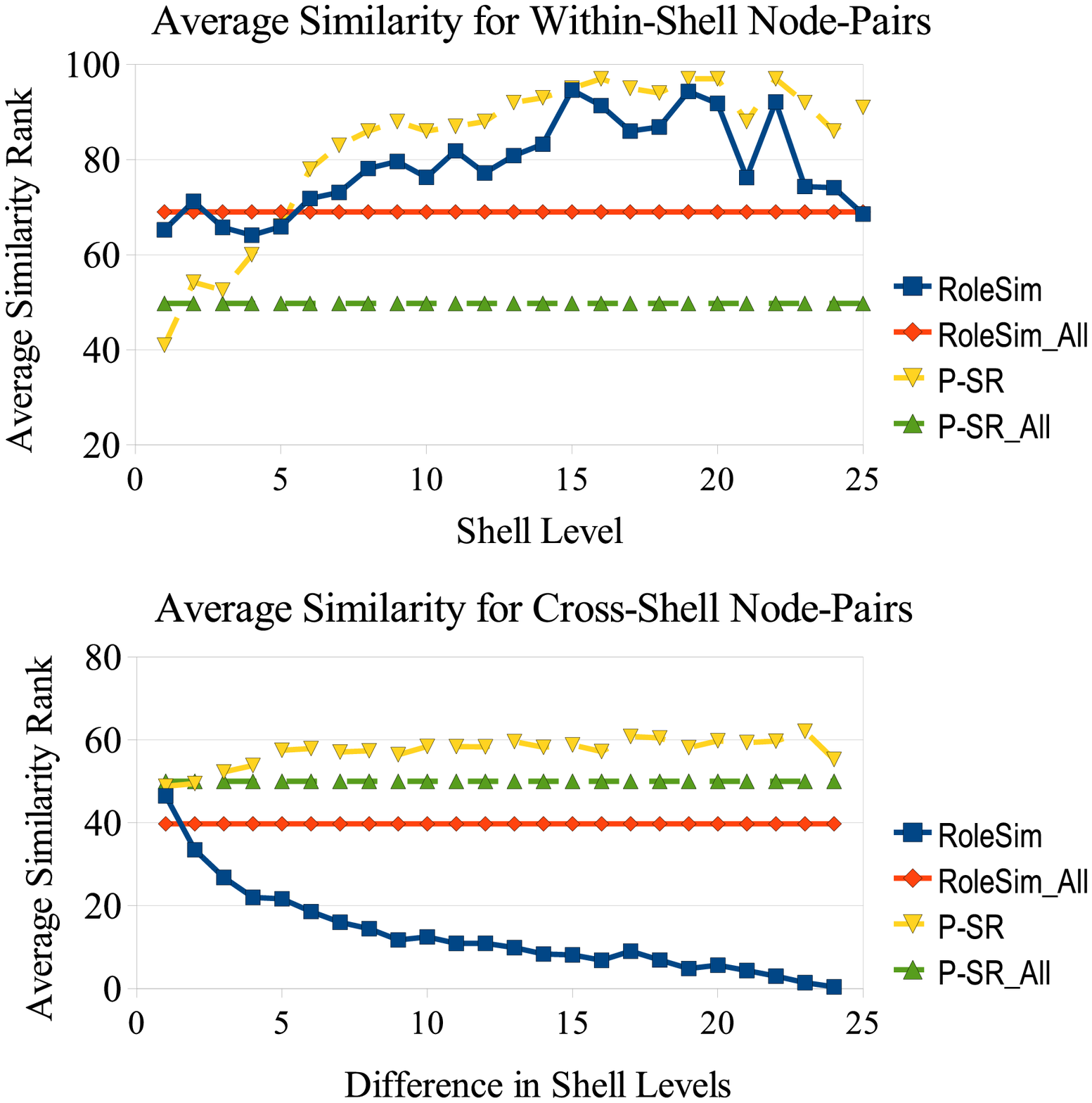,width=3.3in, height = 3.0in}
\caption{Similarity of Authors Grouped By K-Shell}
\vspace*{-1.0ex}
\label{fig:internet-shells}
\end{figure}

Next, we treat $K$-shells the same way that we treated G-index decile bins in the previous test.  See Figure~\ref{fig:internet-shells}. Unlike decile bins, the shells do not have equal sizes.
K-shells 1, 2, and 3 together contain 92\% of all nodes.
To clarify how these three shells dominate, we also show horizontal lines representing the combined weighted average rank of all within-shell comparisons.  RoleSim's within-shell values are consistently high, averaging 70\%. Conversely, P-SimRank finds strong above-average similarity for the small high-K shells, but nearly random similarity for shells 1 to 3, pulling its overall performance down to 50\%.

In cross-shell analysis, RoleSim is able to distinguish different shells very well:
RoleSim approaches zero as shell difference approaches maximum.  On the other hand, P-SimRank shows almost no correlation to shell difference.  Many of its scores are above-average when they should be below-average (dissimilar).  On the whole, it seems that P-SimRank is not detecting role, but something related to connectedness and density.

In all these experiments, we can see that RoleSim provides positive answer to the role similarity ranking: 
1) node-pairs with similar roles have higher RoleSim ranking than node-pairs with dissimilar roles,
and 2) high RoleSim ranking indicates that nodes have similar roles. 
P-SimRank scores, however, do not correlate with network role similarity. 
 
\comment{
For each of the four sets of scores, we mapped each score to a percentile ranking.  For example, the median number of papers would have a ranking of 0.50.  We define an author's productivity rank $PR_a$ as the average and $\sigma(PR_a)$ as the sample standard deviation of her four scores. Our objective is to see how well the similarity of two author's $PR$ scores can be predicted from the network topology alone.
We ran both RoleSim and SimRank on the network (greedy matching, $\beta = 0.1$).  Since we want to compare authors with similar rank, we formed six subsets of authors, each containing 45 authors with $PR$ scores centered around $\{0.5, 0.6, 0.7, 0.8, 0.9, 0.99\}$, $\sigma(PR) < 0.05$. For each subset, we compiled the ${\binom{45}{2}} = 990$ within-group RoleSim and SimRank scores.
}

\comment{
 We computed distributions for two populations: a random sample of 1000 author-pairs and the within-group pairs for the $99^{th}$ percentile authors.
\begin{table}
	\centering
	\small
	\begin{tabular}{c|c|c|c|c} \hline
		&	Top		&			&	Random		&				 \\ \hline
Data	&	mean	&	std dev	&	mean		&	std dev		 \\ \hline
\hline
RoleSim	&	0.361	&	0.161	&	0.233		&	0.104		 \\ \hline
SimRank	&	0.016	&	0.004	&	0.011		&	0.011		 \\ \hline
	\end{tabular}
	\caption{Distribution of direct (unranked) scores} \label{tab:coauthor_comp_dist}
\end{table}

Figure~\ref{fig:coauthor_dist} shows that SimRank is not able to make much distinction among different author-pairs.  It generates a very tight band with nearly all scores in the 0 to 0.1 range, with little difference between random pairs and top author-pairs. RoleSim, however, spreads scores across the available spectrum and has a clear difference between random and top author-pairs.  A Chi-square test compare the RoleSim top-author distribution with the random distribution yields a value of 

Figure~\ref{fig:coauthor_rank_binning} shows more clearly that authors with similar rank are more likely to be ranking higher in structural similarity.  We rated each RoleSim score by its percentile rank and then binned the 990 author-pairs into 20 bins with 5\% ranges.  According to this method, if we consider the full population of author-pairs our 20 bins would each contain exactly 5\% of the pairs.

Thus, the random sample appears as a nearly flat line.  The $99^{th}$ percentile subgroup shows a strong tendency towards high similarity rank.  The $80^{th}$ and $60^{th}$ percentile groups show a less pronounced tendency for high similarity rank. We believe this is because in order to achieve top authorship status, an author must engage in a high number of interactions with other authors who themselves are at least moderately prolific.  This mandates some structural commonality among the elites.  On the other hand, there are many possible patterns of co-authoring to generate a midrange authorship performance; there is not a requirement for structural similarity.
Finally, we validated that if authors from {\em different} subgroups are compared, their role similarity scores are no better than average, with even some tendency to below average similarity (Figure~\ref{fig:coauthor_cross_rank}).

\begin{figure}[bt]
\centering
\epsfig{file=Figures/coauth_sim_dist.eps,width=3.3in, height = 1.7in}
\caption{Distibution of Author-Pair Similarity Scores}
\vspace*{-3.0ex}
\label{fig:coauthor_dist}
\end{figure}



\begin{figure}
\centering
	\subfigure[Within-Group Similarity]{\label{fig:coauthor_rank_binning}
			\includegraphics[width=1.55in,height=1.6in]{Figures/coauth_rank_binning.eps}}
	\subfigure[Cross-Group Similarity]{\label{fig:coauthor_cross_rank}
			\includegraphics[width=1.55in,height=1.6in]{Figures/coauth_cross_rank.eps}}
	\caption{Author-Pair Scores, Binned By Percentile Rank}
	\label{fig:coauthor_dist_rank}
\end{figure}
} 
\comment{
For our second analyis, we examined several so-called food webs from the Pajek Dataset website\footnote{http://vlado.fmf.uni-lj.si/pub/networks/data/}, where each node is a biological species, and a directed edge $A \rightarrow B$ indicates that species $A$ is consumed by species $B$.  Some species play the role of pure predators (sink nodes) or pure prey (source nodes).  The majority, however, serve as both both prey and predator.  We can measure each node's distance from a sink or source node in order to quantify its predator-prey status.  We hypothesize: {\em If a pair of species has a high RoleSim value, then they should play similar predator-prey roles in the network, measured by their source/sink distances.}

A simple way to classify prey role is to compute a node's minimum path length from a source node.  Likewise, its predator role is the path length from it to a sink node.  We used source distance minus sink distance as a simple metric. The pure prey end is labeled level 1; levels increase towards the predators. Pure prey will generally have the smallest values while pure predators will have the largest ones.  We used the directed version of RoleSim and SimRank, weighting out-edges and in-edges equally.  A small network is shown in Figure~\ref{fig:food_small}.  It clearly displays both food chain levels and structural symmetry.  A simple way to summarize the similarity patterns is to treat the similarity matrix as $N$ vectors of length $N$ and then to cluster the vectors using K-means.  RoleSim and SimRank generate slightly different clustering results, but both are credible.  The coloring in Figure~\ref{fig:food_small} indicates RoleSim's clustering.

A much more difficult network to analyze concerns the Everglades food web, with 63 nodes and 1016 edges.  We again applied our layering heuristic, but the results were not so satisfying, because they are typically many different paths coming in and out of each node with quite different lengths.  RoleSim followed by K-means with $k=6$ produced excellent results, though.  When the nodes are arranging into the 6 levels, with a small effort to pick the best sequence for the levels, we produce Figure~\ref{fig:food_everglades}. The wonder of this diagram is that of the 1016 directed edges, almost all pointing left to right.  Only nine are pointing right to left.  A few of the top scoring pairs are \{dollar sunfish, redear sunfish\}, large fish in level 4, \{grebes, bitterns\}, shore birds in level 5, and \{topminnows, killifish\}, small fish in level 3. SimRank did not perform well at all, grouping together source and sink nodes.

\begin{table}
	\centering
	\small
	\begin{tabular}{c|c|c|c|c|c} \hline
Food	& Group	&	RoleSim	&			&	SimRank	&			\\
Level	& Size	&	mean	&	std dev	&	mean	&	std dev	\\ \hline
\hline
L1		&	6	&	0.296	&	0.126	&	0.065	&	0.012	\\ \hline
L2		&	15	&	0.512	&	0.200	&	0.154	&	0.062	\\ \hline
L3		&	33	&	0.501	&	0.170	&	0.150	&	0.041	\\ \hline
L4		&	5	&	0.415	&	0.156	&	0.123	&	0.034	\\ \hline
L5		&	4	&	0.255	&	0.132	&	0.059	&	0.031	\\ \hline
All		&	63	&	0.367	&	0.183	&	0.106	&	0.056	\\ \hline
	\end{tabular}
	\caption{Within-level RoleSim and SimRank values} \label{tab:food_within_values}
\end{table}

\begin{figure}[bt]
\centering
\epsfig{file=Figures/cryc_k5.eps,width=2.2in, height = 1.0in}
\caption{Food Web, Crystal River Creek, Florida}
\label{fig:food_small}
\end{figure}

\begin{figure}[bt]
\centering
\epsfig{file=Figures/everglades_k6.eps,width=3.0in, height = 1.5in}
\caption{Food Web, Everglades Marsh, Florida}
\label{fig:food_everglades}
\end{figure}

}


\comment{
	\subsection{K-Cluster}
	In order to get different similarity matrices, we have four parameters, Depth, which is the depth for initialization in, Method, which decides whether greedy matching or exact matching is used. Denominator, which indicates for the denominator if we should choose the product of two squareroots or the max neighbor size.

	In detail, when $Depth=0$, for the initialization, the similarity between two vertices who have the same number of neighbors is 1 otherwise 0, and when $Depth=1$, the similarity between those two vertices who have the same number of direct neighbors and the two ordered lists consisting of the neighborhood size for each direct neighbor are exactly the same, is 1 otherwise 0.

	When $Method=1$, the program will use greedy algorithm to match related vertices, and when $Method=2$ we choose exact matching algorithm.

	When $Denominator=1$, $RoleSim(u,v)=(1-\beta)\frac{w(\mathcal{M})}{\sqrt{N_u}*\sqrt{N_v}}+\beta$, and when $Denominator=2$, $RoleSim(u,v)=(1-\beta)\frac{w(\mathcal{M})}{\max{ (N_u, N_v)}}+\beta$.

	After we get the similarity matrix, there are three ways to do hierarchical clustering, single linkage, complete linkage and average linkage. Also we allow users to designate the total number of clusters $K$, which forces the clustering program to have $K$ clusters in the end.

	But how to measure the goodness of the above clustering and which one is the best. Regarding to this question, we want to choose the clustering which can maximize the following function $\frac{SimAvg_{innerCluster}}{SimAvg_{interCluster}}$.
} 

\comment{ $  \subsection{Quality Tests}
	\subsection{Quality Tests}

	\begin{itemize}
	\item What is the quality of similarity values computed by RoleSim?
	\item How does RoleSim respond to different initialization schemes?
	\item Does greedy matching indeed perform as well as exact matching?
	\item How is RoleSim affected by changes in decay factor $\beta$?
	\end{itemize}

	{\bf Test Procedure}: For each network, we began by applying degre-based initialization.  We then ran RoleSim interatively until convergence, defined to be when no $R_i(u,v)$ changed by more than $1\%$.  Next we coverted similarity values to distance values according $\Delta(u,v) = 1 - sim(u,v)$.  Finally, we used the distance matrix to cluster nodes hierarchically, pairing the closest nodes together.  For comparison, we ran the same tests on SimRank.
	For each network, we varied the following parameters: Degree-based initialization depth = $\{0,1\}$.  Neighbor matching = $\{greedy,exact\}$.  Decay factor $\beta = \{0,0.1\}$.  We implemented RoleSim in C++ and ran all tests on a 2.0GHz dual-core Opteron CPU with 4.0GB running Linux.

	We define performance in several ways: number of iterations, execution time, and quality of node similarity values.  There is no direct standard for assessing the quality of similar values.  To meet this need, devised the following procedure.  (Something about hierarchical clustering, and then measuring within-cluster and between-cluster variance).

	Figure~\ref{fig:test_small} displays the synthestic networks used.  Each serves a specific purpose.  {\em Tree} presents a regular structure where there is only one path connecting any pair of nodes.  S{\em Odd Distance} is a simple structure that $SimRank$ cannot handle, because the distance between equivalent nodes is odd.  {\em Same Neighbors} checks how structurally equivalence nodes (having the same neighbors) are scored.{\em Cluster-2} exhibits two completely disjoint hubbed components, which follow the same pattern but differ in degree.

	\begin{figure*}
	\centering
		\subfigure[Tree]{\label{tree}\includegraphics[width=1.6in,height=1.2in]{Figures/RoleSim_Tree_Graph.eps}}
		\subfigure[Odd Distance]{\label{odd}\includegraphics[width=1.6in,height=1.2in]{Figures/RoleSim_Symmetric_Odd_Graph.eps}}
		\subfigure[Same Neighbors]{\label{neighbor}\includegraphics[width=1.6in,height=1.2in]{Figures/RoleSim_Neighbors_Graph.eps}}
		\subfigure[Cluster-2]{\label{net2}\includegraphics[width=1.6in,height=1.2in]{Figures/RoleSim_Net2_Graph.eps}}
		\caption{Proof-of-Correctness Test Networks}
		\label{fig:test_small}
	\end{figure*}

	\begin{table*}[bt]
	\subtable[Tree]{
		\centering
		\tiny
		\begin{tabular}{|c|c|c|c|c|} \hline
		Role 	& 	1		 & 	2		& 	 3		& 	 4 	\\ \hline
		Node	&	1		&	2-3		&	4-7		&	8-15	\\ \hline\hline
		
		1		&	1		&	0.485	&	0.695	& 0.433 \\ \hline
		2		&	0.485	&	1		&	0.516	& 0.620 \\ \hline
		3		&	0.695	&	0.516	&	1		& 0.372 \\ \hline
		4		&	0.433	&	0.620	&	0.372	&	1	\\ \hline
		\end{tabular}
	}
	\subtable[Odd Distance]{
		\centering
		\tiny
		\begin{tabular}{|c|c|c|} \hline
		Role	&	1		&	2		\\ \hline
		Node	&	1,4		& 2-6		\\ \hline\hline
		
		1		&	1		&	0.620	 \\ \hline
		2		&	0.620	&	1 		\\ \hline

		\end{tabular}
	}
	\subtable[Neighbors]{
		\centering
		\tiny
		\begin{tabular}{|c|c|c|} \hline
		Role 	& 1 		& 	2		\\ \hline
		Node	&	1,2		&	3-6		\\ \hline\hline
		
		1		&	1		&	0.230	 \\ \hline
		2		&	0.230	&	1 \\ \hline

		\end{tabular}
	}

	\subtable[Clust-2]{
		\centering
		\tiny
		\begin{tabular}{|c|c|c|c|c|c|c|} \hline
		Role	&	1 		&	2		&	3		&	4		&	5		&	6 		\\ \hline
		Node	&	1		&	2-5		&	6-13	&	14		&	15-17	&	18-23		\\ \hline\hline
		
		1		&	1		&	0.772	&	0.631	&	0.739	&	0.700	&	0.574	\\ \hline
		2		&	0.772	&	1		&	0.669	&	0.637	&	0.914	&	0.618	\\ \hline
		3		&	0.631	&	0.669	&	1		&	0.647	&	0.618	&	0.919	\\ \hline
		4		&	0.739	&	0.637	&	0.647	&	1		&	0.720	&	0.713	\\ \hline
		5		&	0.700	&	0.914	&	0.618	&	0.720	&	1		&	0.670	\\ \hline
		6		&	0.574	&	0.618	&	0.919	&	0.713	&	0.670	&	1	\\ \hline
		\end{tabular}
	}
	\caption{Similarity values for concept tests (\small D0 initialization, greedy matching, $\beta=0.1$)} \label{tab:test_small_synth}
	\end{table*}

} 

\subsection{Performance of Iceberg RoleSim}
\begin{figure}[tb]
\centering
	\includegraphics[width=1.05\columnwidth,height=1.4in]{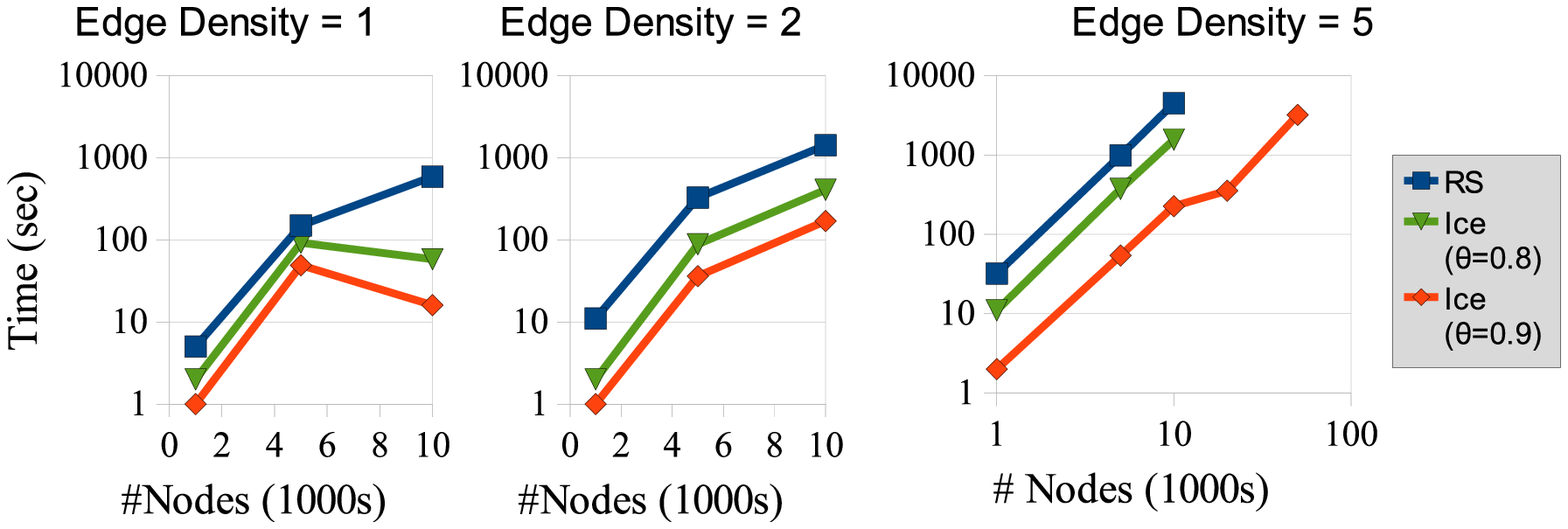}
	\caption{Execution time: Standard vs. Iceberg}
	\label{fig:ice_time}
\end{figure}

In this experiment, we study how Iceberg RoleSim performs in terms of reducing computational time and storage, and its accuracy at approximating the RoleSim score for high similar node-pairs.
Here, we generated $12$ scale-free graphs with up to $100K$ nodes and edge densities of $1$, $2$, and $5$.  
We compared standard RoleSim to Iceberg RoleSim, with $\theta$ values of $0.8$ and $0.9$. 
The parameter $\alpha$, which is the weighting for estimated non-stored values, is set to midpoint $0.5$. 
For the scale-free graphs, the relative scale of the iceberg compared to the full similarity matrix depends on $\theta$ and edge density, but it is almost independent of the number of nodes. Table~\ref{tab:ice_size} shows that the icebergs' hash tables are only $0.15\%$ to $3.5\%$ of the full similarity matrices.  Higher density graphs tend to have more structural variation and thus fewer highly similar node pairs. In Figure~\ref{fig:ice_time}, we see that Iceberg RoleSim is an order of magnitude faster. To check that the ranking has not changed significantly, we computed the Pearson correlation coefficient for each graph's Iceberg RoleSim's rankings vs. the rankings from the corresponding portion of the full similarity matrix.  For $\theta = 0.8$, the average coefficient is 0.823, and for $\theta = 0.9$, it is 0.880. Both show very strong correlation, indicating Iceberg-RoleSim's very good accuracy at ranking role-similarity pairs. 

\begin{table}[tb]
	\centering
	\small
	\begin{tabular}{ | c | p{01.0in} | p{1.0in} | } \hline
		 Edge Density	&	\multicolumn{2}{|c|}{Iceberg Size, as fraction of full matrix}	\\
		 $(|E|/|V|)$	&	$\theta=0.8$	&	$\theta=0.9$	\\ \hline
			1			&	2.77\%			&	1.47\%	\\
			2			&	2.47\%			&	0.63\%	\\
			5			&	3.53\%			&	0.15\%	\\ \hline
	\end{tabular}
	\caption{Iceberg Size Relative to RoleSim Matrix}
	\label{tab:ice_size}
\end{table}

\begin{figure}[tb]
\centering
	\includegraphics[width=1.05\columnwidth,height=1.4in]{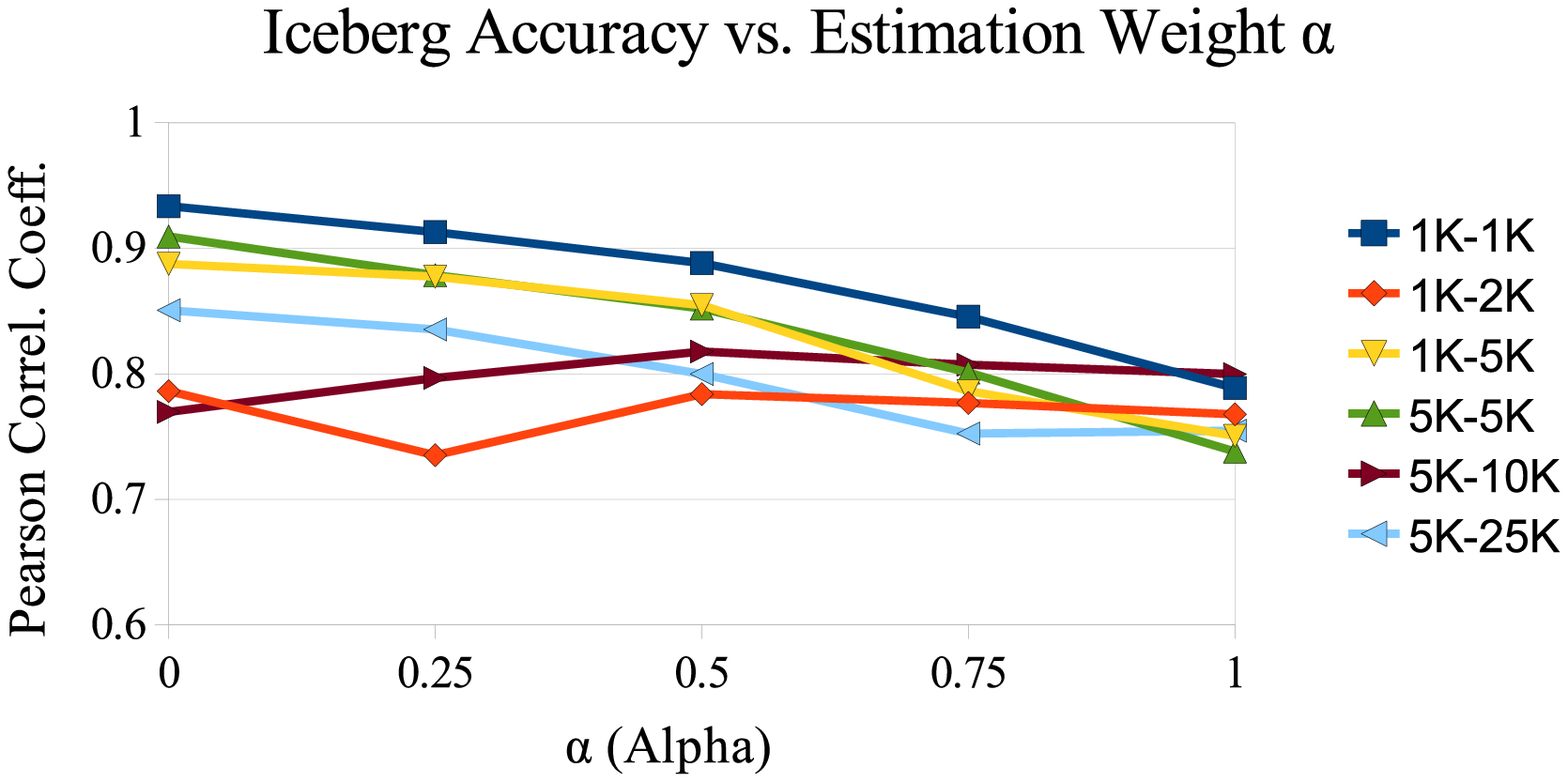}
	\caption{Iceberg Acccuracy vs. $\alpha$}
	\label{fig:iceberg-alpha}
\end{figure}

Next we fixed $\theta$ at $0.9$ and varied $\alpha$ from 0 to 1.0 to see how sensitive is the accuracy of Iceberg RoleSim with respect to $\alpha$. The results from $6$ scale-free grapha are shown in Figure~\ref{fig:iceberg-alpha}. The labels describe the number of nodes and edges of each graph. Most graphs prefer $\alpha=0$, but some prefer a midrange value.  Any value in the lower half seems acceptable.

\section{Related Work}
\label{section:related}

The role similarity problem is a distinct special case of the more general structural or link similarity problems, which find applications in co-citation and bibliographic networks~\cite{Lu01_nodesimilarity}, recommender systems,~\cite{Antonellis08_simrankpp} and Web search~\cite{Haveliwala03_topicpagerank}.
Link similarity means that two objects accrue some amount of similarity if they have similar links. 


Formal definitions of role, which enable a clear idea of what is being measured, arose from the social science community~\cite{Lorrain71,Sailer78,Everett85}.
Block partitioning can be used directly to group nodes into roles~\cite{White76_blockmodel}.  However, block modeling does not produce individual node-pair similarities.  Therefore, it is not useful as a ranking method.


SimRank~\cite{Jeh02_simrank} is the best known algorithm to implement a recursive definition of object similarity: two objects are similar if they relate to similar objects.
SimRank has an elegant random walk interpretation: $SimRank(a,b)$ is the probability that two independent simultaneous random walkers, beginning at $a$ and $b$, will eventually meet at some node. However, the more neighbors that $a$ and $b$ have in common, the less likely that they will both randomly choose the same neighbor.
This then explains SimRank's problem with structural equivalence.
Recently, Zhao~\cite{Zhao09_prank} has pointed out that in-neighbor and out-neighbor SimRank can be used as a univeral framework to describe co-citation (common in-neighbors), bibliographic coupling (commnon out-neighbors), or a weighted combination of the two.  The number of iterations reflects the search radius for discovering similarity.
As we note in Section~\ref{lab:simrank_not}, SimRank has an undesirable trait: its values decrease when the number of common neighbors increases. Several works have tried to address this problem. SimRank++ ~\cite{Antonellis08_simrankpp} adds a so-called {\em evidence} weight which partially compensates for the neighbor matching cardinality problem.
In~\cite{Fogaras05_scaleSim}, they execute Monte Carlo simulations of "intelligent" random walks, where they force the overall probability of $a$ meeting $b$ to be Jaccard coefficient $\frac{|N(u) \cap N(v)|}{|N(u) \cup N(v)|}$.
Recently, MatchSim~\cite{Lin09_matchsim} has also used maximal matching of neighbors to address problems with SimRank's scoring.
However, our formulations have small but important differences. Because they retained SimRank's initialization, their work does not guarantee automorphic equivalence in the final results.
Also, their work is intuition-based, without a theory of correctness. They provide one specific formulation, while we define a theoretical framework for {\em any }admissible measure or metric.  Because RoleSim satisfies the triangle inequality, it is a true metric.
\comment{ 
The work closest to ours is MatchSim~\cite{Lin09_matchsim}, which also uses a maximal matching of neighbors. MatchSim uses the same initialization as RoleSim ($sim(u,v) = 1$ iff $u=v$), which cannot find automorphically equivalent nodes.  Also, we provide a comprehensive axiomatic definition of a role similarity measure and {\em metric}, and then rigorously prove the admissibility of RoleSim and at least two initialization schemes.
} 


\comment{
Several authors have sought to improve upon SimRank's time complexity or to improve the quality of its results.
 In~\cite{Fogaras05_scaleSim}, the authors focus on reducing the storage space and online query time.  The results of precomputed Monte Carlo random walks are stored in an index database of probabilistic fingerprints; at query time, similarity scores are estimated from those fingerprints.  However, the precomputation is still relatively expensive.
The authors of ~\cite{Lizorkin08_accuSimrank} present techniques to estimate the accuracy of iterative SimRank calculations (comparing to the theoretical values obtained when iterations $\rightarrow \infty$). Using this, they offer various computational optimizations which filter out unnecessary computations to achieved an estimated accuracy.
Simrank++~\cite{Antonellis08_simrankpp} extends SimRank to edge-weighted graphs and introduces an "evidence" factor that increases as the number of shared neighbors increases.
}

\comment{
SimTree/LinkClus~\cite{Yin06_simtree} clusters objects into a hierarchical tree. Similarity scores are recorded only between siblings and for parent-child edges; the similarity between an arbitrary pair of nodes is estimated as the product of the edge weights along the connecting path in the tree. However, LinkClus targets bipartite data sets that construct two parallel SimTrees.  It is not clear that it is applicable to monotype data. BlockSim~\cite{Li09_blocksim} first does a $k$-way partitioning~\cite{Karypis09_multilevel_k_partition} of the network, then interatively applies SimRank for both within-block and between-block similarity.  The partitioning reduces the order of complexity.  However, traditional clustering is not appropriate for role analysis.
}

\comment{
Simrank and its variants fall into this category. The key difference is that all the traditional approaches define similarity in terms of a common ancestor or descendent.  Automorphic equivalence does not require common connections; in fact, our algorithm can find similarities between disconnected components.

While SimRank offers an enticingly simple model for link similarity, its native performance characteristics are not so impressive: its time complexity.  The authors suggest operating on a so-called $G^2$ graph which has $n^2$ nodes, yielding $O(n^4)$ time complexity.  Many researchers have built upon SimRank, to specializing its use or to improve the performance.  PSimRank~\cite{Fogaras05_scaleSim} addressed the sub-maximal scores generated by SimRank's paired-random walk model by instead measuring neighborhood similarity with the Jaccard coefficient.

} 

\comment{
\subsection{OLD STUFF}

The SimRank measure and basic iterative algorithm was defined by Jeh and Widom~\cite{jeh02_simrank}.  This algorithm takes $O(n^3)$ per iteration if performed on the original graph $G$.

\subsection{Efficiency Improvements}

There have been several efforts to improve the time and space complexity of SimRank...
Fogaras and R\'acz(WWW05)~\cite{fogaras05} use Monte Carlo methods to generate instances of random walks, summarized as fingerprints.  They construct a DAG which links a vertex $u$ with the first other vertex $v$ that it meets, according to these fingerprints. In fact, the DAG is a forest, so it may be much more compact that the original graph.
Lizorkin {\em et al.}(VLDB08)~\cite{lizorkin08} provide a measure of the accuracy of finite SimRank computations vs. the theoretical point of convergence.  They also provide several optimizations: identifying and using essential nodes,  partial sums, and thresholding-sieved similarities.
Cai {\em et al.}(DASFAA09)~\cite{cai09} develop an adaptive algorithm by avoiding recomputing node-pairs that are likely to have nearly converged.
Jia {\em et al.}(ADMA09)~\cite{jia09} note that in some real-world small world graphs, if a node-pair has zero similarity after a few iterations, it is likely to remain at zero, so it can be ignored, reducing computation.  Their results, however, are purely empirical.
Li {\em et al.}(PAKDD09)~\cite{li09} uses multilevel $k$-way partition of the graph (Karypis 98) to progressively subdivide the graph into blocks.

\subsection{Interpreting and Extending SimRank}

There have also been efforts to modify SimRank.  Each formulation computes some property of the graph.  If the graph is assigned some interpretation, such as a co-citation network, the the similar metric has a corresponding interpretation.  A modification is only justified if it yields a meaningful interpretation.  For practical purposes, we also wish to be able to compute the its values, either exactly or a good approximation, in reasonable time.

First, what is the meaning of SimRank?  Jeh and Widom examined the expected number of backward steps that two synchronized, uniform random walkers engaged in infinite walks would each take until they first meet.  Because this time could be infinite, they mapped the expectation $z$ to $f(z) =  C^z, 0 < C < 1$, because $f(z)$ has values in the finite range $[0,1]$. Therefore, SimRank computes the so-called expected-$f$ distance.  However, they failed to provide an interpretation of their exponential $f$ function.

Fogaras and R\'acz~\cite{fogaras05} extend this to finite walks. That is, if the walkers walk for $t$ steps, then $s_t(u,v)$, the similarity between $u$ and $v$, is their expected-$f$ meeting time.

\subsection{SimRank as probabilty of meeting}
We prove a random walk interpretation that includes the $C$ factor in a more natural way, without mapping to an artificial $f$ space.  

{\bf Theorem:}\\
The $k$-iteration SimRank value $s_k(u,v)$ is equal to the probability that two backwards random walks starting at $u$ and $v$ will meet in no more than $k$ steps, if the following conditions apply:\\
1. At each step, there is a probability $\sqrt{C}$ of exiting the graph (transitioning to a sink node). \\
2. For each vertex $v$, there is an equal probability $(1-\sqrt{C})/|I(v)|$ of traversing any of its in-edges. \\
3. The two walkers move in synchronicity.\\
Note: Since we define SimRank values as probabilities, they naturally fall within the range $[0,1]$, without resolving to an $f$ remapping. At one extreme, $s_k(v,v)$ = 1, because they two walkers have met, and at the other extreme, if they can never meet one another, than $s_l(u,v) = 0.$

{\bf Proof:}\\
{\bf Not complete; needs cleaning up}
Let $S_t$ be the similarity matrix after $t$ iterations.  Note that $S_0 = I$, the $n \times n$ identity matrix.  Let $A$ be the adjacency matrix. Let I(j) be the in-neighbors of vertex $j$, so $|I(j)| = \sum_{i=1}^n a_{ij}$.  Thus, for a uniform random walk travelling backwards on edges, its transition matrix is $P$, where $p_{ij} = a_{ij}/|I(j)|$, the probability of walking from $j$ to $i$. 

A random walk is a Markov chain.  Therefore, the probability of taking a particular path $(u_0,u_1,...,u_{t-1},u_t)$ is $P(u_0)P(u_1|u_0)...P(u_t|u_t-1)$. Consider the incremental change due to the $k^th$ step of the random walks, which is the transition from $S_{k-1}$ to $S_{k}$.

Then, if two independent, simultaneous uniform backwards random walkers start at $u$ and $v$, respectively, the probability that they meet at vertex $w$ after one step is $P(w|u)P(w|v) = d_{wu}d_{wv}$.  The probability of meeting at {\em any} vertex after one step is $\sum_{w=1}^n d_{wu}d_{wv} = \sum_{w=1}^n d^\prime_{uw}d_{wv}$.  In matrix form, $S_1 = D^\prime S_0 D$.

\subsection{Weaknesses, Modifications and Extensions}

Fogaras and R\'acz~\cite{fogaras05} point out an intuitive flaw in SimRank.  Nodes with high in-degree, even if they share all the same in-neighbors with on another, are penalized with lower SimRank.  If two nodes have in-degree $d$, and their full sets of in-neighbors are shared in common, then as $d$ increases, the number of meetings increases linearly, but the number of possible combinations of walks (denominator of SimRank) grows quadratically.
   They propose PSimRank, which replaces the uniform probability of meeting in the next step $\frac{|I(u')|\cap|I(v')|}{|I(u')||I(v')|}$ with the more favorably weighted Jaccard coefficient $\frac{|I(u')|\cap|I(v')|}{|I(u')|\cup|I(v')|}$.
   
Dorow(EACL09)~\cite{dorow09} extends SimRank for edge weights, edge labels, and multiple graphs

Lin {\em et al.}(WIC07)~\cite{lin07} have made the natural extension of SimRank from merely in-neighbors include out-neighbors as well.
Zhao.
Zhao {\em et al.}(CIKM09)~\cite{zhao09} made a minor improvement, adding a parameter to allow the user to choose the balance between in-degree vs. out-degree.  They have dubbed their parameterized in-out recursive ranking as P-Rank.

   Another flaw with SimRank is that similarly is only detected when the two walkers each travel the same distance.  If they travel unequal distances, no similarly is accounted for at all.
   
\subsection{Role Similarity and Social Equivalence Metrics}

Structural Equivalence: $u$ and $v$ are structural equivalent if they share the same exact ne
Automophoric Equivalence\\

Regular Equivalance\\

\subsection{Social Network Analysis}
In the literature, a lot of research have been done recently by analyzing various types of social networks such as Flickr, Youtube, blog network, Orkut, citation, and recommendation networks. Generally, we can categorize them into three categories.
1): the structural or topological properties of social network~\cite{SSSNS,SNSG,MMSNS,SNSK,SNS2,SNS3,SNS4,TopologyBlog:workshop09,TELGRPH}, where the small world phenomena~\cite{SSSNS,MMSNS}, the growth
models have been studied to generate the social network~\cite{SNSG}, the power law property~\cite{MMSNS}, and community discoery~\cite{SNSK,SNS4}.
2): the information cascading patterns~\cite{BlogNetwork:Flow,Kleinberg:BackBone08,OnlineComm:Kleinberg09,DIFFUSION,Dynamics}, where most of them 
focusing on analyzing the specific event related information propagation patterns; only the backbone concept proposed in~\cite{Kleinberg:BackBone08} is similar to part of our research in terms of topological level patterns. Different from our approach, the backbone approach focused on the speed of information flow, not fully explored the micro level of relationships, finally they find the subgraph of social network where the information has the potential to flow quickest.  
3): the dynamics of social network~\cite{BloggingNetwork:Evolution,Nature:GroupEvo,Dynamics,TELGRPH,SNSRC,BlogNetwork:Community}, where they using the 
temporal dimension only in a macro way. That is to partition the social network into different snapshots, and then model the dynamics between different
snapshots or use the time series related analysis. 
Our study in this paper is actually a synergy of the three topics in the literature together with our micro and macro usage of the temporal information.
} 

\section{Conclusion}
We have developed RoleSim, the first real-valued role similarity measure that confirms automorphic equivalence.  We have also presented a set of axioms which can test any future measure to see if it is an admissible measure or metric.  Our experimental tests demonstrate RoleSim's correctness and usefulness on real world data, opening up exciting possibilities for scientific and business applications.  At the same time, we see that other well-known measures, while suitable for other tasks, are not suitable for role similarity.  This axiomatic approach may prove useful for developing and validating solutions to other related tasks.

\label{section:conclusion}


\baselineskip=1.2\normalbaselineskip
\bibliographystyle{plain}
\bibliography{bib/paper,bib/similarity,bib/social,bib/socnet,bib/graphAlg}


\newpage
\appendix
\section{Proofs of Theorems and Lemmas}
\label{proof}

\comment{
\noindent{\bf Proof for Theorm~\ref{GTS} (Generalized Transitive Similarity) }
\bproof 
From triangle inequality, we have $d(a,c) \leq d(a,b)+d(b,c) \leq d(b,c)$ and $d(b,c) \leq d(b,a)+d(a,c) \leq d(a,c)$ ($d(a,b)=0$). 
Thus, $d(a,c)=d(b,c)$. 
Similarly, $d(a,d)=b(b,d)$, $d(c,a)=d(d,a)$, and $d(d,a)=d(d,b)$. 
Put together, we have $sim(a,c)=sim(a,d)=sim(b,c)=sim(b,d)$. 
\eproof
}
\comment{
\noindent{\bf Proof for Theorem~\ref{MWM} (Maximal Weighted Matching)}
We need to show that Equations~\eqref{eqn:rolesim} and~\eqref{eqn:max_weight_match} are equivalent.
Without loss of generality, let $N_u\geq N_v$. 
First, we show that {\em the cardinality of the maximal weighted matching $|\mathcal{M}| = \min{(N_u,N_v)}=N_v$ }. It cannot be greater, because there are insufficient elements in $N_v$.
It cannot be smaller, because if it were, there must exist an available edge between an uncovered node in $N_u$ with one in $N_v$.  Adding this edge would increase the matching (every edge has weight $\geq \beta$). If $|\mathcal{M}| = \min{(N_u,N_v)}$, it follows that
${N_u+N_v-|M|} = \max{ (N_u, N_v)}$.  Thus, for a given pair of nodes $u$ and $v$, the denominators in Equations~\eqref{eqn:rolesim} and~\eqref{eqn:max_weight_match} are constant and identical.  It is then a trivial observation that the numerators are in fact the same calcuation.  Therefore, the maximal value for the entire Equation~\eqref{eqn:rolesim} is the same as the value in ~\eqref{eqn:max_weight_match}.
\eproof
}

\noindent{\bf Proof for Theorem~\ref{thm:convergence} (RoleSim Convergence)}
Let the difference of $RoleSim(u,v)$ scores between iterations $k$ and $(k-1)$ be 
$\delta^k(u,v)=RoleSim^{k}(u,v)-RoleSim^{k-1}(u,v)$. 
Also, let $D_k=\max_{(u,v)} |\delta^k(u,v)|$ be the maximal absolute difference across all $u$ and $v$ in iteration $k$.
To prove converge, we will show that $D_k$ is monotonically decreasing, i.e., 
$D_{k+1} < D_k$. 
For any node pair $(u,v)$, let the maximal weighted matching between $N(u)$ and $N(v)$ computed at iteration $k+1$ be
$\mathcal{M}^{k+1}$. 
Note that its weight is $w(\mathcal{M}^{k+1})=\sum_{(x,y)\in \mathcal{M}^{k+1}} RoleSim^{k}(x,y)$.
Without loss of generality, assume $N_u \leq N_v$, so that $max(N_u,N_v) = N_v$ and $|\mathcal{M}| = N_u$.
Given this, we observe that 

\begin{eqnarray*}
w(\mathcal{M}^{k+1}) - (N_v \cdot D_k)\! &\!\leq& \\
w(\mathcal{M}^{k+1}) - |\mathcal{M}|\cdot D_k\! &\!\leq\!	w(\mathcal{M}^{k})\! &\!\leq\! w(\mathcal{M}^{k+1}) + |\mathcal{M}|\cdot D_k\\
											& 					&\!\leq\! w(\mathcal{M}^{k+1}) + (N_v \cdot D_k)
\end{eqnarray*}

Therefore, $|w(\mathcal{M}^{k+1}) - w(\mathcal{M}^k)| \leq N_v \times D_k$. Then,


\begin{eqnarray*}
|\delta^{k+1}(u,v)|	&=& |RoleSim^{k+1}(u,v)-RoleSim^{k}(u,v)| \\
		&=& |(1-\beta) \frac{w(\mathcal{M}^{k+1})} {N_v}  - 
			(1-\beta) \frac{w(\mathcal{M}^{k})}{N_v}  | \\
		&=& \frac{(1-\beta)}{N_v} |w(\mathcal{M}^{k+1})-w(\mathcal{M}^{k})| \\
		&\leq&  \frac{(1-\beta)}{N_v} N_v \times D^k < D^k  
\end{eqnarray*}
Therefore, $D^{k+1}=\max_{(u,v)} |\delta^{k+1}(u,v)| < D^k$, and therefore, $RoleSim^k$ will converge. 
\eproof

\comment{
\noindent{\bf  Proof for Lemma~\ref{lem:invariant3} (Automorphic Equivalence Invariant)} 
\bproof
We show that for any $x \equiv y$ where $RoleSim^{k-1}(x,y)=1$, then $RoleSim^k(x,y)=1$.
By the definition of graph automorphism (Section~\ref{section:equivalence}), if $u \equiv v$, there is a permutation $\sigma$ of vertex set $V$, such that $\sigma(u)=v$, and any edge $(u,x) \in E$ iff $(v, \sigma(x)) \in E$.
This indicates that $\sigma$ provides a one-to-one equivalence between nodes in $N(u)$ and $N(v)$.
Also, $u$ and $v$ have the same number of neighbors, i.e., $N_u=N_v$. 
So, it is clear that the maximal weighted matching  $\mathcal{M}$ in the bipartite graph $(N(u) \cup N(v), N(u) \times N(v))$ selects $N_u = N_v$ pairs of weight 1 each.
Thus, $RoleSim^{k+1}(u,v)=(1-\beta)\frac{w(\mathcal{M})}{\max{(N_u,N_v)}}+\beta=1$.
\comment{By the definition of automorphic equivalence, if $a \equiv b$, then their neighbors are equivalent, meaning there is a mapping of equivalent pairs.  
Furthermore, there is a mapping of weights such that $W_{xu} = W_{yv}$.  Clearly, if we find the optimal neighbor matching, then $RoleSim_{k+1} = 1$.  However, even greedy matching will generate maximal similarity:
Suppose that the greedy matching algorithm pairs $x$ in true autmorphic class $C(x)$ with a node $y$ in a different class $C(y)$.  This selection will only occur if classes $C(x)$ and $C(y)$ are grouped in the same superclass, with similarity 1. Since we made a "wrong" selection, we have not yet used a neighbor in class $C(x)$ and another in $C(y)$ which can be paired to get similarity 1.}
\eproof
} 

\comment{
\noindent{\bf  Proof for Lemma~\ref{lem:invariant4} (Transitive Similarity Invariant)}  

\bproof
We prove this by contradiction. 
For any $a \equiv b$, $c \equiv d$, we claim $RoleSim^{k}(a,c) = RoleSim^{k}(b,d)$, but $RoleSim^{k+1}(a,c)\neq RoleSim^{k+1}(b,d)$.
Without loss of generality, assume $RoleSim^{k+1}(a,c) > RoleSim^{k+1}(b,d)$.
Denote the maximal weighted matching between $N(a)$ and $N(c)$ as $\mathcal{M}$.
Since there is a one-to-one equivalence correspondence $\sigma$ between $N(a)$ and $N(b)$
and a one-to-one equivalence correspondence $\sigma^\prime$ between $N(c)$ and $N(d)$, 
we can construct a matching $\mathcal{M}^\prime$ between $N(b)$ and $N(d)$ as follows: $\mathcal{M}^\prime=\{(\sigma(x),\sigma^\prime(y))| (x,y) \in \mathcal{M}\}$. 
Since the transitive similarity property holds for $RoleSim^{k}$, we have 
$RoleSim^{k}(x,y)=RoleSim^{k}(\sigma(x),\sigma^\prime(y))$. 
Thus, $w(\mathcal{M}^\prime)=w(\mathcal{M})$, and we find the following contradiction:
\begin{eqnarray*}
RoleSim^{k+1}(b,d) &=& (1-\beta)\frac{w(\mathcal{M}^\prime)}{\max{(N_b,N_d)}}+\beta  \\
				&=& (1-\beta)\frac{w(\mathcal{M})}{\max{(N_a,N_c)}}+\beta \\
= RoleSim^{k+1}(a,c) &>& RoleSim^{k+1}(b,d),
\end{eqnarray*}
which is contradictory.
\comment{
$M1$ and $M2$ be the mappings of neighbors of $a$ to $c$ and $a$ to $b$, respectively. $RoleSim_{k+1}(a,c) = RoleSim_{k+1}(a,c)$ if $M1$ and $M2$ select equal values from the $RoleSim_k$ matrix. Since $a \equiv b$, they have neighbors which are currently believed to be equivalent.  By the transitive similarity of $RoleSim_k$, for every neighbor $x \in N^a, y \in N^b, z \in N^c$, we have $(x,z) = (y,z)$.  Since the set of similarity values are equal, if $M1$ was the selected matching from $a$ to $c$, whether optimal or greedy, then $M2$ should be the selected matching from $b$ to $c$.
}
\eproof

} 

\noindent{\bf Proof for Lemma~\ref{lem:invariant5} (Triangle Inequality Invariant)}
For iteration $k$, for any nodes $a$, $b$, and $c$, $d^k(a,c) \leq d^k(a,b)+d^k(b,c)$, where $d^k(a,b)=1 - RoleSim^k(a,b)$. We must prove that this inequality still holds for the next iteration: 
$d^{k+1}(a,c) \leq d^{k+1}(a,b)+d^{k+1}(b,c)$. 
To facilitate our discussion, we abbreviate $RoleSim^k(u,v)$ as $r(u,v)$ , and without loss of generality, let $N_a \leq N_c$.

We utilize the following observation: 
{\em if there is a matching $M$ between $N(a)$ and $N(c)$ which satisfies
$1-((1-\beta)\frac{w(M)}{N_c}+\beta) \leq d^{k+1}(a,b)+d^{k+1}(b,c)$, then $d^{k+1}(a,c) \leq d^{k+1}(a,b)+d^{k+1}(b,c)$.}
This is because $\frac{w(M)}{N_c}  \leq \frac{w(\mathcal{M})}{N_c}$, where $\mathcal{M}$ is the maximal weighted matching between $N(a)$ and $N(c)$, and thus, 
$1-((1-\beta)\frac{w(M)}{N_c}+\beta) \geq 1-((1-\beta)\frac{w(\mathcal{M})}{N_c}+\beta)=d^{k+1}(a,c)$. 

In addition, we also denote the maximal weighted matching between $N(a)$ and $N(b)$ as $\mathcal{M}(a,b)$, and
the maximal weighed matching between $N(b)$ and $N(c)$ as $\mathcal{M}(b,c)$.  
Now, we consider three cases characterizing the relationship between $N(a)$, $N(b)$, and $N(c)$.  

\noindent{\bf Case 1 ($N_b \leq N_a \leq N_c$):}
In this case, we observe $|\mathcal{M}(a,b)|=|\mathcal{M}(b,c)|=N_b$. 
Given this, we consider the following matching $M$ between $N(a)$ and $N(c)$: 
\beqnarr
M=\{(x,z) | (x,y) \in \mathcal{M}(a,b) \wedge (y,z) \in \mathcal{M}(b,c)\}, |M| = N_b \nonumber
\eeqnarr
Then, we have the following relationships: 
{\small
\begin{eqnarray*}
d^{k+1}(a,b)+d^{k+1}(b,c)- (1-(1-\beta)\frac{w(M)}{N_c}-\beta)  \\
= (1-\beta)[- \frac{w(\mathcal{M}(a,b))}{N_a} - \frac{w(\mathcal{M}(b,c))}{N_c} + \frac{w(M)}{N_c}] + 1 - \beta \\
= (1-\beta)[\frac{N_b-w(\mathcal{M}(a,b))}{N_a} - \frac{N_b}{N_a}
				+ \frac{N_b-w(\mathcal{M}(b,c))}{N_c} -\frac{N_b}{N_c} \\
				- \frac{N_b-w(M)}{N_c} + \frac{N_b}{N_c}] + 1 - \beta  \\
\geq (1-\beta) [1-\frac{N_b}{N_a} + \frac{\sum_{(x,y)\in \mathcal{M}(a,b)} (1-r(x,y))}{N_c} \\
\end{eqnarray*}
\begin{eqnarray*}
+ \frac{\sum_{(y,z) \in \mathcal{M}(b,c)}(1-r(y,z))}{N_c} -
                                            \frac{\sum_{(x,z) \in M} (1-r(x,z))} {N_c}] \\
\geq (1-\beta)[\frac{\sum_{(x,y,z)} (d^k(x,y)+d^k(y,z)-d^k(x,z))}{N_c}] \geq 0  \\
\mbox{where }(x,y) \in \mathcal{M}(a,b), (y,z) \in \mathcal{M}(b,c), (x,z) \in M
\end{eqnarray*} 
}

\noindent{\bf Case 2 ($N_a \leq N_b \leq N_c$):}
In this case, we observe $|\mathcal{M}(a,b)|=N_a$ and $|\mathcal{M}(b,c)|=N_b$.  It follows that there is a subset $n(b)$ of $N(b)$ of size $N_a$ that participates in both $\mathcal{M}(a,b)$ and $\mathcal{M}(b,c)$:
$n(b) = \{y|(y,z) \in \mathcal{M}(b,c) \backslash \{(y,z)|\not \exists (x,y) \in \mathcal{M}(a,b)\}\}$.
Given this, we consider the following matching $M$ between $N(a)$ and $N(c)$:
\beqnarr
M=\{(x,z) | (x,y) \in \mathcal{M}(a,b) \wedge (y,z) \in \mathcal{M}(b,c)\}, \nonumber
\eeqnarr 
$|M| = N_a$. Then, we have the following relationships: 
{\small
\begin{eqnarray*}
d^{k+1}(a,b)+d^{k+1}(b,c)- (1-(1-\beta)\frac{w(m)}{n_c}-\beta) \\
= (1-\beta)[-\frac{w(\mathcal{M}(a,b))}{n_b} -  \frac{w(\mathcal{M}(b,c))}{n_c}
		+ \frac{w(m)}{n_c}] + 1 - \beta \\
= (1-\beta)[\frac{n_a-w(\mathcal{M}(a,b))}{n_b} - \frac{n_a}{n_b}
			+ \frac{n_a-w(\mathcal{M}(b,c))}{n_c} -\frac{n_a}{n_c}  \\
			- \frac{n_a-w(m)}{n_c} + \frac{n_a}{n_c}] + 1 - \beta \\
\geq (1-\beta)[1-\frac{n_a}{n_b} + \frac{\sum_{(x,y)\in \mathcal{M}(a,b)} (1-r(x,y))}{n_c} \\
	+ \frac{\sum_{(y,z) \in \mathcal{M}(b,c)\backslash \{(y,z)|\not \exists (x,y) \in \mathcal{M}(a,b)\}}(1-r(y,z))}{n_c}   \\ 
	- \frac{n_b-n_a}{n_c} - \frac{\sum_{(x,z) \in m} (1-r(x,z))} {n_c}]  \\
\geq (1-\beta)[1- \frac{n_a}{n_b} - \frac{n_b-n_a}{n_c} \\
	+ \frac{\sum_{(x,y,z)} (d^k(x,y)+d^k(y,z)-d^k(x,z))}{n_c}]   \\
\mbox{where } (x,y) \in \mathcal{M}(a,b), (y,z) \in \mathcal{M}(b,c), (x,z) \in m 
\end{eqnarray*}
\begin{eqnarray*}
\geq (1-\beta)[1-\frac{n_a}{n_b} - \frac{n_b}{n_c} + \frac{n_a}{n_c}] \\
= (1-\beta)\frac{n_b n_c - n_a n_c - n_b^2 + n_a n_b }{n_b n_c} \\
= (1-\beta)\frac{(n_b - n_a) (n_c-n_b)} {n_b n_c} \geq 0
\end{eqnarray*}
}

\noindent{\bf Case 3 ($N_a \leq N_c \leq N_b$):} 
In this case, we observe $|\mathcal{M}(a,b)|=N_a$ and $|\mathcal{M}(b,c)|=N_c$. 
Given this, we consider the following matching $M$ between $N(a)$ and $N(c)$: 
\beqnarr
M=\{(x,z) | (x,y) \in \mathcal{M}(a,b) \wedge (y,z) \in \mathcal{M}(b,c)\} \nonumber
\eeqnarr
In addition, we define: 
\beqnarr
M_1=\{(x,y) | (x,y) \in \mathcal{M}(a,b) \wedge \not \exists (y,z) \in \mathcal{M}(b,c)\} \nonumber \\
M_2=\{(y,z) | (y,z) \in \mathcal{M}(b,c) \wedge \not \exists (x,y) \in \mathcal{M}(a,b)\} \nonumber  
\eeqnarr
In other words, $M_1 \subset \mathcal{M}(a,b)$ and $M_2 \subset \mathcal{M}(b,c)$ do not link to each other using intermediate node $y \in N(b)$.  
We further denotes $m_1=|M_1|$, $m_2=|M_2|$, $m_3=|M|$. 
Note that $m_1=N_a-m_3$, $m_2=N_c-m_3$, and $N_b \geq m_1+m_2+m_3$. 

Then, we have the following relationships: 
{\small
\beqnarr
d^{k+1}(a,b)+d^{k+1}(b,c)- (1-(1-\beta)\frac{w(M)}{N_c}-\beta) \geq \nonumber \\
d^{k+1}(a,b)+d^{k+1}(b,c)- (1-(1-\beta)\frac{w(M)}{N_b}-\beta) \geq  \nonumber \\
1-\beta - (1-\beta)(\frac{w(\mathcal{M}(a,b))}{N_b}+\frac{w(\mathcal{M}(b,c))}{N_b} -\frac{w(M)}{N_b} )= \nonumber \\
(1-\beta)(1 + \frac{m_3-w(\mathcal{M}(a,b))}{N_b} - \frac{m_3}{N_b} + \frac{m_3-w(\mathcal{M}(b,c))}{N_b}  
-\frac{m_3}{N_b} - \nonumber \\
 \frac{m_3-w(M)}{N_b} + \frac{m_3}{N_b}) \geq  \nonumber \\
(1-\beta)(1-\frac{m_3}{N_b} + \frac{\sum_{(x,y)\in \mathcal{M}(a,b) \backslash M_1} (1-r(x,y))}{N_b} - \frac{m_1}{N_b}+ \nonumber \\
\frac{\sum_{(y,z) \in \mathcal{M}(b,c)\backslash M_2}(1-r(y,z))}{N_b} - \frac{m_2}{N_b} \nonumber  \\ 
- \frac{\sum_{(x,z) \in M} (1-r(x,z))} {N_b}) \geq \nonumber  \\
(1-\beta)(1- \frac{m_3}{N_b} - \frac{m_1}{N_b} - \frac{m_2}{N_b}+ \nonumber \\ 
\frac{\sum_{(x,y,z)} (d^k(x,y)+d^k(y,z)-d^k(x,z))}{N_b}) \geq  \nonumber  \\
((x,y) \in \mathcal{M}(a,b), (y,z) \in \mathcal{M}(b,c), (x,z) \in M) \nonumber \\
(1-\beta)(1-\frac{m_1+m_2+m_3}{N_b}) \geq 0 \nonumber
\eeqnarr 
}

\eproof

\section{SimRank and other structural similarity measures}
\label{section:similarity}

\subsection{Non-iterative Predecessors of SimRank}
\label{sec:predecessors}

{\em Bibliographical coupling}~\cite{Kessler63_coupling} measures the similarity between two research publications by counting the number of works that are listed in both of their bibliographies.  {\em Co-citation}~\cite{Small73_cocitation} turns this around by counting the number of later works that cite both of the two original documents.  As the size of a work's bibliography increases, the likelihood that it will contain a particular work increases.  Therefore, a common normalization of these two measures is to divide the count by the number of distinct works cited.

We can form a {\em citation graph}, where each vertex is a document and a directed edge $(a,b)$ means that document $a$ cites document $b$.  Let $I(a)$ and $O(a)$ be the in-neighbor set and out-neighbor set of $a$, respectively.  Let $I_a$ and $O_b$ be the in-degree and out-degree of $a$.  Then, the normalized bibliographic coupling index is
\begin{align}
\label{biblio-coupling}
S_{bc}(a,b) = \frac{ |O(a) \cap O(b)| }{ |O(a) \cup O(b)| },
\end{align}
and the normalized co-citation index is
\begin{align}
\label{co-citation}
S_{cc}(a,b) = \frac{ |I(a) \cap I(b)| }{ |I(a) \cup I(b)| }.
\end{align}
These are simply the Jaccard coefficients of the out-neighbor sets and in-neighbors sets, respectively.

These two are suitable for unweighted and directed graphs.  If a graph is undirected, then the two measures are the same. Suppose we have a weighted graph, though.  This could be an author-collaboration graph, where edge $(a,b)$ counts how many times author $a$ has worked with author $b$.  Or, it could be a bipartite document-term graph, where edge $(d_a, t_b)$ counts the number of times that document $a$ uses term $b$. Assign to each vertex a feature vector.  For the homogeneous co-authorship graph, each author is a feature dimension; its feature vector is the set of edge weights to every other author.  For the document-term graph, a document has a term vector, weighted according to term frequencies of the document.
Then the cosine between two objects is a convenient and meaningful measure.  Identical documents have cosine of 1, and documents with no features in common are orthogonal with cosine of 0.
\begin{align}
\label{cosine}
S_{cos}(a,b) = \frac{ A \cdot B }{ ||A||~||B|| },
\end{align}

where $A$ is the feature vector of vertex $a$.  A small modification to the denominator, attributed to Tanimoto~\cite{Tanimoto} maintains the overall behavior of the similarity function while aligning it with the Jaccard coefficient when the feature vectors are binary-valued:
\begin{align}
\label{tanimoto}
S_{tani}(a,b) = \frac{ A \cdot B }{ ||A||^2 + ||B||^2 - A \cdot B},
\end{align}

Schultz~\cite{Schultz99_topic} adapted the well-known TF-IDF query-document similarity measure to produce a term-weighted document-document similarity measure.  Here, $A(t)$ is the frequency of term $t$ for object $a$, and $idf(t)$ is the inverse document frequency for term $t$.  More generally, it is the significance or importance of term $t$ appearing in a document.
\begin{align}
\label{schultz99}
S_{wcos}(a,b) = \frac{\sum_{t \in T} A(t)B(t)idf(t) }{ ||A||~||B|| }
\end{align}

\subsection{SimRank and Simple Generalizations}
\label{sec:simrank}

Jeh and Widom~\cite{Jeh02_simrank} realized that a more general way to attack the object similarity problem was to not only look for shared neighbors, that is, neighbors that are {\em identical}, but to look for neighbors that are {\em similar}.  This produces the recursive statement, "Two objects are similar if they are related to similar objects."~\cite{Jeh02_simrank}  Formally, their SimRank measure is defined as follows:
\begin{align}
\label{simrank}
sim_{sr}(a,b) = \frac{c}{|I(a)||I(b)|}\sum_{x \in I(a)}\sum_{y \in I(b)}sim_{sr}(x,y)
\end{align}
if $a \neq b$.  If $a = b$, then $sim_{sr}(a,b) = 1$.  $c$ is a constant $0 < c < 1$. Also, for SimRank and all its variants, if either $a$ or $b$ has no neighbors, then $sim(a,b) = 0$. SimRank can be computed iteratively by initializing the matrix of $sim(.)$ values, hereafter called the $S$ matrix, to the identity matrix.

Obviously, we can add the effects of in-neighbors and out-neighbors to produce a more comprehensive measure of the neighbor similarity between two objects.  Several authors have proposed this~\cite{lin07_extend, Zhao09_prank}.

\subsection{Improving the SimRank's Computational Performance}
\label{sec:fast_simrank}

SimRank can be described as a recursive extension of the co-citation index.  An important difference between the non-iterative algorithms in Section~\ref{sec:predecessors} and SimRank is that the earlier algorithms can be computed locally with a minimum of computational effort.  With SimRank, however, to compute the similarity of even a single pair of objects, one has to consider the entire graph.  This increases the computational requirements by a factor of $n^2k$, where $k$ is the number of iterations. Consequently, several authors~\cite{Lizorkin08_accuSimrank, jia09_swsimrank, cai09_asimrank, Li09_blocksim} have worked to reduce both the computational and memory requirements for SimRank, for general and specific applications.

\subsection{Meaningful Extensions and Alternative to SimRank}
\label{sec:alt_simrank}

In addition to concerns about the computational efficiency of the original SimRank formula, there are some structural flaws which mar its elegance.  First, SimRank scores sometimes decrease when we would ituitively expect them to increase.  Suppose we have an object-pair that has all neighbors in common.  Then $sim_{sr}(a,b) = c/d$, $d$ is the degree of $a$ or $b$.  As $d$ increases, this should means stronger ties between $a$ and $b$, but clearly $sim_{sr}$ actually decreases.

\subsubsection{SimRank++}
Antonellis et al.~\cite{Antonellis08_simrankpp} partially compensates for this unwanted decrease by inserting an {\em evidence} factor.  The more neighbors in common, the higher the evidence of similarity.  They define evidence as
\begin{align}
\label{evidence}
ev(a,b) = \sum_{i=1}^{|N(a) \cap N(b)|} \frac{1}{2^i},
\end{align}
where $N(a)$ is the undirected neighbor set of $a$.  If $a$ and $b$ have only one neighbor in common, $ev = 1/2$.  As the number of neighbors increases, $ev \rightarrow 1$. This yields to following similarity definition:
\begin{align}
\label{evidence_simrank}
sim_{ev}(a,b) = ev(a,b)\cdot c\sum_{x=1}^{N(a)}\sum_{y=1}^{N(b)}sim_{ev}(x,y)
\end{align}
The very narrow range $[0.5,1]$ of the evidence factor, however, leads to the problem that $sim_{ev}(.)$ values are no longer bounded to a maximum of 1 or even to a constant.  Instead, the maximum depends on the maximum value of $||N(a)||\cdot||N(b)||$ for the graph.
The authors make one more extension to support edge-weighted graphs.  Their final measure is called SimRank++:
\begin{align}
\label{simrankpp}
sim_{spp}(a,b) = ev(a,b)\cdot c\sum_{x=1}^{N(a)}\sum_{y=1}^{N(b)}w_{ab}w_{by}sim_{spp}(x,y)
\end{align}

\subsubsection{PSimRank}
Fogaras and R\'{a}cz~\cite{Fogaras05_scaleSim} realize that the cause of improper weighted of neighbor-matching in SimRank is due to the paired-random walk model.  Ignoring the decay constant $c$ for the moment, SimRank values are equal to the probability that two simultaneous random walkers, starting at vertices $a$ and $b$, will encounter each other eventually.  Even if $a$ and $b$ have all $N_a = N_b$ neighbors in common, the probability that the two walkers will happen to choose the same neighbor is $1/N_a$, which decreases as the degree increases.  To emend this situation, Fogaras and R\'{a}cz introduce coupled random walks.  They partition the event space into three cases:
\begin{enumerate}
\item $P_1 = P$($a$ and $b$ step to the same vertex) $= \frac{ |I(a)\cap I(b)| }{ |I(a)\cup I(b)|}$
\item $P_2 = P$($a$ steps to a vertex in $I(a)\backslash I(b)$) $= \frac{ |I(a)\backslash I(b)| }{ |I(a)\cup I(b)|}$
\item $P_3 = P$($b$ steps to a vertex in $I(b)\backslash I(a)$) $= \frac{ |I(b)\backslash I(a)| }{ |I(a)\cup I(b)|}$
\end{enumerate}
Note that case 1, which we would consider the direct similarity of $a$ and $b$, is described by the Jaccard Coefficient. As required, the sum of these probabilities equals 1.  We can then compute a similarity measure which takes the general form
\[
sim_{ps}(a,b) = \sum_{i=1}^3 P_i \cdot sim(\mbox{neighbors in Case $i$}).
\]
Noting that there are $\frac{1}{|I(a)\backslash I(b)||I(b)|}$ neighbor-pairs in Case 2 and $\frac{1}{I(b)\backslash I(a)||I(a)|}$ in Case 3, this produces the logical but somewhat unwieldly formula:
\begin{align}
\label{psimrank}
sim_{ps}&(a,b)= c\cdot [~P_1\cdot 1 \nonumber \\
+& \frac{P_2}{|I(a)\backslash I(b)||I(b)|}\sum_{\substack{x \in I(a)\backslash I(b) \\ y \in I(b)}} sim_{ps}(x,y) \nonumber \\
+& \frac{P_3}{|I(b)\backslash I(a)||I(a)|}\sum_{\substack{x^\prime \in I(b)\backslash I(a) \\ y^\prime \in I(a)}} sim_{ps}(x^\prime,y^\prime)~].
\end{align}

\subsubsection{MatchSim}
The authors of MatchSim~\cite{Lin09_matchsim} take this emendment of random walking to its limit.  They observe that when a human compares the features of two objects, a human does not select random features to see if they match. Rather, people look to see if there exists an alignment of features that produces a perfect or near-perfect matching.  Therefore, their similarity measure discards the idea of random walk and replaces it with "the average similarity of the maximal matching between their neighbors."~\cite{Lin09_matchsim}:
\begin{align}
\label{matchsim}
sim_{ms}(a,b) = \frac{\sum_{(x,y) \in m^\star_{ab}}sim_{ms}(x,y)}{max(|I(a)|,|I(b)|)},
\end{align}
where $m^\star$ represents the maximal matching.
MatchSim omits the usual decay factor $c$, but this seems to be an idealization rather than a necessary alteration.
Note that the size of the maximal matching is $min(|I(a)|,|I(b)|)$.  Without loss of generality, assume $a$ has fewer neighbors than $b$.  The upper bound for $sim_{ms}(a,b)$ occurs when every neighbor of $a$ is also a neighbor of $b$.  In this special case, $max(sim_{ms}(a,b)) = max(\frac{ min(|I(a),I(b)| }{ max(|I(a),I(b)| }) = \frac{|I(a) \cap I(b)|}{|I(a)\cup I(b)|}$, which is the Jaccard coefficient.

\subsubsection{PageSim}
All of the previous works are modifications of the original SimRank measure and principles.  We now consider two measures that are markedly different than SimRank.  We first consider PageSim, which not only borrows the entire PageRank computation as a starting point, but also borrows the meaning of PageRank's iterative computation to devise a related computation. The canonical interpretation of PageRank is that for each step, each page sends out an equal fraction of its own importance to each of its neighbors.  Its importance for the next step is the sum of the fractional importance it received from its in-neighbors.  PageSim also uses this spreading or propagating mechanism; however, rather than there being a universal importance feature which can be summed, each node begins with a distinct self-feature, which is orthogonal to every other vertex feature. The authors describe the propagation process as occurring over distinct paths, and they sum the contributions of each path to compute the total distribution.  As long as we permit self-intersecting paths, this is equivalent to measuring for each vertex is the random walk distribution after $k$ steps.
PageSim follows a multi-step procedure:
\begin{enumerate}
\item For each vertex $a$, define feature vector $FV(a)$.  $FV_b(a)$ is the $b^{th}$ element of $FV(a)$.
\item Initialize all vectors: $FV^0_a(a) = PageRank(a)$.  $FV^0_b(a) = 0, b \neq a$.
\item For $t=1$ to $k$ iterations, $FV^t = c\cdot \sum_{a \in V} \frac{FV^{t-1}(a)}{|O(a)|}$
\item Measure the similarity between pairs of feature vectors.  In their original paper~\cite{Lin06_pagesim}, the similarity measure is defined thus:

\begin{align}
\label{pagesim1}
sim_{pg1}(a,b) = \sum_{i=1}^n\frac{min(FV_i(a),FV_i(b))^2}{max(FV_i(a),FV_i(b))}
\end{align}
In an expanded work~\cite{lin07_extend}, they modify the formula to more closely resemble the Jaccard coefficient:
\begin{align}
\label{pagesim2}
sim_{pg2}(a,b) = \frac{\sum_{i=1}^n min(FV_i(a),FV_i(b))}{\sum_{i=1}^n max(FV_i(a),FV_i(b))}
\end{align}

\end{enumerate}

\begin{table*}[t]
	\large
	\centering
	\begin{tabular*}{0.95\textwidth} {p{.80in} | l} \hline
	measure					&	formula	\\ 	\hline
	\hline	
	bibliographic coupling	& $ S_{bc}(a,b) = \frac{ |O(a) \cap O(b)| }{ |O(a) \cup O(b)| }	$ \\ \hline
	co-citation				& $ S_{cc}(a,b) = \frac{ |I(a) \cap I(b)| }{ |I(a) \cup I(b)| }	$ \\ \hline
	cosine 					& $ S_{cos}(a,b) = \frac{ A \cdot B }{ ||A||~||B|| }			$ \\ \hline
	Tanimoto				& $ S_{tani}(a,b) = \frac{ A \cdot B }{ ||A||^2 + ||B||^2 - A \cdot B} $ \\ \hline
	weighted cosine			& $ S_{wcos}(a,b) = \frac{\sum_{t \in T} A(t)B(t)idf(t) }{ ||A||~||B|| } $ \\ \hline
	SimRank					& $ sim_{sr}(a,b) = \frac{c}{|I(a)|I(b)|}\sum_{x \in I(a)}\sum_{y \in I(b)}sim_{sr}(x,y) $ \\ \hline
	SimRank++	& $ sim_{spp}(a,b) = \sum_{i=1}^{|N(a) \cap N(b)|} \frac{1}{2^i}\cdot c\sum_{x=1}^{N(a)}\sum_{y=1}^{N(b)}w_{ab}w_{by}sim_{spp}(x,y) $ \\ \hline
	PSimRank	& $ sim_{ps}(a,b)=c\cdot[ \frac{ |I(a)\cap I(b)| }{ |I(a)\cup I(b)| } $ 
				 $+ \frac{ \sum_{x \in I(a)\backslash I(b), y \in I(b)} sim_{ps}(x,y) }{ |I(a)\cup I(b)||I(b)| } $ 
				 $+ \frac{ \sum_{x^\prime \in I(b)\backslash I(a), y^\prime \in I(a)} sim_{ps}(x^\prime,y^\prime) }{ |I(b)\cup I(a)||I(a)| } ] $ \\ \hline
	MatchSim				& $ sim_{ms}(a,b) = \frac{\sum_{(x,y) \in m^\star_{ab}}sim_{ms}(x,y)}{max(|I(a)|,|I(b)|)} $ \\ \hline
	PageSim~\cite{lin07_extend}	& $ sim_{pg2}(a,b) = \frac{\sum_{i=1}^n min(FV_i(a),FV_i(b))}{\sum_{i=1}^n max(FV_i(a),FV_i(b))} $ \\ \hline
	VertexSim				& $ DS_vD = \frac{c}{\lambda_1}A (DS_vD)+ I $ \\ \hline
	\end{tabular*}
	\caption{Structural Similarity Measures} \label{tab:sim_measures}
\end{table*}

\subsubsection{Vertex Similarity in Networks}
The last measure that we consider addresses the other major weakness of SimRank: it considers only equal-length paths of similarity.  As stated earlier, a SimRank value equals the probability that a given pair of vertices will meet {\em if they take steps simultaneously with the other.}  That is, it would not count a case where Walker $a$ takes 3 steps to reach $c$, and Walker $b$ takes 4 steps to reach $c$.  To address this limitation, Leicht et al.~\cite{Leicht06_vertexsim} formulate their measure from the following maxim: "Vertex $a$ is similar to $b$ if $a$ has any neighbor $c$ this is itself similar to $b$."  On one hand, this statement explicitly supports asymmetrical pairs of paths.  On the other hand, it makes a questionable leap by assuming that being neighbors implies similarity.

Coming from the network science community rather than the data mining community, the authors did not give a catchy or convenient name to their measure.  For convenience, we will call it VertexSim (notated $sim_v$ or $S_v$).  The initial version of VertexSim, written in matrix form is
\begin{align}
\label{eq:vertexsim_raw}
\bf{ S_v= \phi AS_v + I },
\end{align}
where $A$ is the adjacency matrix and $\phi$ is a parameter to be determined.  Solving for $\bf{S_v}$ and performing a power series expansion, we get
\[ \bf{S_v = I + \phi A + \phi^2 A^2 + \cdots}. \]
After normalizing for the expected number of paths from $a$ to $b$ and some simplifying approximations, they authors finally derive the following:
\begin{align}
\label{vetexsim}
\bf{ S_v = D^{-1}\left( I -\frac{c}{\lambda_1}A \right)^{-1} D^{-1} },
\end{align}
where $\lambda_1$ is the largest eigenvalue of $A$, and $D$ is the degree matrix ($d_{ii}$ = degree of vertex $i$; all other $d_{ij} = 0$).  Here we have a closed form solution, which seems convenient, but we also need to invert two matrices.  An iterative computation process being simpler, the authors rewrite the equation this way:
\begin{align}
\label{vetexsim_iter}
\bf{ DS_vD = \frac{c}{\lambda_1}A (DS_vD)+ I },
\end{align}
which we see resembles Eq.~\ref{eq:vertexsim_raw}.  The authors claim $\bf{DS_vD}$ can be initialized to any values such as $\bf{0}$ and will converge after 100 iterations or fewer.

\subsection{Summary}
We summarize the foregoing structural similarity measures in Table~\ref{tab:sim_measures}.

\end{document}